\documentclass[preprint,10pt]{elsarticle}
\usepackage{lineno}
\modulolinenumbers[5]
\usepackage{hyperref} 
\hypersetup{ 
  colorlinks   = true, %Colors links instead of ugly boxes 
  urlcolor     = blue, %Color for external hyperlinks 
  linkcolor    = blue, %Color of internal links 
  citecolor    = blue, %Color of Bibliography  citations 
} 
\usepackage[fleqn]{amsmath}
\usepackage{amsfonts}
\usepackage{amssymb}
\usepackage{amsthm}
\usepackage{pifont}
\usepackage{natbib}
\usepackage{geometry}
\usepackage{graphicx}
\usepackage{txfonts}
\usepackage{setspace}
\usepackage[utf8]{inputenc}
\usepackage{caption}
\usepackage{subcaption}
\graphicspath{ {images/} }
\usepackage[usenames, dvipsnames]{color}
\usepackage{color,soul}
\setstcolor{red}
\usepackage[dvipsnames]{xcolor}
\journal{Journal of Computational Physics}

\biboptions{sort&compress}
\begin{document}

\begin{frontmatter}

%\title{Elsevier \LaTeX\ template\tnoteref{mytitlenote}}
\title{A Momentum Balance Correction to the Non-Conservative One-Fluid Formulation in Boiling Flows using Volume-of-Fluid}

\author{Jordi Poblador-Ibanez\fnref{myfootnote1}\corref{mycorrespondingauthor}}
\ead{J.PobladorIbanez@tudelft.nl}
\author{Nicol\'{a}s Valle\fnref{myfootnote2}}
\author{Bendiks Jan Boersma\fnref{myfootnote3}}
\address{Delft University of Technology, Delft, 2628 CD, The Netherlands}
% \author{Fazle Hussain\fnref{myfootnote4}}
% \address{Texas Tech University, Lubbock, TX 79409, United States}
\fntext[myfootnote1]{Postdoctoral Researcher, Department of Maritime and Transport Technology.}
\fntext[myfootnote2]{Assistant Professor, Department of Maritime and Transport Technology, Department of Process and Energy.}
\fntext[myfootnote3]{Full Professor, Department of Maritime and Transport Technology.}
% \fntext[myfootnote4]{Professor, Department of Mechanical Engineering, Texas Tech University.}


\cortext[mycorrespondingauthor]{Corresponding author}
%\ead{poblador@uci.edu}


\begin{abstract}
A proven methodology to solve multiphase flows is based on the one-fluid formulation of the governing equations, which treats the phase transition across the interface as a single fluid with varying properties and adds additional source terms to satisfy interface jump conditions, e.g., surface tension and mass transfer. Used interchangeably in the limit of non-evaporative flows, recent literature has formalized the inconsistencies that arise in the momentum balance of the non-conservative one-fluid formulation compared to its conservative counterpart when phase change is involved. This translates into an increased sensitivity of the numerical solution to the choice of formulation. Motivated by the fact that many legacy codes using the non-conservative one-fluid formulation have been extended to phase-change simulations, the inclusion of two corrective forces at the interface and a modification of the pressure-velocity solver with an additional predictor-projection step are shown to recover the exact momentum balance in the evaporative non-conservative one-fluid framework for low-viscosity incompressible flows. This has direct implications for obtaining a physically meaningful pressure jump across the interface and is seen to affect the dynamics of two-phase flows. In the high-viscosity domain, the discretization of the viscous term introduces a momentum imbalance which is highly dependent on the chosen method to model the phase transition. In the context of film boiling, this imbalance affects the time scales for the instability growth. Lastly, the need to develop sub-models for heat and mass transfer and for surface tension becomes evident since typical grid resolutions defined as ``resolved" in the literature may not be enough to capture interfacial phenomena.
\end{abstract}


\end{frontmatter}

% \linenumbers

%%%%%%%%%%%%%%%%%%%%%%%%%%%%%%%%%%%%%%%%%%%%%%%%%%%%%%%%%%
% TO FIX MANUALLY SPACES BETWEEN TEXT AND EQN BOXES
\setlength\abovedisplayshortskip{0pt}
\setlength\belowdisplayshortskip{-5pt}
\setlength\abovedisplayskip{-5pt}
%\setlength\belowdisplayskip{0pt}
%%%%%%%%%%%%%%%%%%%%%%%%%%%%%%%%%%%%%%%%%%%%%%%%%%%%%%%%%%

\section{Introduction} 
\label{sec:intro}

Multiphase flows undergoing significant phase change are found in many engineering applications including the evaporation of liquid fuels during injection \cite{2022_JFM_Gao,2023_IJMF_Gaballa}, the boiling of coolants used in heat exchangers for thermal management \cite{2020_inventions_Giustini,2023_NED_Komen}, or the formation of hydrogen bubbles in water electrolysis \cite{2016_CJCE_Liu,2020_fluids_Taqieddin}. Regardless of the mechanisms driving the phase transition, these flows typically involve a variety of scales ranging from system sizes of \(\mathcal{O}(10^{-1})\) to \(\mathcal{O}(10^{-3})\) meters with droplets or bubbles usually in the micro-scale, i.e., \(\mathcal{O}(10^{-5})\) to \(\mathcal{O}(10^{-6})\) meters. This multiscale nature is demanding for both experimental and numerical studies. Nonetheless, numerical computations have become a key tool to improve the physical understanding of relevant multiphase flows as shown by the aforementioned works. \par 

The development of accurate numerical tools for multiphase flows is a challenging on-going field of study \cite{2019_ARFM_Elghobashi}. A common approach is to use the so-called one-fluid formulation to solve the Navier-Stokes equations. That is, the two phases are treated as a single fluid whose properties vary across the interface weighted by some function, e.g., smoothed Heaviside or volume average, and which is driven by a single velocity field. The interface between both phases is still considered and the jump conditions across it are satisfied by including additional source terms in the governing equations only active at the interface by means of a Dirac delta function \(\delta_\Gamma\) \cite{2013_JCP_Sato,2014_JCP_Dodd,2019_JCP_Palmore,2020_JCP_Scapin,2021_IJHMT_Dodd,2021_JCP_Malan,2021_IJMF_Trujillo,2022_CES_Gennari,2022_PoF_Poblador,2023_CaF_Boyd}. \par

Conservative formulations of the one-fluid momentum equation, coupled to momentum-consistent advection schemes, successfully deal with flows that traditionally induced numerical difficulties (e.g., high density ratio flows) \cite{2019_JCP_Palmore,2021_CaF_Arrufat,2023_CaF_Boyd}. Yet, non-conservative formulations of the one-fluid momentum equation are commonly used in the literature. Mainly out of numerical convenience, they have been extended to incompressible two-phase flows undergoing phase change \cite{2013_JCP_Sato,2020_JCP_Scapin,2021_IJHMT_Dodd,2021_JCP_Malan,2021_IJMF_Trujillo,2022_CES_Gennari}. Although mathematically equivalent to the conservative form in the continuous phase, the analytical derivation of the jump conditions for the non-conservative formulation differs from the exact jumps obtained with the conservative form when phase change occurs and, therefore, are physically inconsistent across the interface \cite{2021_IJMF_Trujillo}. In other words, numerical integration of the non-conservative forms leads to an ill-defined momentum balance expressed in terms of a non-physical pressure field. \par

This inconsistency may seem negligible for practical purposes when convective speeds are large compared to the Stefan flow or when surface tension dominates over the momentum jump caused by the change of phase; thus, it is usually ignored. However, it may become important in, e.g., pressurized systems with fluids undergoing phase change close to the critical point as the surface tension coefficient and latent heat of vaporization drop \cite{propertiesofgasesandliquids}. These configurations may include droplet-laden or bubbly flows. Despite the correct velocity jump may be recovered, the momentum imbalance leads to a pressure solution that compromises the dynamical evolution of the flow and analyses based on pressure data, such as determining turbulence statistics (i.e., pressure fluctuations), studying the vorticity dynamics (i.e., calculating the baroclinic torque), or calculating forces around bodies (e.g., drag force on a droplet). These issues challenge the use of the one-fluid framework in such cases. \par

This work proposes a correction for the non-conservative form of the one-fluid momentum equation in terms of additional body forces in the context of the Continuum Surface Force (CSF) model \cite{1992_JCP_Brackbill} and a modified predictor-projection method to recover the correct momentum balance when phase change occurs and, consequently, obtain a physically meaningful pressure jump. Moreover, this paper is also intended as a reflective exercise to highlight how the choice of numerical methods in two-phase solvers with phase change drastically affects the numerical solution. \par

The paper is structured as follows: Section \ref{sec:math} introduces the governing equations for two-phase flows undergoing phase change and the basis to define new body forces to correct the momentum imbalance in non-conservative formulations; Section \ref{sec:flow_solver} describes the discretization of the governing equations, the flow solver and its implementation; Section \ref{sec:validation} validates the momentum balance corrections in canonical flows with constant vaporization flux; Section \ref{sec:val_withenergy} extends the validation to fully-coupled systems involving energy transport; and Section \ref{sec:concl} summarizes the major findings and contributions of this work. \par

\section{Mathematical Description}
\label{sec:math}

\subsection{Governing Equations}
\label{subsec:goveqs}

The governing equations for the motion and evaporation (or condensation) of a single-component fluid are the Navier-Stokes equations, given by the mass conservation or continuity equation (\ref{eqn:cont}), the momentum equation (\ref{eqn:mom}), and the energy transport equation (\ref{eqn:ene}). The variables \(\rho\), \(\mathbf{u}=(u,v,w)\), \(p\), \(\mu\), and \(\mathbf{g}\) are, respectively, the density, velocity, pressure, dynamic viscosity and gravity, while \(h\), \(k\), and \(T\) are the specific enthalpy, thermal conductivity and temperature. A Newtonian fluid is assumed under Stokes hypothesis; thus, the viscous stress tensor is \(\mathbf{\tau}=\mu\big[\nabla\mathbf{u}+(\nabla\mathbf{u})^T -\tfrac{2}{3}(\nabla\cdot\mathbf{u})\mathbf{I}\big]\), where \(\mathbf{I}\) is the identity tensor. \par

\begin{equation}
\label{eqn:cont}
\frac{\partial \rho}{\partial t} + \nabla\cdot(\rho\mathbf{u})=0
\end{equation}

\begin{equation}
\label{eqn:mom}
\frac{\partial \rho \mathbf{u}}{\partial t} + \nabla\cdot(\rho\mathbf{u}\mathbf{u})=-\nabla p + \nabla\cdot\bigg(\mu\big[\nabla\mathbf{u}+(\nabla\mathbf{u})^T -\tfrac{2}{3}(\nabla\cdot\mathbf{u})\mathbf{I}\big]\bigg) + \rho\mathbf{g}
\end{equation}

\begin{equation}
\label{eqn:ene}
\frac{\partial \rho h}{\partial t} + \nabla\cdot(\rho h\mathbf{u})= \nabla\cdot(k\nabla T)
\end{equation}

Convective speeds tend to be small in boiling flows or droplet-laden flows in most configurations of interest \cite{2021_IJHMT_Dodd,2021_IJMF_Nemati} except in, e.g., jet atomization \cite{2022_JFM_Gao,2023_IJMF_Gaballa}. Accordingly, the pressure term and viscous dissipation term that would appear in the energy equation are neglected. Further, the flow velocities approach the incompressible limit. Henceforth, the mathematical formulation and numerical approach are simplified to incompressible flows, except for the local volume dilatation at the vaporizing interface enforced in the continuity equation. However, many practical applications involve high pressures, large temperature variations and multi-component mixtures. In such environments, fluid properties may vary strongly and a low-Mach number formulation is more suitable. The framework outlined in this work can easily be extended to such scenarios, similar to previous studies \cite{2021_IJHMT_Dodd,2022_PoF_Poblador}. \par

Eqs. (\ref{eqn:cont})-(\ref{eqn:ene}) are valid within each phase, while jump conditions link both phases across the liquid-gas interface, herein denoted by \(\Gamma\), based on conservation relations. Denoting the interface normal unit vector as \(\mathbf{n}_\Gamma\), which is defined from liquid to gas, the jump conditions normal to the interface for Eqs. (\ref{eqn:cont})-(\ref{eqn:ene}) are given, respectively, by \cite{2021_IJMF_Trujillo}

\begin{equation}
\label{eqn:cont_jump}
\dot{m}''=\rho_L(\mathbf{u}_L-\mathbf{u}_\Gamma)\cdot\mathbf{n}_\Gamma=\rho_G(\mathbf{u}_G-\mathbf{u}_\Gamma)\cdot\mathbf{n}_\Gamma
\end{equation}

\begin{equation}
\label{eqn:mom_jump}
[\rho_G\mathbf{u}_G(\mathbf{u}_G-\mathbf{u}_\Gamma)-\rho_L\mathbf{u}_L(\mathbf{u}_L-\mathbf{u}_\Gamma)]\cdot\mathbf{n}_\Gamma=-(p_G-p_L)\mathbf{n}_\Gamma + [\mathbf{\tau}_G-\mathbf{\tau}_L]\cdot\mathbf{n}_\Gamma - \sigma\kappa\mathbf{n}_\Gamma
\end{equation}

\begin{equation}
\label{eqn:ene_jump}
[\rho_G h_G(\mathbf{u}_G-\mathbf{u}_\Gamma)-\rho_L h_L(\mathbf{u}_L-\mathbf{u}_\Gamma)]\cdot\mathbf{n}_\Gamma= [k_G\nabla T_G-k_L\nabla T_L]\cdot\mathbf{n}_\Gamma
\end{equation}
\noindent
where \(\dot{m}''\) is the mass flux per unit area across the interface (\(\dot{m}''>0\) for vaporization), \(\mathbf{u}_\Gamma = \mathbf{u}_G-\tfrac{\dot{m}''}{\rho_G}\mathbf{n}_\Gamma = \mathbf{u}_L-\tfrac{\dot{m}''}{\rho_L}\mathbf{n}_\Gamma \) the interface velocity, \(\sigma\) the surface tension coefficient (assumed constant) and \(\kappa=\nabla\cdot\mathbf{n}_\Gamma\) the interface curvature. Note the subscripts \(L\) and \(G\) indicate liquid and gas values at the interface. Further, \(L\) and \(G\) are also used to refer to the fluid properties in each single-phase region since they are assumed constant for the purpose of this work. Tangent to the interface, the velocity and shear stress are continuous. Rearranging Eq. (\ref{eqn:cont_jump}), the velocity jump is given by \((\mathbf{u}_G-\mathbf{u}_L)\cdot\mathbf{n}_\Gamma=\dot{m}''(\rho_G^{-1}-\rho_L^{-1})\).
Then, substitution of Eq. (\ref{eqn:cont_jump}) into Eqs. (\ref{eqn:mom_jump}) and (\ref{eqn:ene_jump}) results in an equation for the pressure jump across the interface and an equation for the mass flux, given by

\begin{equation}
\label{eqn:mom_jump2}
p_L-p_G=\sigma\kappa + ([\mathbf{\tau}_L-\mathbf{\tau}_G]\cdot\mathbf{n}_\Gamma)\cdot\mathbf{n}_\Gamma+(\dot{m}'')^2(\rho_G^{-1}-\rho_L^{-1})
\end{equation}

\begin{equation}
\label{eqn:ene_jump2}
\dot{m}''=\frac{[k_G\nabla T_G-k_L\nabla T_L]\cdot\mathbf{n}_\Gamma}{h_{LV}}
\end{equation}

\noindent
with \(h_{LV}=h_G-h_L\) being the latent heat of vaporization. \par

\subsection{Interface Capture with Volume-of-Fluid}
\label{subsec:intcap}

The interface between liquid and gas is captured using the Volume-of-Fluid (VOF) method \cite{1981_JCP_Hirt}. Despite other interface capturing techniques could be considered (e.g., Level-Set \cite{1994_JCP_Sussman,1998_CaF_Sussman,2001_JCP_Osher} or Phase-Field \cite{2022_JCP_Jain}), VOF provides a good balance between the physical representation of the interface (e.g., sharpness, evaluation of the thermodynamic state) and numerical behavior (e.g., mass conservation to machine error with a divergence-free velocity field \cite{2014_CaF_Baraldi}). In the VOF method, an indicator function \(X(\mathbf{x},t)\) identifies the reference phase (i.e., the liquid) with \(X(\mathbf{x},t)=1\) and the gas phase with \(X(\mathbf{x},t)=0\). Accordingly, a volume fraction is defined in each cell as \(C=\tfrac{1}{V_0}\int_{V_0} XdV\), where \(V_0\) is the cell volume. \par

A geometrical VOF framework where the interface is reconstructed at each cell is considered. A Piecewise Linear Interface Construction (PLIC) is implemented where \(\mathbf{n}_\Gamma\) and \(\kappa\) are obtained, respectively, from the Mixed-Youngs-Center (MYC) method \cite{2007_JCP_Aulisa} and the Height-Function (HF) method \cite{2010_JCP_Lopez} following \cite{2014_CaF_Baraldi}. Analytical relations are used to calculate the cell volume cut by the reconstruction plane \cite{2000_JCP_Scardovelli}. Further, the Interface Reconstruction Library\footnote[1]{The Interface Reconstruction Library, R. Chiodi and F. Evrard, \url{https://github.com/robert-chiodi/interface-reconstruction-library/tree/paraboloid_cutting}} \cite{2022_JCP_Chiodi,2023_SIAM_Evrard} is included to enhance the calculations of geometric fluxes and interface curvature by fitting a paraboloid to the PLIC reconstruction, i.e., via a Piecewise Paraboloid Interface Construction (PPIC) \cite{2019_CaMA_Jibben}. Although the HF method is more robust in static tests, curvatures obtained from PPIC do not deteriorate as quickly under interface displacement due to the combined effect of reduced geometrical errors during advection and the derivation of curvature directly from the paraboloid itself. Therefore, surface-tension related spurious currents are reduced compared to PLIC \cite{2024_JCP_Han}. In this work, the PPIC enhancements are used but default back to PLIC wherever the paraboloid fitting is locally ill-defined. Note that PPIC is about six times as costly as PLIC \cite{2023_SIAM_Evrard}. \par 

The indicator function \(X(\mathbf{x},t)\) is transported with the material interface (i.e., \(DX/Dt=0\)), which in the case of \(\dot{m}''=0\) is advected with the fluid. The discontinuous velocity when \(\dot{m}''\neq0\) requires some rethinking of the advection approach to obtain consistent interface displacements. Some works advect the interface directly with \(\mathbf{u}_\Gamma\) \cite{2019_JCP_Palmore,2020_JCP_Scapin,2023_hal_Cipriano}. In contrast, the transport equation for \(X(\mathbf{x},t)\) is defined here by Eq. (\ref{eqn:VOFindicator}) using a liquid velocity \(\mathbf{u}_L\) and a source term to account for the liquid volume subtracted (or added) due to the vaporizing (condensing) interface. This term is described by a mass flux per unit volume \(\dot{m}'''=\dot{m}''\delta_\Gamma\), active only at interface cells \cite{2021_JCP_Malan,2022_PoF_Poblador,2023_CaF_Boyd}. \(\mathbf{u}_L\) is a divergence-free velocity field representative of the liquid phase obtained by subtracting the Stefan flow from \(\mathbf{u}\) by solving a Poisson equation efficiently using an FFT-based solver \cite{2020_JCP_Scapin}. Alternatively, one may use an iterative solver to solve the Poisson equation only in a narrow band around the interface \cite{2021_JCP_Malan}. The use of \(\mathbf{u}_L\) and \(\rho_L\) in Eq. (\ref{eqn:VOFindicator}) is applicable to droplet-laden flows, whereas the subtraction of the Stefan flow results in a divergence-free gas phase velocity \(\mathbf{u}_G\) in bubbly flows \cite{2020_JCP_Scapin}. Consequently, \(\mathbf{u}_G\) and \(\rho_G\) replace \(\mathbf{u}_L\) and \(\rho_L\) in Eq. (\ref{eqn:VOFindicator}). \par

\begin{equation}
\label{eqn:VOFindicator}
\frac{\partial X}{\partial t} + \nabla\cdot(X\mathbf{u}_L)-X\nabla\cdot\mathbf{u}_L=-\frac{\dot{m}'''}{\rho_L}
\end{equation}

The geometric advection of \(X\) (i.e., \(C\)) is performed in two steps similar to \cite{2021_JCP_Malan,2023_CaF_Boyd}. First, an intermediate volume fraction field is obtained by shifting the interface plane along the direction of \(\mathbf{n}_\Gamma\) to account for the phase change source term in Eq. (\ref{eqn:VOFindicator}) (see \cite{2021_JCP_Malan}). Overshoots and undershoots may appear, which are redistributed to acceptable neighboring cells \cite{2022_PoF_Poblador}. Next, a conservative three-step split advection is used \cite{2000_JCP_Sussman}. Conservation is only ensured for a divergence-free velocity, i.e., \(\nabla\cdot\mathbf{u}_L=0\). Note \(X\nabla\cdot\mathbf{u}_L\) in Eq. (\ref{eqn:VOFindicator}) is not removed since it must be included during the directional split advection steps. However, the inclusion of a phase change source term has implications for mass conservation (see Section \ref{subsubsec:staticdrop_error}). \par

\subsection{Non-Conservative One-Fluid Formulation of the Navier-Stokes Equations}
\label{subsec:onefluid}

The non-conservative one-fluid formulation of the Navier-Stokes equations are given by

\begin{equation}
\label{eqn:cont_1f}
\nabla\cdot\mathbf{u}=\dot{m}'''(\rho_G^{-1}-\rho_L^{-1})=\dot{m}''(\rho_G^{-1}-\rho_L^{-1})\delta_\Gamma
\end{equation}

\begin{equation}
\label{eqn:mom_1f}
\rho\frac{\partial \mathbf{u}}{\partial t} + \rho(\mathbf{u}\cdot\nabla\mathbf{u})=-\nabla p + \nabla\cdot\bigg(\mu\big[\nabla\mathbf{u}+(\nabla\mathbf{u})^T\big]\bigg) + \rho\mathbf{g} + \mathbf{f}_\sigma
\end{equation}

\noindent
where the continuity equation is replaced by Eq. (\ref{eqn:cont_1f}) and the momentum equation in its non-conservative form by Eq. (\ref{eqn:mom_1f}). Note that despite \(\nabla\cdot\mathbf{u}\neq 0\) at the interface, the divergence is removed from the viscous term \cite{2020_JCP_Scapin,2021_JCP_Malan,2023_CaF_Boyd}. \(\mathbf{f}_\sigma=- \sigma\kappa\mathbf{n}_\Gamma\delta_\Gamma\) is a localized body force to include the effects of surface tension. In the context of VOF, any generic fluid property, \(\phi\), is locally weighted by the cell's volume fraction as \(\phi=C\phi_l + (1-C)\phi_g\), with \(\phi\) being, e.g., the density \(\rho\) or viscosity \(\mu\). \par 

A methodology to properly derive jump conditions in one-fluid formulations is given by \cite{2021_IJMF_Trujillo}. The momentum jump across the interface resulting from Eq. (\ref{eqn:mom_1f}) \cite{2021_IJMF_Trujillo} is 

\begin{equation}
\label{eqn:mom_1f_jump}
\rho(\mathbf{u}_L-\mathbf{u}_G)\mathbf{u}_\Gamma\cdot\mathbf{n}_\Gamma+\rho\mathbf{u}\cdot(\mathbf{u}_G-\mathbf{u}_L)\mathbf{n}_\Gamma=-(p_G-p_L)\mathbf{n}_\Gamma + [\mathbf{\tau}_G-\mathbf{\tau}_L]\cdot\mathbf{n}_\Gamma - \sigma\kappa\mathbf{n}_\Gamma
\end{equation}

\noindent
where \(\rho\) and \(\mathbf{u}\) are the one-fluid density and velocity, respectively. The terms on the left-hand side (LHS) of Eq. (\ref{eqn:mom_1f_jump}) differ from Eq. (\ref{eqn:mom_jump}), demonstrating the ill-defined momentum balance introduced by the numerical integration of Eq. (\ref{eqn:mom_1f}) if \(\dot{m}''\neq 0\). Note that the analysis presented in \cite{2021_IJMF_Trujillo} is analytical in nature and does not include issues that may arise from a given discretization approach. Similarly, a different momentum balance is obtained if the non-linear term \(\mathbf{u}\cdot\nabla\mathbf{u}\) is replaced with \(\nabla\cdot(\mathbf{uu})-\mathbf{u}(\nabla\cdot\mathbf{u})\) (\(\equiv\nabla\cdot(\mathbf{uu})\) in an incompressible flow), which also does not satisfy the form of Eq. (\ref{eqn:mom_jump}). Still, the same steps outlined in the following lines to correct the momentum balance for the non-conservative form could be implemented to find the correction terms of this other form of the momentum equation. \par

The imbalance is addressed by aiming at the replacement of the LHS of Eq. (\ref{eqn:mom_1f_jump}) with that of Eq. (\ref{eqn:mom_jump}). The first term on the LHS of Eq. (\ref{eqn:mom_1f_jump}), arising from the time derivative of Eq. (\ref{eqn:mom_1f}) due to the displacement of the interface, is taken care of during the predictor-projection method in the flow solver (see Section \ref{subsec:navierstokes_solve}). Then, two body forces similar to \(\mathbf{f}_\sigma\) are added to the right-hand side (RHS) of Eq. (\ref{eqn:mom_1f}): (1) \(\mathbf{f_{\text{NC}}}\) to cancel the momentum jump error introduced by the convective term in the non-conservative (NC) formulation, i.e., the second term on the LHS of Eq. (\ref{eqn:mom_1f_jump}); and (2) \(\mathbf{f}_{\dot{m}''}\) to impose the exact momentum jump induced by phase change, i.e., the LHS of Eq. (\ref{eqn:mom_jump}), similar to \cite{2021_IJHMT_Dodd}. These are given by

\begin{equation}
\label{eqn:bodyforce2}
\mathbf{f_{\text{NC}}} = \rho\dot{m}''(\rho_G^{-1}-\rho_L^{-1})(\mathbf{u}\cdot\mathbf{n}_\Gamma)\mathbf{n}_\Gamma\delta_\Gamma
\end{equation}

\begin{equation}
\label{eqn:bodyforce1}
\mathbf{f}_{\dot{m}''} = -(\dot{m}'')^2(\rho_G^{-1}-\rho_L^{-1})\mathbf{n}_\Gamma\delta_\Gamma
\end{equation}

\noindent
where \(\mathbf{f_{\text{NC}}}\) is obtained by rewriting \(\rho\mathbf{u}\cdot(\mathbf{u}_G-\mathbf{u}_L)\mathbf{n}_\Gamma\) using the relation \((\mathbf{u}_G-\mathbf{u}_L)=\dot{m}''(\rho_G^{-1}-\rho_L^{-1})\mathbf{n}_\Gamma\). Thus, Eq. (\ref{eqn:mom_1f}) becomes 

\begin{equation}
\label{eqn:mom_1f_full}
\rho\frac{\partial \mathbf{u}}{\partial t} + \rho(\mathbf{u}\cdot\nabla\mathbf{u})=-\nabla p + \nabla\cdot\bigg(\mu\big[\nabla\mathbf{u}+(\nabla\mathbf{u})^T\big]\bigg) + \rho\mathbf{g} + \mathbf{f}_\sigma + \mathbf{f}_{\dot{m}''} + \mathbf{f_{\text{NC}}}
\end{equation}

If the convective term is written as \(\nabla\cdot(\mathbf{uu})\), \(\mathbf{f}_{\dot{m}''}\) is still identical but not \(\mathbf{f_{\text{NC}}}\). For the term \(\nabla\cdot(\mathbf{uu})\), the jump across \(\Gamma\) is given by \(\rho(\mathbf{u}_G\mathbf{u}_G-\mathbf{u}_L\mathbf{u}_L)\cdot\mathbf{n}_\Gamma\) \cite{2021_IJMF_Trujillo}, which can be used to derive the necessary body force that cancels the momentum balance error at the interface, similar to \(\mathbf{f_{\text{NC}}}\). A validation of whether the correction is achieved at a discrete level needs to be performed, which is show in Section \ref{subsec:1D_analysis} for the non-conservative form. Note that the imbalance caused by the temporal term across the interface is not corrected by means of a body force, i.e., it is not included in \(\mathbf{f_{\text{NC}}}\). A correction at a discrete level using such approach is not straightforward given the tight coupling with the predictor-projection method. \par

In this work, it is assumed that the pressure jump due to viscous stresses is small compared to the effects of surface tension or phase change. Thus, regularization of viscous forces is enough to recover the effects of the viscous term \cite{2015_JCP_Lalanne}. From a numerical standpoint, the jump introduced by the one-fluid velocity in the discretization of \(\nabla\mathbf{u}\) in the viscous term, which scales with the inverse of the mesh size, is also deemed negligible. Nonetheless, this assumption merits further analysis and the proposed methods could be extended to include viscosity effects explicitly in the momentum jump, if necessary. This numerical issue is independent of the form of the momentum equation used in the one-fluid framework. The problematic discretization of the viscous term dominates the discussion in Section \ref{subsec:2dfilmboiling}. \par 

The aforementioned inconsistencies could be mitigated by using phase-wise velocities (i.e., \(\mathbf{u}_l\) and \(\mathbf{u}_g\)) and recovering the exact momentum jump by means of body forces, e.g., Eq. (\ref{eqn:bodyforce1}), or by implementing a Ghost Fluid Method (GFM) \cite{2018_JCP_Anumolu}. However, precise control of the numerically induced jump may be lost with the first approach and consistent extrapolation of variables across the interface or the implementation of GFM are prone to increased numerical instabilities due to the non-smooth \(\mathbf{n}_\Gamma\) distributions in the context of VOF. Thus, the one-fluid formulation with corrective terms may be preferable while seeking a final outcome similar to GFM. In contrast, GFM may be more suited for diffuse methods such as Level-Set or Phase-Field. \par

Possible inconsistencies must be carefully addressed in conservative formulations as well. Despite the analytical derivation of the jump conditions result in the exact momentum balance given by Eq. (\ref{eqn:mom_jump}), the numerical approach can introduce issues. For example, during temporal integration or in the implementation of momentum-consistent advection schemes when dealing with a sharp velocity jump across the interface. \par

\subsection{Phase-Wise Modeling of the Energy Equation}
\label{subsec:phasewise_energy} 

The phase-wise energy equation given in terms of temperature in its non-conservative form is

\begin{equation}
\label{eqn:ene_pw}
\rho_f c_{p,f}\bigg(\frac{\partial T}{\partial t} + \mathbf{u}_f\cdot\nabla T\bigg)= \nabla\cdot(k_f\nabla T)
\end{equation}

\noindent
where \(c_p\) is the isobaric specific heat. Here, phase-wise fluid properties and velocities are used and are denoted by the subscript \(f\), which switches phase depending on the value of \(C\) (i.e., \(f=g,G\) if \(C<0.5\) and \(f=l,L\) if \(C\geq0.5\)). The details on the evaluation of phase-wise velocities are given in \ref{apn:A}. \par

Instead of pursuing a one-fluid formulation for the energy equation which includes additional source terms \cite{2020_JCP_Scapin,2021_IJHMT_Dodd,2021_JCP_Malan,2021_IJMF_Trujillo,2023_hal_Cipriano}, Eq. (\ref{eqn:ene_pw}) is solved separately in a phase-wise manner in each phase with the interface acting as a moving boundary \cite{2013_JCP_Sato,2018_JCP_Anumolu,2022_PoF_Poblador,2023_CaF_Boyd}. Both options are commonly used in the literature, but in the latter the thermodynamic equilibrium at the interface defines an interface temperature (e.g., saturation temperature), which is embedded in the discretization of convective and diffusive terms. Generally, the interface temperature can vary depending on the local equilibrium conditions. In this work, the analysis is limited to saturated conditions with a uniform interface temperature. The interface state is then implicitly accounted for; thus, no additional source terms are required. The phase-wise approach allows for a sharp treatment of the energy jump condition and a reduction of the smearing of the temperature field compared to the one-fluid formulation. Albeit out of scope, a similar strategy can be used to solve the species transport equations in multi-component multiphase flows. Note that a one-fluid formulation of the energy equation could be developed similar to the proposed momentum equation to ensure that the energy balance is satisfied \cite{2021_IJMF_Trujillo}. However, the sharpness of the phase-wise approach has been favored in this work. \par

\section{Flow Solver Algorithm}
\label{sec:flow_solver}

The details of the algorithm used to solve the multiphase problem described in Section \ref{sec:math} are presented in this section in the corresponding step order. It is inferred that the workflow described below starts from the initial condition or solution at the previous time step \(n\). As a summary, the steps in each iteration are:

\begin{enumerate}
    \item Obtain phase-wise velocities \(\mathbf{u}_{l}^{n}\) and \(\mathbf{u}_{g}^{n}\) for energy transport equation (see \ref{apn:A})
    \item Solve energy transport equation, Eq. (\ref{eqn:energy_integration}), and obtain \(T^{n+1}\)
    \item Calculate divergence-free velocity \(\mathbf{u}_{L}^{n}\) or \(\mathbf{u}_{G}^{n}\) from \(\mathbf{u}^n\) for interface advection (see Section \ref{subsec:intcap})
    \item Solve VOF transport equation and obtain \(C^{n+1}\) (see Section \ref{subsec:intcap})
    \item Update fluid properties based on \(C^{n+1}\)
    \item Calculate new \(\dot{m}''\) based on \(T^{n+1}\) and \(C^{n+1}\)
    \item Calculate \(\mathbf{f}_\sigma\) and \(\mathbf{f}_{\dot{m}''}\) based on new \(\dot{m}''\) and \(C^{n+1}\)
    \item Solve first predictor-projection step to shift Stefan flow and obtain \(\mathbf{u}^*\) (Eqs. (\ref{eqn:predictor_step1}) and (\ref{eqn:predictor_step1PPE}))
    \item Calculate \(\mathbf{f}_{NC}\) based on new \(\dot{m}''\), \(C^{n+1}\) and \(\mathbf{u}^*\)
    \item Solve second predictor step to obtain \(\mathbf{u}^{**}\) (Eq. (\ref{eqn:predictor_step2})), i.e., momentum equation
    \item Solve pressure Poisson equation to obtain \(p^{n+1}\) (Eq. (\ref{eqn:projection_step2PPE})) and \(\mathbf{u}^{n+1}\) (Eq. (\ref{eqn:projection_step1VEL}))    
    \item Calculate time step \(\Delta t\) for next iteration
\end{enumerate}

\begin{figure}
\centering
\includegraphics[width=0.6\linewidth]{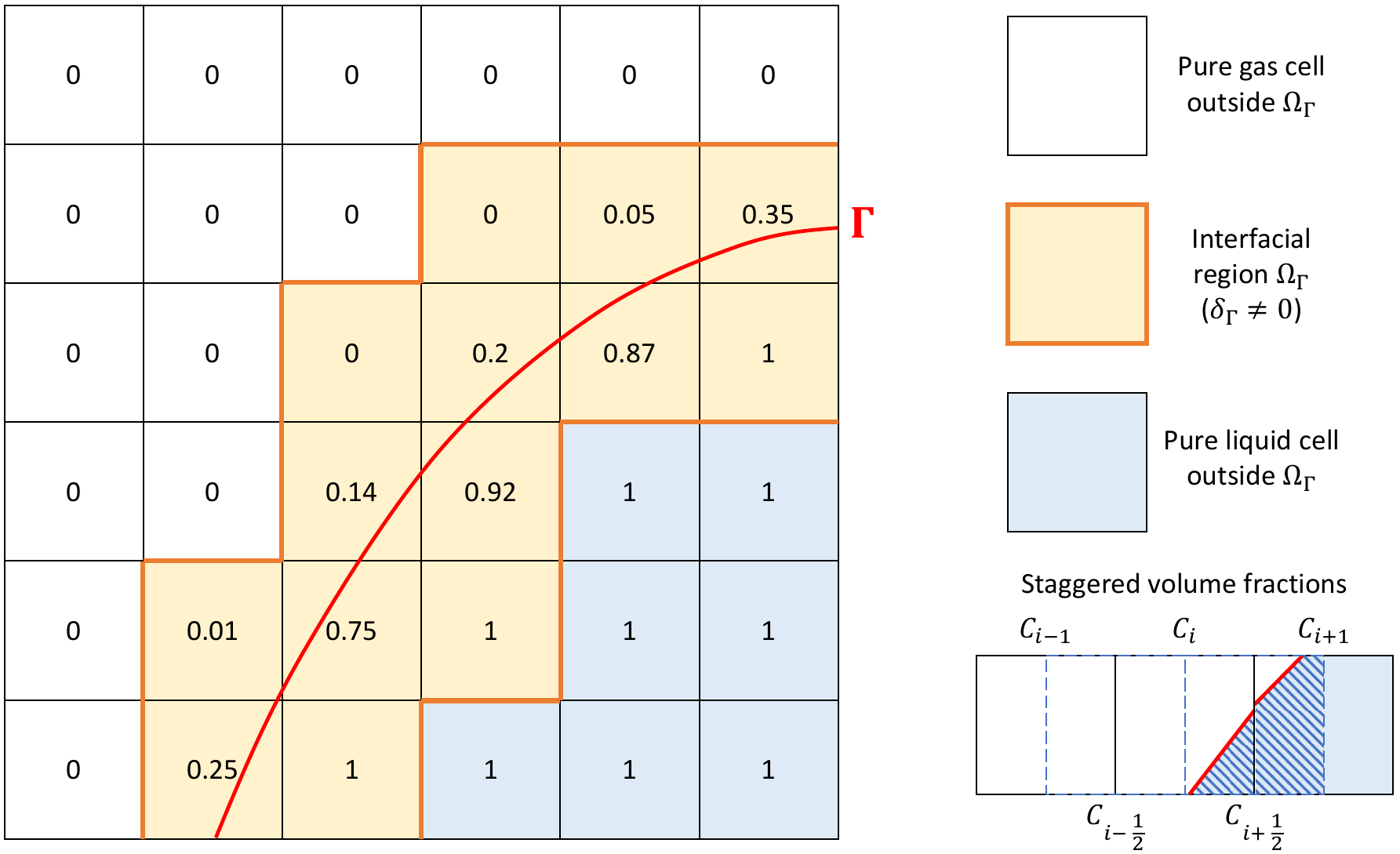}
\caption{Definition of the interfacial region \(\Omega_\Gamma\) in terms of \(\delta_\Gamma=||\nabla C||\neq 0\) and calculation of staggered volume fractions from the interface reconstruction (e.g., PLIC). A volume fraction value is shown in each cell.}
\label{fig:interfacial_region}
\end{figure}

For the discretization of the governing equations, a Cartesian uniform mesh is adopted with a staggered configuration for the velocity components \cite{1965_PoF_Harlow}. That is, scalar variables such as pressure, density or volume fraction are stored in cell centers, whereas the velocity components are stored at the cell faces. The notation in this work follows, e.g., \(p_i\), \(p_{i+1}\) and \(p_{i+2}\) for variables at cell centers and \(u_{i-\frac{1}{2}}\), \(u_{i+\frac{1}{2}}\) and \(u_{i+\frac{3}{2}}\) for variables at cell faces. A sketch is shown in Figure \ref{fig:interfacial_region} highlighting the definition of an interfacial region \(\Omega_\Gamma\) in cells where \(\delta_\Gamma\neq 0\) (see Section \ref{subsec:navierstokes_solve} for more details on the calculation of \(\delta_\Gamma\)) and the definition of staggered volume fractions, both cornerstones of the proposed flow solver. Lastly, the time step \(\Delta t\) is determined by a CFL condition for multiphase flows \cite{2000_JSC_Kang}, described in \ref{apn:C}. \par

\subsection{Solution of the Energy Equation}
\label{subsec:energy_solve}

The phase-wise energy equation, Eq. (\ref{eqn:ene_pw}), is solved using the Adams-Bashforth (AB) explicit temporal integration scheme with variable time stepping \cite{2020_JCP_Scapin} as

\begin{equation}
\label{eqn:energy_integration}
\frac{T^{n+1}-T^n}{\Delta t^{n+1}} = \Bigg(1+\frac{1}{2}\frac{\Delta t^{n+1}}{\Delta t^n}\Bigg)\text{RT}^n - \Bigg(\frac{1}{2}\frac{\Delta t^{n+1}}{\Delta t^n}\Bigg)\text{RT}^{n-1}
\end{equation}

\noindent
where \(\Delta t^{n+1}=t^{n+1}-t^n\) and \(\Delta t^{n}=t^{n}-t^{n-1}\). The term \(\text{RT}^n\) is given by

\begin{equation}
\label{eqn:energy_RT}
\text{RT}^n = -\mathbf{u}_{f}^{n}\cdot\nabla T^n + \frac{1}{\rho_{f}^{n}c_{p,f}^n}\nabla\cdot\big(k_{f}^{n}\nabla T^n\big)
\end{equation}

AB is used everywhere except in the interfacial cells and cells neighboring the interface that define \(\Omega_\Gamma\). There, a first-order forward Euler explicit scheme is used. Note that the use of variable time stepping in an AB scheme may deviate from formal second order integration in time. However, consecutive time steps are expected to be very similar and the method provides flexibility to adjust \(\Delta t\) accordingly if, e.g., flow velocities change over time or, in the case of momentum, surface tension increases locally. \par

Outside \(\Omega_\Gamma\), the convective term is discretized with a third-order WENO scheme \cite{2000_SIAM_Jiang} and the diffusive term with second-order central differences. Inside \(\Omega_\Gamma\), the proximity of the interface is accounted for by embedding the interface solution into the numerical stencils used to discretize Eq. (\ref{eqn:energy_RT}) whenever necessary. Moreover, a first-order upwind scheme is used for the convective term. Details on the interface embedding in the discretization are provided in \ref{apn:B}. \par

\subsection{Calculation of Interfacial Mass Flux}
\label{subsec:normalprobe}

Different approaches have been implemented in the literature to obtain \(\dot{m}''\) from Eq. (\ref{eqn:ene_jump2}), focusing on the calculation of the temperature gradients in each phase, e.g., \cite{2013_JCP_Sato,2018_JCP_Anumolu,2019_JCP_Palmore,2021_IJHMT_Bures,2021_JCP_Malan,2022_JFM_Gao,2023_CaF_Boyd,2023_hal_Cipriano}. Here, a normal stencil or probe is extended from the interface plane centroid into each phase along \(\mathbf{n}_\Gamma\) \cite{1999_JCP_Udaykumar,2015_IUTAM_Tryggvason,2022_PoF_Poblador}. Then, scalar quantities from the grid nodes, i.e., temperature, are linearly interpolated on two probing points on each side. To avoid using values from the opposite phase during interpolation, the first point has to be placed sufficiently far away from the interface. One can determine that the minimum distance to avoid this conflict is the cell diagonal (i.e., \(\sqrt{2}\Delta x\) in two dimensions and \(\sqrt{3}\Delta x\) in three dimensions). Consequently, the first point is placed at a distance of \(\Delta x^{(1)}=1.75\Delta x\) from the interface. Additionally, a second point is placed at \(\Delta x^{(2)}=\Delta x\) from the first point (i.e., \(\Delta x^{(1)}+\Delta x^{(2)}=2.75\Delta x\)). Although this is not problematic in the problems analyzed in this work, one may shorten the stencil and use a single point if it extends into a conflicting interface, e.g., two bubbles coming very close to each other. \par

\begin{figure}
\centering
\includegraphics[width=0.4\linewidth]{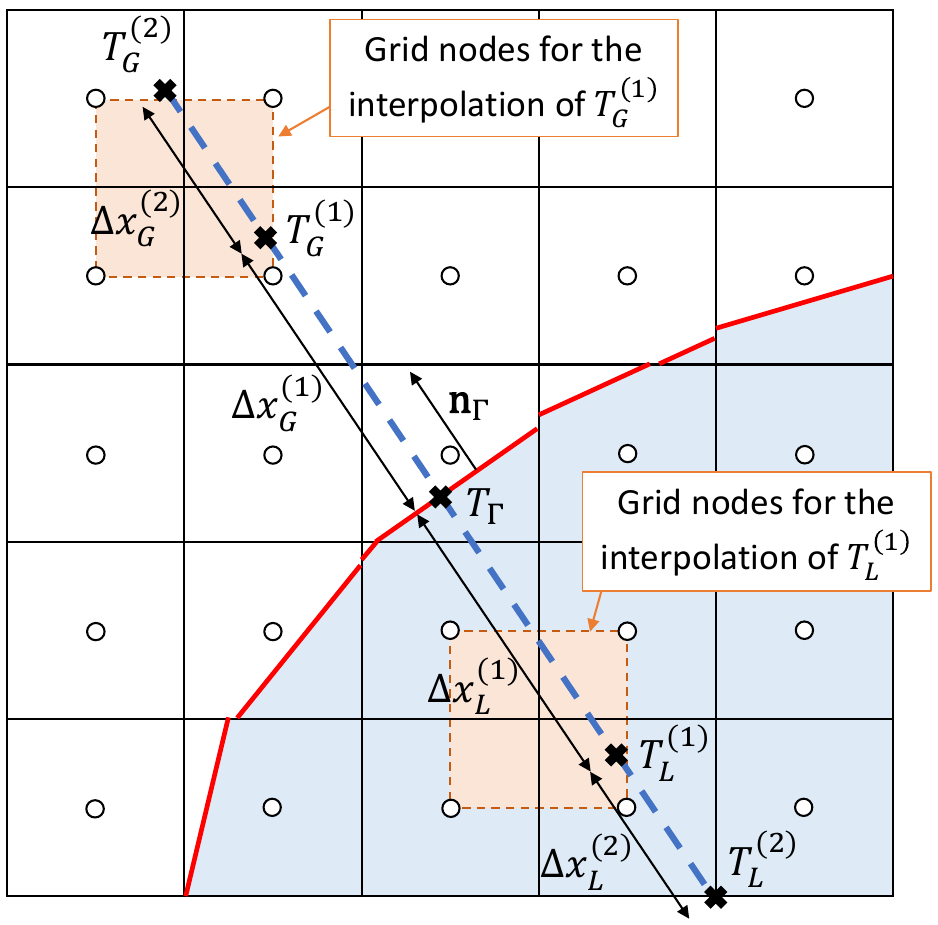}
\caption{Two-dimensional sketch showing the normal probe technique to calculate the normal gradients of scalar quantities, i.e., temperature. The PLIC interface is represented by a solid red line and the liquid volume is colored in blue. The extension to three dimensions is straightforward.}
\label{fig:normal_probe}
\end{figure}

A two-dimensional sketch is shown in Figure \ref{fig:normal_probe} for a non-conflicting scenario. One-sided second-order finite differences with non-uniform spacing can be used to calculate 

\begin{equation}
\label{eqn:normal_gradients}
\begin{split}
    &\nabla T_L\cdot\mathbf{n}_\Gamma \approx \frac{\partial T_L}{\partial n} = -\frac{(\Delta x_{L}^{(1)})^2\big[T_{L}^{(1)}-T_{L}^{(2)}\big]+(\Delta x_{L}^{(2)})^2\big[T_{L}^{(1)}-T_\Gamma\big]+2\Delta x_{L}^{(1)}\Delta x_{L}^{(2)}\big[T_{L}^{(1)}-T_\Gamma\big]}{\Delta x_{L}^{(1)}\big[(\Delta x_{L}^{(2)})^2+\Delta x_{L}^{(1)}\Delta x_{L}^{(2)}\big]} \\
    &\nabla T_G\cdot\mathbf{n}_\Gamma \approx \frac{\partial T_G}{\partial n} = \frac{(\Delta x_{G}^{(1)})^2\big[T_{G}^{(1)}-T_{G}^{(2)}\big]+(\Delta x_{G}^{(2)})^2\big[T_{G}^{(1)}-T_\Gamma\big]+2\Delta x_{G}^{(1)}\Delta x_{G}^{(2)}\big[T_{G}^{(1)}-T_\Gamma\big]}{\Delta x_{G}^{(1)}\big[(\Delta x_{G}^{(2)})^2+\Delta x_{G}^{(1)}\Delta x_{G}^{(2)}\big]}\\
\end{split}
\end{equation}

\noindent
where, oftentimes, \(T_\Gamma\) equals the saturation temperature at the given system pressure \(T_{sat}\). An inherent issue of the calculation of \(\nabla T_G\cdot\mathbf{n}_\Gamma\) and \(\nabla T_L\cdot\mathbf{n}_\Gamma\) is its effect on the smoothness of the \(\dot{m}''\) distribution. As such, some works perform weighted averages of the temperature gradients around the interface cell to obtain a more uniform distribution \cite{2019_JCP_Palmore,2021_JCP_Malan,2023_CaF_Boyd}. Despite a smoother \(\dot{m}''\) distribution is observed with the proposed gradient calculation compared to other works for similar grid resolutions, variations in \(\dot{m}''\) still affect the long-term evolution of canonical configurations such as static evaporating droplets or growing bubbles. Therefore, once Eq. (\ref{eqn:ene_jump2}) is evaluated, the value of \(\dot{m}''\) is recalculated at each interface cell by averaging \(\dot{m}''\) from interface cells contained in a stencil of size \(3\times 3\times 3\). As shown in Section \ref{subsec:bubblegrowth}, even though the first probing point extends almost two grid sizes into each phase, the second-order one-sided scheme results in \(\dot{m}''\) errors and convergence rates similar to other works. \par 

\subsection{Solution of the Navier-Stokes Equations}
\label{subsec:navierstokes_solve}

The Navier-Stokes equations are solved using a modification of the predictor-projection method \cite{1968_MC_Chorin} together with machine-accurate direct solvers for Poisson-type equations based on Fast Fourier Transforms (FFT) \cite{1988_JCP_Schumann,2018_CMA_Costa}, which have been incorporated in the solver using the \textit{vfftpack} library\footnote[2]{A vectorized package of Fortran subprograms for the fast Fourier transform of multiple real sequences, R.A. Sweet, L.L. Lindgren and R.F. Boisvert, \url{https://www.netlib.org/vfftpack/}}. \par

Similar to other works \cite{2014_JCP_Dodd,2021_JCP_Malan}, the advection of the interface from \(t^n\) to \(t^{n+1}\) is performed prior to the solution of the Navier-Stokes equations. As a result, fluid properties and interfacial terms belong to \(t^{n+1}\). Following the same idea, two predictor-projection steps are used where, first, an intermediate velocity \(\mathbf{u^*}\) is calculated as an adjustment to the Stefan flow from \(t^n\) to \(t^{n+1}\) using a velocity potential \(\psi\) as

\begin{equation}
\label{eqn:predictor_step1}
\mathbf{u}^* = \mathbf{u}^n - \nabla\psi
\end{equation}

\noindent
Then, Eq. (\ref{eqn:predictor_step1PPE}) is a Poisson equation for \(\psi\) obtained by taking the divergence of Eq. (\ref{eqn:predictor_step1}), which is solved using the FFT-based or direct solver with the same boundary conditions as the pressure field. As shown later in Section \ref{subsec:1D_analysis}, a pressure jump that does not correspond to the expected physical jump is introduced if this first predictor-projection step is not included (i.e., if \(\mathbf{u}^*=\mathbf{u}^n\) remains true), resulting from the different Stefan flow from \(t^n\) to \(t^{n+1}\) and directly linked to the first term on the LHS of Eq. (\ref{eqn:mom_1f_jump}). \par

\begin{equation}
\label{eqn:predictor_step1PPE}
\nabla^2\psi = \nabla\cdot\mathbf{u}^n-\nabla\cdot\mathbf{u}^*=\dot{m}''(\rho_G^{-1}-\rho_L^{-1})\delta_\Gamma\big|^n-\dot{m}''(\rho_G^{-1}-\rho_L^{-1})\delta_\Gamma\big|^{n+1}
\end{equation}

The use of FFT implies a discretization based on second-order central differences of Eq. (\ref{eqn:predictor_step1PPE}) with a direct computation of the discrete velocity divergence at the cell from the staggered mesh as

\begin{equation}
\nabla\cdot\mathbf{u}_{i,j,k}\approx(u_{i+\frac{1}{2}}-u_{i-\frac{1}{2}})/\Delta x+(v_{j+\frac{1}{2}}-v_{j-\frac{1}{2}})/\Delta y+(w_{k+\frac{1}{2}}-w_{k-\frac{1}{2}})/\Delta z
\end{equation}

\noindent
with, e.g., \(\nabla\psi_{i+\frac{1}{2}}\approx (\psi_{i+1}-\psi_i)/\Delta x\) at the cell faces. Note some indices have been dropped for convenience. Here, \(\delta_\Gamma=||\nabla C||\) is evaluated from staggered volume fractions to minimize the smearing of the velocity jump when compared to using central differences \cite{2017_PhD_Dodd}. For example, Figure \ref{fig:interfacial_region} shows how \(C_{i+\frac{1}{2}}\) is obtained from adding the enclosed volumes on the right half of \(C_i\) and on the left half of \(C_{i+1}\). This evaluation of \(\delta_\Gamma\) results in a more uniform and smoother distribution of the volumetric source term around curved interfaces compared to obtaining \(\delta_\Gamma\) from the surface area density, e.g., \cite{2019_JCP_Palmore,2022_PoF_Poblador}, especially at lower interface resolutions. This preserves the symmetries of the flow around, e.g., an evaporating droplet better while maintaining a reasonably sharp phase jump across a region of thickness \(\sim2\Delta x\). \par

Next, the second predictor step calculates a velocity \(\mathbf{u}^{**}\) from the momentum equation, Eq. (\ref{eqn:mom_1f_full}), without the pressure gradient, \(\mathbf{f}_\sigma\), \(\mathbf{f}_{\dot{m}''}\) and \(\mathbf{f}_{NC}\) using the AB scheme

\begin{equation}
\label{eqn:predictor_step2}
\frac{\mathbf{u}^{**}-\mathbf{u}^*}{\Delta t^{n+1}} = \Bigg(1+\frac{1}{2}\frac{\Delta t^{n+1}}{\Delta t^n}\Bigg)\textbf{RU}^n - \Bigg(\frac{1}{2}\frac{\Delta t^{n+1}}{\Delta t^n}\Bigg)\textbf{RU}^{n-1}
\end{equation}

\noindent
where \(\textbf{RU}^n\) is given by

\begin{equation}
\label{eqn:predictor_step2RU}
\textbf{RU}^n = -\mathbf{u}^*\cdot\nabla\mathbf{u}^* + \frac{1}{\rho^{n+1}}\nabla\cdot\bigg(\mu^{n+1}\big[\nabla\mathbf{u}^*+(\nabla\mathbf{u}^*)^T\big]\bigg) + \mathbf{g}
\end{equation}

As done for the energy equation, AB is used everywhere outside \(\Omega_\Gamma\), while a classic first-order scheme is used in \(\Omega_\Gamma\). The viscous term is discretized with second-order central differences, and the convective term is discretized with the third-order WENO scheme in cells not belonging to \(\Omega_\Gamma\) and second-order central differences in \(\Omega_\Gamma\). These modifications around the interface are necessary to aim for the cancellation of the numerical jump caused by \(\mathbf{u}^*\cdot\nabla\mathbf{u}^*\) with \(\mathbf{f}_{NC}\). Note the choice of WENO in the momentum equation is to maintain consistency with the discretization of the energy equation. For practical purposes, no major differences have been observed between using WENO or central differences in the problems considered in this work. \par

The variable-coefficient pressure Poisson equation (PPE) resulting from the projection step is converted into a constant-coefficient PPE via the approximate substitution \cite{2012_JCP_Dong,2014_JCP_Dodd}

\begin{equation}
\label{eqn:split_gradP}
\frac{1}{\rho^{n+1}}\nabla p^{n+1}\rightarrow\frac{1}{\rho_0}\nabla p^{n+1}+\bigg(\frac{1}{\rho^{n+1}}-\frac{1}{\rho_0}\bigg)\nabla\hat{p}
\end{equation}

\noindent
which splits the gradient \(\tfrac{1}{\rho^{n+1}}\nabla p^{n+1}\) into a constant-coefficient implicit term and a variable-coefficient explicit term. For accuracy and stability purposes, \(\rho_0=\text{min}(\rho_g,\rho_l)\) and \(\hat{p}=[1+(\Delta t^{n+1}/\Delta t^n)]p^{n}-(\Delta t^{n+1}/\Delta t^n)p^{n-1}\) is a linear extrapolation in time of the pressure field \cite{2012_JCP_Dong,2014_JCP_Dodd}. Then, a direct pressure solver can be used, resulting in a computationally more efficient solver than iterative methods \cite{2014_JCP_Dodd,2019_ICHMT_Talebanfard,2020_JCP_Scapin,2021_IJHMT_Dodd,2022_PoF_Poblador}. \par 

However, pressure-splitting introduces considerable time-stepping restrictions for stability purposes when the pressure jump across the interface is large (e.g., large surface tension) or when the density ratio is large (i.e., \(\nabla\hat{p}\) has a greater contribution in the PPE) \cite{2014_JCP_Dodd}. In other words, smoothness in \(\hat{p}\) may be lost and substantial pressure oscillations occur as the interface moves across grid cells \cite{2019_JCP_Cifani,2021_CaF_Turnquist}. Such limitation could defy the implementation of the gradient splitting technique and reduce the benefits of a direct solver. This problem is mitigated by explicitly accounting for the pressure jump in the PPE \cite{2019_JCP_Cifani}, similar to a GFM. Thus, \(\mathbf{f}_\sigma\), \(\mathbf{f}_{\dot{m}''}\) and \(\mathbf{f}_{NC}\) are removed from Eq. (\ref{eqn:mom_1f_full}) during the predictor step in Eqs. (\ref{eqn:predictor_step2}) and (\ref{eqn:predictor_step2RU}) and, instead, are added in the projection step. Within the context of the gradient splitting technique, this is equivalent to the temporal extrapolation of the pressure jump instead of the pressure itself, i.e., see Eq. (\ref{eqn:projection_step1VEL}). \par

As a result of pressure-splitting, the velocity at the new time step \(\mathbf{u}^{n+1}\) is obtained from the projection step as

\begin{equation}
\label{eqn:projection_step1VEL}
\mathbf{u}^{n+1} = \mathbf{u}^{**} - \Delta t^{n+1}\Bigg[\frac{1}{\rho_0}\bigg(\nabla p^{n+1}-\mathbf{f}_\sigma^{n+1}-\mathbf{f}_{\dot{m}''}^{n+1}\bigg) + \bigg(\frac{1}{\rho^{n+1}}-\frac{1}{\rho_0}\bigg)\bigg(\nabla \hat{p}-\hat{\mathbf{f}}_\sigma-\hat{\mathbf{f}}_{\dot{m}''}\bigg) - \frac{1}{\rho^{n+1}}\mathbf{f}_{NC}^{n+1}  \Bigg]
\end{equation}

\noindent
which defines a constant-coefficient PPE by taking the divergence of Eq. (\ref{eqn:projection_step1VEL}), i.e.,

\begin{equation}
\label{eqn:projection_step2PPE}
\nabla^2p^{n+1} = \nabla\cdot\bigg[\bigg(1-\frac{\rho_0}{\rho^{n+1}}\bigg)\bigg(\nabla \hat{p}-\hat{\mathbf{f}}_\sigma-\hat{\mathbf{f}}_{\dot{m}''}\bigg)+\mathbf{f}_\sigma^{n+1}+\mathbf{f}_{\dot{m}''}^{n+1}+\frac{\rho_0}{\rho^{n+1}}\mathbf{f}_{NC}^{n+1}\bigg] + \frac{\rho_0}{\Delta t^{n+1}}\big[\nabla\cdot\mathbf{u}^{**}-\nabla\cdot\mathbf{u}^{n+1}\big]
\end{equation}

\noindent
where \(\nabla\cdot\mathbf{u}^{n+1}=\dot{m}''(\rho_G^{-1}-\rho_L^{-1})\delta_\Gamma\big|^{n+1}\) ensures that the correct velocity jump given by Eq. (\ref{eqn:cont_1f}) is imposed. It is important to highlight that the splitting operator is not applied to \(\mathbf{f}_{NC}^{n+1}\) since the variable density \(\rho^{n+1}\) effectively cancels with the density term inside \(\mathbf{f}_{NC}^{n+1}\). Then, Eq. (\ref{eqn:projection_step2PPE}) is solved using the FFT-based or direct solver. Another advantage of using the direct solver relates to stiffness problems that might arise as \(\nabla\cdot\mathbf{u}^{n+1}\) increases, which could affect the convergence rate of iterative solvers \cite{2013_JCP_Sato,2020_JCP_Scapin}. \par  

The forces \(\mathbf{f}_\sigma\), \(\mathbf{f}_{\dot{m}''}\) and \(\mathbf{f}_{NC}\) must be calculated at cell faces to correspond with the evaluation of \(\nabla p\). In \(\mathbf{f}_\sigma\) and \(\mathbf{f}_{\dot{m}''}\), \(\kappa\) and \(\dot{m}''\) are averaged at cell faces \cite{2014_JCP_Dodd} and \(\mathbf{n}_\Gamma\delta_\Gamma=-\nabla C\) following the CSF approach. \(\nabla C\) is evaluated from cell values, e.g., \(\partial C/\partial x|_{i+1/2} \approx (C_{i+1}-C_i)/\Delta x\). In contrast, \(\mathbf{f}_{NC}\) is discretized differently for smoothness purposes in order to minimize geometrical errors that may arise from an ill-defined discrete force at the interface given the sensitivity of the term \(\mathbf{u}^*\cdot\mathbf{n}_\Gamma\). Note that the corrective nature of \(\mathbf{f}_{NC}\) to counteract the error introduced by \(\mathbf{u}^*\cdot\nabla\mathbf{u}^*\) in \(\Omega_\Gamma\) implies that \(\mathbf{u}^*\) substitutes \(\mathbf{u}\) in Eq. (\ref{eqn:bodyforce2}). In this work, \(\mathbf{u}^*\) is linearly averaged at cell faces while \(\mathbf{n}_\Gamma\) is obtained at staggered interface cells (e.g., \(0<C_{i+1/2}<1\)) using the MYC method with staggered volume fractions. Then, a narrow band of staggered cells around the interface is populated with a weighted average of the normal unit vectors obtained in the previous step with the method described in \ref{apn:A}. Lastly, discretizing \(\mathbf{n}_\Gamma\delta_\Gamma\) with \(\delta_\Gamma\) obtained in staggered locations as, e.g., \(\delta_{\Gamma,i+1/2}=\tfrac{1}{2}(\delta_{\Gamma,i}+\delta_{\Gamma,i+1})\) improves the smoothness of \(\mathbf{f}_{NC}\) compared to the substitution \(\mathbf{n}_\Gamma\delta_\Gamma=-\nabla C\). As shown in Section \ref{subsec:1D_analysis}, both \(\delta_{\Gamma,i+1/2}\) or \(\nabla C\) evaluated with staggered central differences, e.g., \(\partial C/\partial x|_{i+1/2} \approx (C_{i+3/2}-C_{i-1/2})/(2\Delta x)\), result in the correction of the error introduced by \(\mathbf{u}^*\cdot\nabla\mathbf{u}^*\) at the discrete level for a well-posed one-dimensional case. \par

Lastly, it is worth discussing the computational cost of the flow solver, in particular the inclusion of additional Poisson-type equations. Besides the PPE, a Poisson equation is solved to calculate the divergence-free velocity for the advection of \(C\) in Section \ref{subsec:intcap} and to calculate \(\mathbf{u}^*\) via Eqs. (\ref{eqn:predictor_step1}) and (\ref{eqn:predictor_step1PPE}). Depending on the set tolerance and type of, e.g., multigrid iterative solver, direct solvers may provide a speed-up between 3 and 14 times in typical two-phase flow implementations \cite{2014_JCP_Dodd,2020_CaF_Ahmed}. This translates into the PPE representing only about 58\% of the wall-clock time per time step in parallel computing implementations of non-evaporative flows \cite{2014_JCP_Dodd}. However, many configurations of interest involve disperse bubbly or droplet-laden flows and load balancing easily deteriorates when using typical domain decompositions, e.g., pencils. That is, only a few processes might take care of geometric reconstructions, evaluating interfacial heat and mass transfer, or calculating localized forces. As such, a behavior closer to what is reported in \cite{2020_JCP_Scapin} is observed where solving each additional Poisson equation represents about 5\% of the cost per time step in a configuration like the static evaporating droplet in Section \ref{subsec:droplet}. \par

\subsection{Formal Analysis of the Modified Predictor-Projection Method}
\label{subsec:LUdecomp}

The modified predictor-projection method or Fractional Step Method (FSM) described in Section \ref{subsec:navierstokes_solve} is analyzed following \cite{1993_JCP_Perot,1995_JCP_Perot}. The discrete Navier-Stokes equations are recast in matrix-vector form \(\mathbf{A}\mathbf{x}=\mathbf{b}\) as

\begin{equation}
\label{eqn:LU_1}
    \begin{pmatrix}
        \text{I} & \Delta t \frac{1}{\rho} \text{G} \\
        \text{D} & 0
    \end{pmatrix}
    \begin{pmatrix}
        \mathbf{u}^{n+1} \\
        p^{n+1}_{\text{FSM}}
    \end{pmatrix}=
    \begin{pmatrix}
        \Delta t \mathbf{R} + \mathbf{u}^n \\
        \text{M}^{n+1}
    \end{pmatrix}
\end{equation}

\noindent
where I, G and D are the discrete identity, gradient and divergence operators, respectively. Further, \(\mathbf{R}\) corresponds to all the discretized terms on the RHS of the momentum equation, Eq. (\ref{eqn:mom_1f_full}), plus the convective term. The various body forces are included here. That is, for the purpose of this analysis, no gradient splitting operator is implemented nor the body forces are moved to the projector step. Moreover, the temporal scheme for \(\mathbf{R}\) is left unspecified and could be obtained from, e.g., an explicit forward Euler or from the AB scheme. The volumetric expansion at the interface is given by \(\text{M}^{n+1}=\dot{m}''(\rho_G^{-1}-\rho_L^{-1})\delta_\Gamma|^{n+1}\). Note that the predictor-projection steps can be obtained by performing the LU decomposition of \(\mathbf{A}\mathbf{x}=\mathbf{b}\) \cite{1993_JCP_Perot,1995_JCP_Perot}, i.e., a Poisson-type equation emerges naturally. \par

However, the direct solution of Eq. (\ref{eqn:LU_1}) suffers from numerical inaccuracies arising from the abrupt shift of \(\mathbf{u}\) in the vicinity of the interface, which translates into the introduction of a time-oscillatory pressure field \(p^{n+1}_{FSM}\) that may impact the momentum balance across \(\Omega_\Gamma\). Such oscillation originates at the projection step of the naive FSM as

\begin{equation}
    \label{eqn:FSM-naive-projection}
    \text{D} \frac{1}{\rho} \text{G} p^{n+1}_{FSM} = 
    \text{D} \mathbf{u^p} - \text{M}^{n+1} = \Delta t \text{D} \mathbf{R} + \text{D} \mathbf{u^n} - \text{M}^{n+1}
\end{equation}

\noindent
where the two rightmost terms are equal to \(\text{M}^n - \text{M}^{n+1}\), which is indeed time-oscillatory due to its close relation to the surface interface transport. However, in the context of an incompressible flow, the pressure field is defined instantaneously to impose the divergence of the velocity field, and thus any time dependence is inconsistent with the original method. One can see how the naive approach does, inadvertently, break the original formulation of the problem. These oscillations are observed in Sections \ref{subsec:1D_liquid} and \ref{subsec:bubblerising} when looking at the evolution of the pressure jump over time in different two-phase flows. \par

To improve the numerical solution of Eq. (\ref{eqn:LU_1}), \(\mathbf{u^n}\) is regularized by introducing a Stefan flow shift before proceeding to the FSM, as discussed in Section \ref{subsec:navierstokes_solve}. Such a shift removes the dependence of pressure with the Stefan flow of the previous time step by introducing an intermediate velocity field, \(\mathbf{u}^*\), which corrects the velocity field \(\mathbf{u}^n\) to satisfy the volumetric expansion at the next time step \(\text{M}^{n+1}\). The shift is defined implicitly by introducing an intermediate pressure, \(p^*\), resulting in an additional linear system of equations given by

\begin{equation}
\label{eqn:LU_2}
    \begin{pmatrix}
        \text{I} & \Delta t \frac{1}{\rho} \text{G} \\
        \text{D} & 0
    \end{pmatrix}
    \begin{pmatrix}
        \mathbf{u}^* \\
        p^*
    \end{pmatrix}=
    \begin{pmatrix}
        \mathbf{u}^n \\
        \text{M}^{n+1}
    \end{pmatrix}
\end{equation}

The connection between the original FSM given by Eq. (\ref{eqn:LU_1}), the Stefan flow shift given by Eq. (\ref{eqn:LU_2}), and the FSM step used to solve the Navier-Stokes equations in the flow solver materializes by rewriting Eq. (\ref{eqn:LU_1}) in the form \(\mathbf{A}\mathbf{x}=\mathbf{b}_0 + \mathbf{b}_1\) as

\begin{equation}
\label{eqn:LU_7}
    \begin{pmatrix}
        \text{I} & \Delta t \frac{1}{\rho} \text{G} \\
        \text{D} & 0
    \end{pmatrix}
    \begin{pmatrix}
        \mathbf{u}^{n+1} \\
        p^{n+1}_{\text{FSM}}
    \end{pmatrix}=
    \begin{pmatrix}
        \Delta t \mathbf{R} \\
        0
    \end{pmatrix}+
    \begin{pmatrix}
        \mathbf{u}^n \\
        \text{M}^{n+1}
    \end{pmatrix}=
     \begin{pmatrix}
        \Delta t \mathbf{R} \\
        0
    \end{pmatrix}+
    \begin{pmatrix}
        \text{I} & \Delta t \frac{1}{\rho} \text{G} \\
        \text{D} & 0
    \end{pmatrix}
    \begin{pmatrix}
        \mathbf{u}^*\\
        p^*
    \end{pmatrix}
\end{equation}

\noindent
where the vector \(\mathbf{b_1}\) is, in fact, the RHS of Eq. (\ref{eqn:LU_2}), such that the same system of equations is being solved. Now, manipulation of Eq. (\ref{eqn:LU_7}) results in

\begin{equation}
\label{eqn:LU_8}
    \begin{pmatrix}
        \text{I} & \Delta t \frac{1}{\rho} \text{G} \\
        \text{D} & 0
    \end{pmatrix}
    \begin{pmatrix}
        \mathbf{u}^{n+1} -\mathbf{u}^* \\
        p^{n+1}_{\text{FSM}}-p^*
    \end{pmatrix}=
    \begin{pmatrix}
        \Delta t \mathbf{R} \\
        0
    \end{pmatrix}
    \rightarrow
    \begin{pmatrix}
        \text{I} & \Delta t \frac{1}{\rho} \text{G} \\
        \text{D} & 0
    \end{pmatrix}
    \begin{pmatrix}
        \mathbf{u}^{n+1} \\
        p^{n+1}
    \end{pmatrix}=
    \begin{pmatrix}
        \Delta t \mathbf{R} + \mathbf{u}^* \\
        \text{M}^{n+1}
    \end{pmatrix}
\end{equation}

\noindent
which is essentially the second FSM step used in the flow solver described in Section \ref{subsec:navierstokes_solve}. One can easily check that the second projection step now implies

\begin{equation}
        \text{D} \frac{1}{\rho} \text{G} p^{n+1} = 
    \text{D} \mathbf{u^{**}} - \text{M}^{n+1} = \Delta t \text{D} \mathbf{R} + \text{D} \mathbf{u^*} - \text{M}^{n+1}
\end{equation}

\noindent
where the two rightmost terms, which were the problematic ones as identified in Eq. (\ref{eqn:FSM-naive-projection}), are identical now, removing the oscillatory behavior. \par

Note that Eq. (\ref{eqn:predictor_step1}) as a function of \(\Psi\) has been replaced in the analysis by \(\mathbf{u}^*=\mathbf{u}^n-\Delta t \frac{1}{\rho}\nabla p^*\) in Eq. (\ref{eqn:LU_2}). However, the actual implementation of the solver does indeed use \(\Psi\) to avoid having to solve another variable-coefficient Poisson equation. Since knowing the actual value of \(p^*\) is not necessary, only \(\mathbf{u}^*\) is obtained. \par

\begin{figure}
\centering
\includegraphics[width=0.45\linewidth]{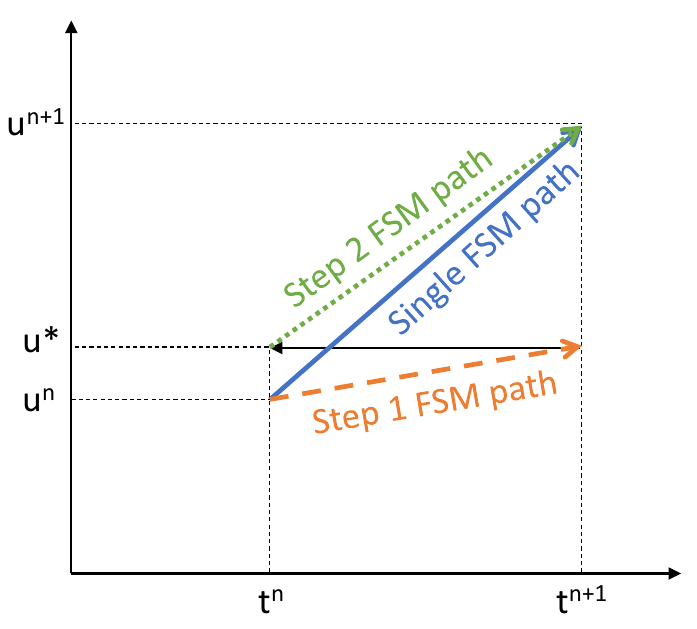}
\caption{Schematic comparing the integration from \(\mathbf{u}^n\) to \(\mathbf{u}^{n+1}\) with a single predictor-projection step or Fractional Step Method (FSM) step and the two FSM steps used in the flow solver described in Section \ref{subsec:navierstokes_solve} and formalized in Section \ref{subsec:LUdecomp}.}
\label{fig:FSM_steps}
\end{figure}

Various details of the algorithm have become clear with the previous analysis and are summarized here. First, the flow solver has been split into two predictor-projection or FSM steps rather than one. By doing so, one integrates the velocity first from \(\mathbf{u}^n\) to \(\mathbf{u}^*\) purely based on mass conservation constraints due to the moving interface, absorbing the time oscillations in the pressure term \(p^*\); then, one integrates momentum from \(\mathbf{u}^*\) to \(\mathbf{u}^{n+1}\), as visualized in Figure \ref{fig:FSM_steps}. Second, by using the first FSM step, one eliminates the divergence constraint in \(\Omega_\Gamma\) and obtains a pressure field \(p^{n+1}\) free of oscillations caused by a time-dependent stiff source term, resembling the implementation of the FSM in an incompressible single-phase problem. Third and last, the recasting of Eq. (\ref{eqn:LU_8}) shows that the second FSM step is indeed solving for a new pressure \(p^{n+1}=p^{n+1}_{\text{FSM}}-p^*\) that differs from the pressure field that one would obtain from directly solving the system given by Eq. (\ref{eqn:LU_1}). In other words, the numerical pressure \(p^{n+1}_{\text{FSM}}\) obtained with the traditional formulation of the FSM includes an additional pressure \(p^*\) resulting from the temporal change of the Stefan flow, which carries the problematic pressure behavior across \(\Omega_\Gamma\) caused by the inconsistent temporal integration between the one-fluid Navier-Stokes equations and the interface transport when \(\dot{m}''\neq 0\). \par

Some limitations arise with the proposed two-step predictor-projection method. As shown in Eq. (\ref{eqn:predictor_step2RU}), \(\mathbf{R}\) is discretized with \(\mathbf{u}^*\) instead of \(\mathbf{u}^n\). Thus, the impact on the evaluation of \(\mathbf{R}\) and how it might affect the temporal integration of momentum must be understood. This is left for future work, but preliminary analysis following the static droplet evaporation case presented in Section \ref{subsec:droplet} suggests that the convective and diffusive terms differ by less than 0.03\% given the considered \(\Delta t\) restrictions. Further, by removing the information carried by \(p^*\) regarding unsteady effects of the Stefan flow, one effectively imposes that the interface movement with respect to the fluid is a quasi-steady state process. Therefore, the pressure \(p^{n+1}\) obtained with the modified FSM is only physical if the quasi-steady assumption holds despite the correct pressure jump and volumetric expansion may be recovered across \(\Omega_\Gamma\) when solving the Navier-Stokes equations. While this assumption may not be problematic in cases where the flow dynamics are driven by other forces (e.g., gravity), differences become more evident if the flow field is dominated by the evaporation process. \par

\subsection{One-dimensional Discrete Analysis of Momentum Balance Corrections}
\label{subsec:1D_analysis}

\begin{figure}
\centering
\includegraphics[width=0.8\linewidth]{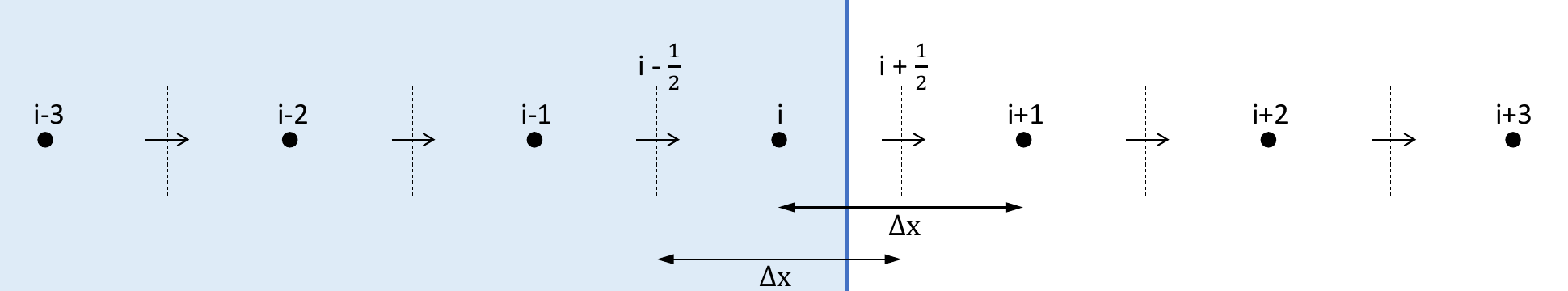}
\caption{Discretization of the one-dimensional evaporating liquid surface with a uniform mesh \(\Delta x\). The interface sits inside cell \(i\), with the liquid on the left and the gas on the right.}
\label{fig:1D_sketch}
\end{figure}

A one-dimensional discrete analysis of the momentum balance corrections is provided in this section to visualize the modifications introduced in the flow solver discussed in Sections \ref{subsec:navierstokes_solve} and \ref{subsec:LUdecomp}. An evaporating liquid film sitting on a fixed wall is analyzed where the liquid is at rest and the flat surface does not deform. Similar to the Stefan problem but without coupling to the energy equation \cite{2021_JCP_Malan,2023_CaF_Boyd,2023_hal_Cipriano}, this configuration results in the one-dimensional evolution of an evaporating flow. The analytical solution is straightforward, where the velocities in the liquid and gas phases are constant and given by \(u_l=0\) and \(u_g=\dot{m}''(\rho_G^{-1}-\rho_L^{-1})\). Thus, the viscous term is identically zero and the quasi-steady assumption of the interface regression with respect to the fluid has no impact beyond \(\Omega_\Gamma\). Here, phase change causes a pressure jump across the interface given by \(\Delta p = (\dot{m}'')^2(\rho_G^{-1}-\rho_L^{-1})\). The discretization of the one-dimensional domain is shown in Figure \ref{fig:1D_sketch}. \par 

From this setup one can directly show the effects of the two modifications proposed to the flow solver, namely the intermediate step to calculate \(\mathbf{u}^*\) and the addition of \(\mathbf{f}_{\dot{m}''}\) and \(\mathbf{f}_{NC}\). Combining Eqs. (\ref{eqn:predictor_step2RU}) and (\ref{eqn:projection_step1VEL}) in \(\Omega_\Gamma\) where the forward Euler scheme is used for the temporal integration, the pressure gradient across a velocity node, e.g., \(i\)+\(\tfrac{1}{2}\), is given by 

\begin{equation}
\label{eqn:1D_gradp}
\frac{dp^{n+1}}{dx}= \frac{\rho_0}{\Delta t^{n+1}}(u^*-u^{n+1})-\rho_0 u^* \frac{du^*}{dx} + \bigg(1-\frac{\rho_0}{\rho^{n+1}}\bigg)\bigg(\frac{d\hat{p}}{dx}-\hat{f}_{\dot{m}'',x}\bigg) + f_{\dot{m}'',x}^{n+1} + \frac{\rho_0}{\rho^{n+1}} f_{NC,x}^{n+1}
\end{equation}

The pressure field obtained from predictor-projection methods is essentially a mathematical operation to enforce the divergence-free condition of the velocity field in incompressible flows. As long as \(\mathbf{u}^{**}\) is well posed, the pressure is physically meaningful. However, this is not the case with the one-fluid formulation introduced in Section \ref{subsec:onefluid} across \(\Omega_\Gamma\) if \(\dot{m}'''\neq0\). Fundamentally, the PPE is used to overwrite the velocity field in \(\Omega_\Gamma\) to satisfy Eq. (\ref{eqn:cont_1f}) regardless of \(\mathbf{u}^{**}\). Thus, the correct velocity field may be obtained but with an ill-defined momentum balance. The introduction of \(\mathbf{u}^*\), \(\mathbf{f}_{NC}\) and \(\mathbf{f}_{\dot{m}''}\) aim to solve this issue by cancellation and correction. Similar to the modeling of surface tension in \cite{2019_JCP_Cifani}, the discretization of \(f_{\dot{m}'',x}^{n+1}\) and \(\hat{f}_{\dot{m}'',x}\) consistent with the pressure gradient ensures that the exact pressure jump due to phase change is recovered across \(\Omega_\Gamma\) if no other jumps are introduced numerically. \par

In Figure \ref{fig:1D_sketch}, the volumetric expansion occurs in cells \(i\) and \(i\)+1 where \(\delta_{\Gamma,i}\) and \(\delta_{\Gamma,i+1}\) are non-zero. Thus, one can show that from the staggered discretization of Eq. (\ref{eqn:1D_gradp}) using central differences, \(u^* \frac{du^*}{dx}\) is non-zero only at velocity nodes \(i+\frac{1}{2}\) and \(i+\frac{3}{2}\). In particular,

\begin{equation}
\label{eqn:1D_ududx1}
u^* \frac{du^*}{dx}\approx 
    \begin{cases}
        u_{i+\frac{1}{2}}^*\frac{u_{i+\frac{3}{2}}^*-u_{i-\frac{1}{2}}^*}{2\Delta x} & \text{if $i+\frac{1}{2}$}\\
        u_{i+\frac{3}{2}}^*\frac{u_{i+\frac{5}{2}}^*-u_{i+\frac{1}{2}}^*}{2\Delta x} & \text{if $i+\frac{3}{2}$}\\
    \end{cases}
\end{equation}

\noindent
where \(u_{i-\frac{1}{2}}^*=u_l=0\) and \(u_{i+\frac{5}{2}}^*=u_g=\dot{m}''(\rho_G^{-1}-\rho_L^{-1})\). Further, from \(\nabla\cdot\mathbf{u}^*\), one obtains 

\begin{equation}
\label{eqn:1D_ududx3}
\begin{split}
    &u_{i+\frac{1}{2}}^*=\dot{m}''(\rho_G^{-1}-\rho_L^{-1})\delta_{\Gamma,i}\Delta x\\
    &u_{i+\frac{3}{2}}^*=u_{i+\frac{1}{2}}^*+\dot{m}''(\rho_G^{-1}-\rho_L^{-1})\delta_{\Gamma,i+1}\Delta x\\
    &u_{i+\frac{5}{2}}^*=u_{i+\frac{3}{2}}^*+\dot{m}''(\rho_G^{-1}-\rho_L^{-1})\delta_{\Gamma,i+2}\Delta x
\end{split}
\end{equation}

\noindent
where \(\delta_{\Gamma,i}=|C_{i+\frac{1}{2}}-C_{i-\frac{1}{2}}|\Delta x^{-1}\), \(\delta_{\Gamma,i+1}=|C_{i+\frac{3}{2}}-C_{i+\frac{1}{2}}|\Delta x^{-1}\) and \(\delta_{\Gamma,i+2}=|C_{i+\frac{5}{2}}-C_{i+\frac{3}{2}}|\Delta x^{-1}\). Then, substitution into Eq. (\ref{eqn:1D_ududx1}) results in

\begin{equation}
\label{eqn:1D_ududx2}
u^* \frac{du^*}{dx}\approx 
    \begin{cases}
        u_{i+\frac{1}{2}}^*\dot{m}''(\rho_G^{-1}-\rho_L^{-1})\frac{1}{2}(\delta_{\Gamma,i}+\delta_{\Gamma,i+1})=-u_{i+\frac{1}{2}}^*\dot{m}''(\rho_G^{-1}-\rho_L^{-1})\bigg(\frac{C_{i+\frac{3}{2}}-C_{i-\frac{1}{2}}}{2\Delta x}\bigg) & \text{if $i+\frac{1}{2}$}\\
        u_{i+\frac{3}{2}}^*\dot{m}''(\rho_G^{-1}-\rho_L^{-1})\frac{1}{2}(\delta_{\Gamma,i+1}+\delta_{\Gamma,i+2})=-u_{i+\frac{3}{2}}^*\dot{m}''(\rho_G^{-1}-\rho_L^{-1})\bigg(\frac{C_{i+\frac{5}{2}}-C_{i+\frac{1}{2}}}{2\Delta x}\bigg) & \text{if $i+\frac{3}{2}$}\\
    \end{cases}
\end{equation}

\noindent
which is equivalent to the discretization of \(\frac{1}{\rho^{n+1}}f_{NC,x}^{n+1}\) in the same nodes (note \(\mathbf{u}^*\cdot\mathbf{n}_\Gamma=u^*\)) using either \(\mathbf{n}_\Gamma\delta_\Gamma\) or \(-\nabla C\). \par 

Despite \(\frac{1}{\rho^{n+1}}f_{NC,x}^{n+1}\equiv u^* \frac{du^*}{dx}\) ensures the cancellation of both terms in the RHS of Eq. (\ref{eqn:1D_gradp}), the term \(\frac{\rho_0}{\Delta t^{n+1}}(u^*-u^{n+1})\) still introduces a pressure jump due to the adjustment of the velocity to satisfy mass conservation constraints. If Eq. (\ref{eqn:predictor_step1}) is not solved, then \(u^*=u^n\) and the velocity difference \(u^n-u^{n+1}\) may be sufficiently large in \(\Omega_\Gamma\) to cause a significant pressure jump as the interface moves. By including the calculation of \(\mathbf{u}^*\) as an intermediate step to account for the adjustment of the Stefan flow from \(t^n\) to \(t^{n+1}\) before solving the Navier-Stokes equations, one effectively cancels \(\frac{\rho_0}{\Delta t^{n+1}}(u^*-u^{n+1})\) in \(\Omega_\Gamma\) if \(u^*=u^{n+1}\), such as in this one-dimensional problem, or reduces the jump if \(u^*\approx u^{n+1}\). \par 

The adjustment of the Stefan flow with the first predictor-projection step eliminates the pressure jump introduced by the time derivative in Eqs. (\ref{eqn:mom_1f}) or (\ref{eqn:mom_1f_full}), i.e., the first term on the LHS of Eq. (\ref{eqn:mom_1f_jump}). Without correction, the temporal term in Eq. (\ref{eqn:1D_gradp}) causing a pressure jump becomes \(\frac{\rho_0}{\Delta t^{n+1}}(u^n-u^{n+1})\). In, e.g., node \(i\)+\(\tfrac{1}{2}\), it is given by 

\begin{equation}
\label{eqn:1D_tempjump}
\frac{\rho_0}{\Delta t^{n+1}}(u^n-u^{n+1})=-\frac{\rho_0}{\Delta t^{n+1}}\dot{m}''(\rho_G^{-1}-\rho_L^{-1})(\delta_{\Gamma,i}^{n+1}-\delta_{\Gamma,i}^{n})\Delta x
\end{equation}

\noindent
Note that if \(\rho(\mathbf{u}_L-\mathbf{u}_G)(\mathbf{u}_\Gamma\cdot\mathbf{n}_\Gamma)\) from Eq. (\ref{eqn:mom_1f_jump}) is rewritten as a localized body force similar to \(\mathbf{f}_{NC}\), it could be approximated at \(i\)+\(\tfrac{1}{2}\) as \(-\rho_0\dot{m}''(\rho_G^{-1}-\rho_L^{-1})u_{\Gamma,i+\frac{1}{2}}^{n+1}\delta_{\Gamma,i+\frac{1}{2}}^{n+1}\). Thus, the effect of the interface shift over time on the momentum balance of the non-conservative formulation, driven by \(u_{\Gamma,i+\frac{1}{2}}^{n+1}\delta_{\Gamma,i+\frac{1}{2}}^{n+1}\), is approximated in the predictor-projection method with \((\delta_{\Gamma,i}^{n+1}-\delta_{\Gamma,i}^{n})\frac{\Delta x}{\Delta t^{n+1}}\). That is, the interface velocity is approximated by the shift in \(\delta_\Gamma\) from \(t^n\) to \(t^{n+1}\), which is eliminated by considering the intermediate velocity \(\mathbf{u}^*\) instead. \par

\section{Validation with Fixed Mass Flux}
\label{sec:validation}

The numerical approach and proposed momentum balance corrections are validated against fabricated solutions of evaporating flows with constant \(\dot{m}''\). This effectively decouples the flow solver from the energy equation. Thus, a better assessment of the handling of phase change in the advection of the interface and the Navier-Stokes equations can be made. Moreover, gravity is neglected. \par 

To minimize the need to significantly reduce the time step for stability purposes due to the gradient splitting operator in Eq. (\ref{eqn:projection_step2PPE}) and the handling of phase change, conditions relevant to high-pressure boiling flows are considered since density and viscosity ratios are small. The properties of each phase are set to \(\rho_G=100\) kg/m\textsuperscript{3}, \(\rho_L=500\) kg/m\textsuperscript{3}, \(\mu_G=25\) \(\mu\)Pa\(\cdot\)s and \(\mu_L=50\) \(\mu\)Pa\(\cdot\)s, resulting in a density ratio \(\frac{\rho_L}{\rho_G}=5\) and viscosity ratio \(\frac{\mu_L}{\mu_G}=2\). At these conditions, a characteristic value of the surface tension coefficient is \(\sigma=1\) mN/m and, given the reduced latent heat of vaporization at such high pressures, \(\dot{m}''\) may vary substantially depending on the heat flux into the interface \cite{propertiesofgasesandliquids}. Thus, values ranging from \(0.1\) kg/(m\textsuperscript{2}s) to \(10\) kg/(m\textsuperscript{2}s) are considered in Section \ref{subsec:droplet}. \par 

Note that \(\dot{m}''\) values are pushed to what could be considered upper limits in evaporative flows. Since the modifications to what is considered a standard two-phase flow solver disappear as \(\dot{m}''\rightarrow 0\), it is out of scope of this section to show results approaching non-evaporative conditions. Further, in this section the CFL constraints on the time step are driven by the interface evaporation. As such, a restrictive CFL condition on the interface shift limits the maximum shift per time step to \((\Delta\text{d}_{\text{max}}/\Delta x)=0.001\) to minimize the geometrical errors introduced by the respective term in Eq. (\ref{eqn:VOFindicator}) and focus on the behavior of the additional body forces \(\mathbf{f}_{\dot{m}''}\) and \(\mathbf{f}_{NC}\). \par 

\subsection{One-dimensional Evaporating Liquid Surface}
\label{subsec:1D_liquid}

\begin{figure}
\centering
\begin{subfigure}{0.45\textwidth}
  \centering
  \includegraphics[width=1.0\linewidth]{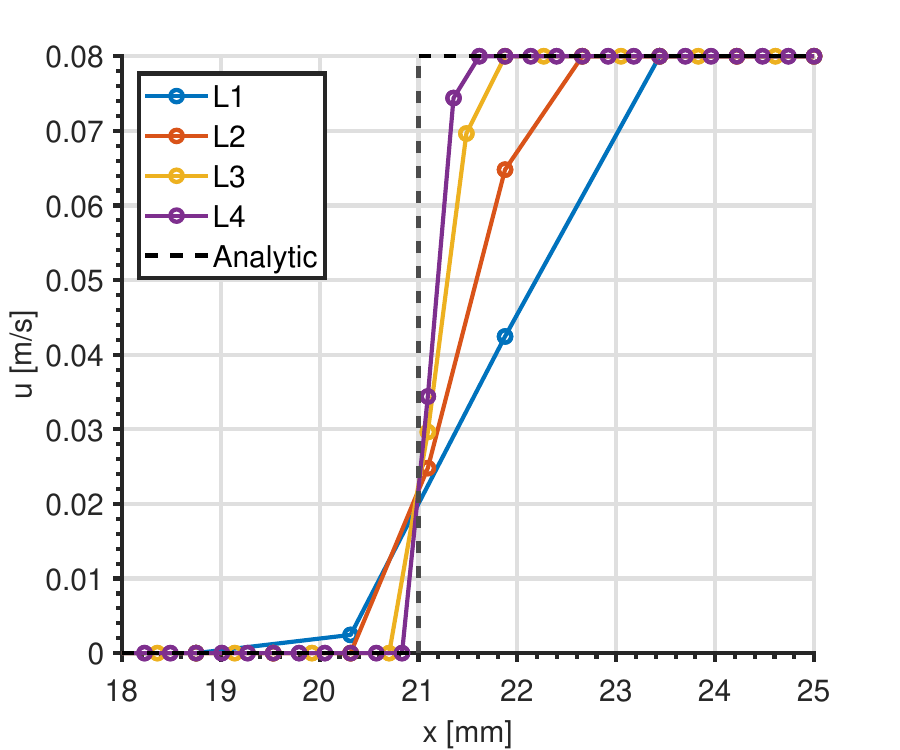}
  \caption{} % No text so (a) appears
  \label{subfig:1D_ujump_200ms}
\end{subfigure}%
\begin{subfigure}{0.45\textwidth}
  \centering
  \includegraphics[width=1.0\linewidth]{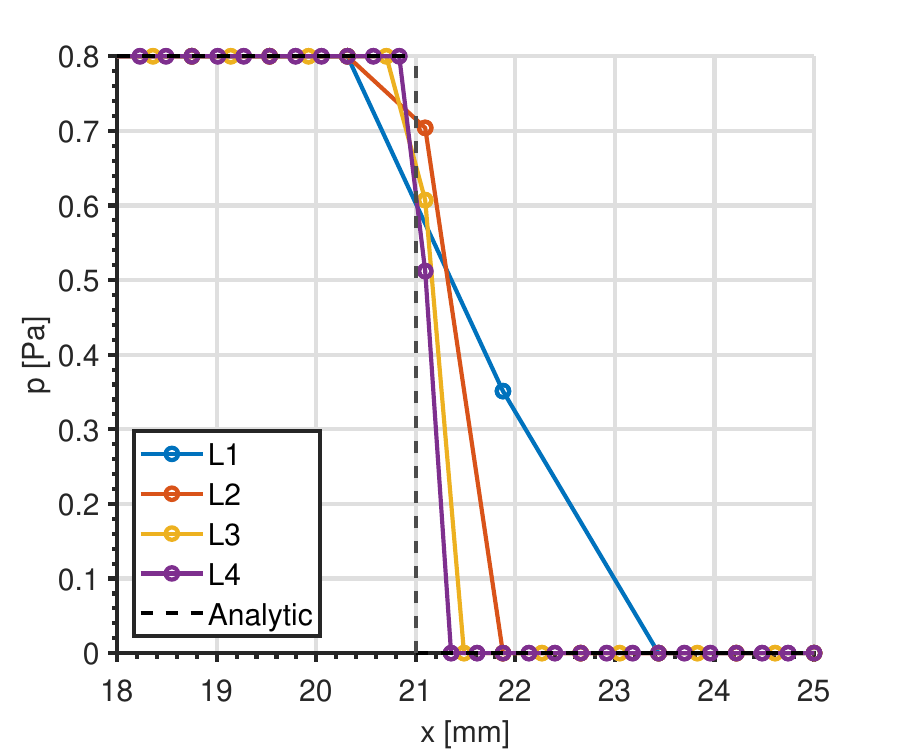}
  \caption{} % No text so (a) appears
  \label{subfig:1D_pjump_200ms}
\end{subfigure}%
\caption{Velocity and pressure jumps across \(\Omega_\Gamma\) at \(t=0.2\) s in the one-dimensional evaporating liquid problem. (a) velocity; (b) pressure.}
\label{fig:1D_jumps200ms}
\end{figure}

A numerical test following the one-dimensional liquid surface configuration described in Section \ref{subsec:1D_analysis} shows the recovery of the correct pressure and velocity jumps across \(\Omega_\Gamma\) in the one-fluid formulation, regardless of the mesh resolution, as seen in Figure \ref{fig:1D_jumps200ms}. The computational domain is defined with the wall at \(x=0\), the liquid-gas interface initially at \(x=25\) mm and an open boundary at \(x=150\) mm where the reference pressure is set at 0 Pa. The mass flux is set to \(\dot{m}''=10\) kg/(m\textsuperscript{2}s), resulting in \(p_l = 0.8\) Pa and \(u_g=0.08\) m/s. Various mesh resolutions are tested, which are given in Table \ref{tab:mesh_resolution} (i.e., L1, L2, L3 and L4). Although the error in the momentum balance introduced by the one-fluid velocity in the viscous term is negligible (i.e., well below 1\% for the tested grid sizes and fluid properties), the results in this section have been obtained by neglecting viscous effects in the one-fluid momentum equation. \par

\begin{table}
\begin{center}
\begin{tabular}{|c|c|} 
\hline
Mesh & Grid size [mm] \\
\hline
L1 & 150/96 \\ 
\hline
L2 & 150/192 \\
\hline
L3 & 150/384 \\
\hline
L4 & 150/576 \\
\hline
\end{tabular}
\caption{Grid size for each level of mesh refinement L1, L2, L3 and L4 used in the validation cases with fixed mass flux.}
\label{tab:mesh_resolution}
\end{center}
\end{table}

\begin{figure}
\centering
\begin{subfigure}{0.33\textwidth}
  \centering
  \includegraphics[width=1.0\linewidth]{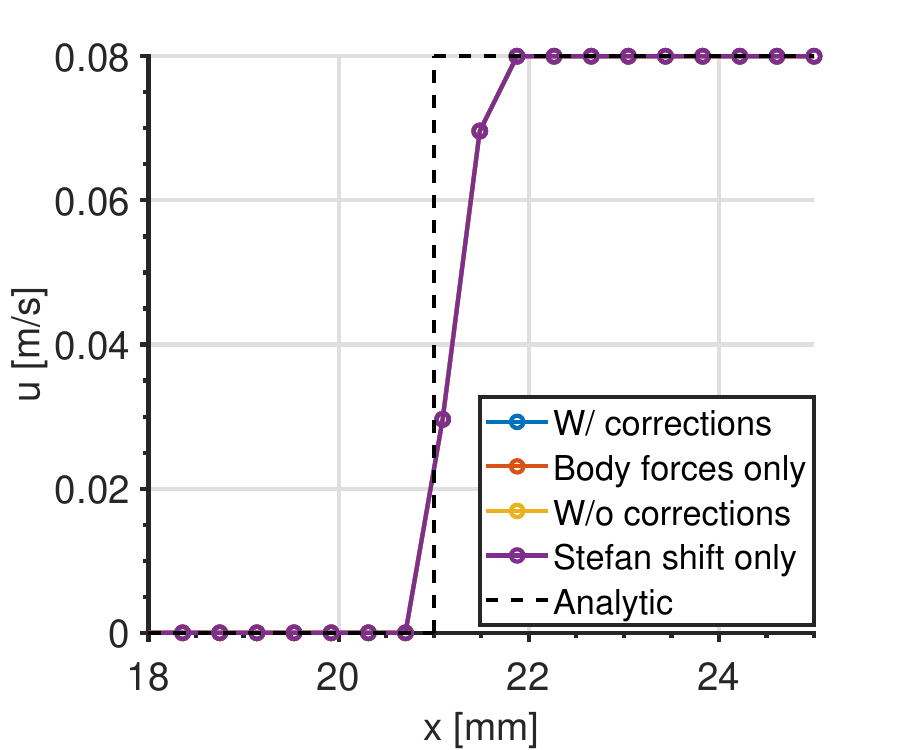}
  \caption{} % No text so (a) appears
  \label{subfig:1D_ujump_200ms_L3}
\end{subfigure}%
\begin{subfigure}{0.33\textwidth}
  \centering
  \includegraphics[width=1.0\linewidth]{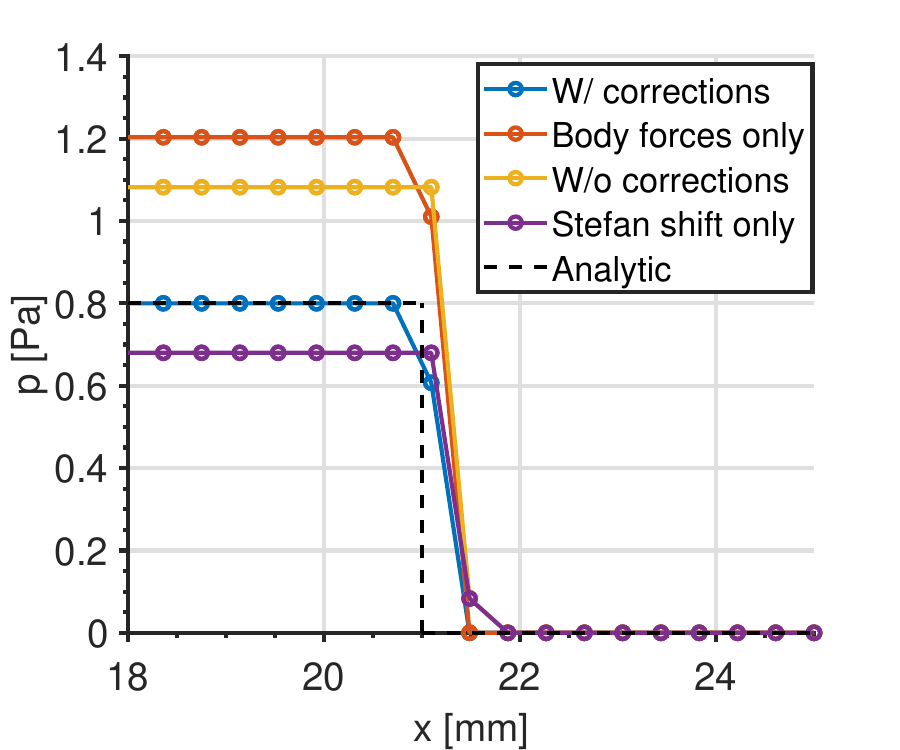}
  \caption{} % No text so (a) appears
  \label{subfig:1D_pjump_200ms_L3}
\end{subfigure}%
\begin{subfigure}{0.33\textwidth}
  \centering
  \includegraphics[width=1.0\linewidth]{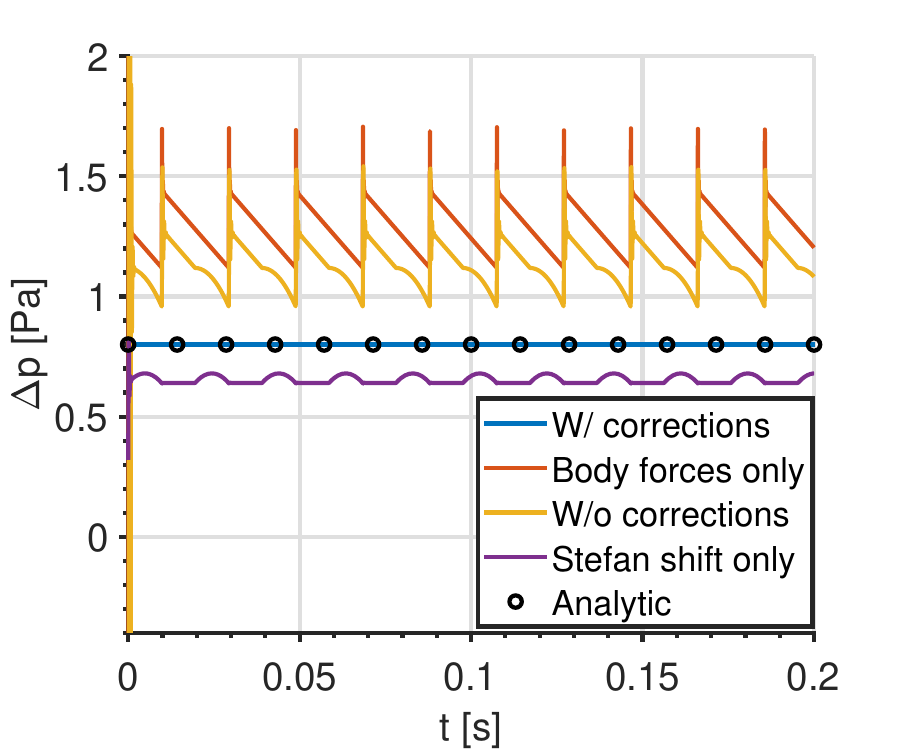}
  \caption{} % No text so (a) appears
  \label{subfig:1D_pjump_vs_time_L3}
\end{subfigure}%
\caption{Pressure and velocity jumps across \(\Omega_\Gamma\) in the one-dimensional evaporating liquid problem using mesh L3. The results with the flow solver corrections are compared against the results without corrections (i.e., traditional one-fluid approach), the results only adding the corrective body forces, and the results obtained by only considering the Stefan flow shift. (a) velocity at \(t=0.2\) s; (b) pressure at \(t=0.2\) s; (c) pressure jump vs. time.}
\label{fig:1D_jumps200ms_L3}
\end{figure}

The errors introduced by the one-fluid formulation without corrections (i.e., \(\mathbf{f}_{NC}=0\), \(\mathbf{f}_{\dot{m}''}=0\) and \(\mathbf{u}^{*}=\mathbf{u}^n\)) are visualized in Figure \ref{fig:1D_jumps200ms_L3} by comparing the results obtained with L3. Further, the figure also analyzes the individual effect of each correction. As expected, a pressure field is obtained that corrects the predicted velocity to satisfy \(\nabla\cdot\mathbf{u}^{n+1}\). Therefore, the velocity jump is the same with and without corrections (see Figure \ref{subfig:1D_ujump_200ms_L3}). However, the pressures are different. For example, the pressure difference \(p^*\) between the pressures \(p^{n+1}_{\text{FSM}}\) (W/o corrections) and \(p^{n+1}\) (Stefan shift only), as defined in Section \ref{subsec:LUdecomp}, can be as large as 1 Pa or 25\% more than the actual pressure jump. Moreover, \(p^*\) varies over time as the interface moves across cells, as seen in Figures \ref{subfig:1D_pjump_200ms_L3} and \ref{subfig:1D_pjump_vs_time_L3}. For this particular configuration, the momentum imbalance caused by the convective term does not vary much over time as the interface crosses grid cells; thus, the addition of \(\mathbf{f}_{\dot{m}''}\) and \(\mathbf{f}_{NC}\) mainly correct the pressure offset necessary to obtain the exact \(\Delta p\) with \(p^{n+1}\). In this case, the pressure jump is underestimated by more than 12.5\% without the additional body forces. In contrast, the Stefan flow shift is responsible for the elimination of the large temporal pressure oscillations that take some time to decrease before re-emerging once a new cell is suddenly identified as belonging to \(\Omega_\Gamma\). That is, \(\Delta p\) spikes as the one-fluid suddenly accelerates due to the volumetric expansion caused by phase change when \(\Omega_\Gamma\) moves across cells. \par 

\subsection{Static Evaporating Droplet}
\label{subsec:droplet}

The flow solver is tested for a static droplet evaporating under constant \(\dot{m}''\) to analyze the physical consistency of the solution and stability of the proposed method. An analytical solution to a simplified version of this problem can be obtained by decoupling the velocity calculation from the pressure. Under spherical symmetry, the radial velocity \(u_r\) is obtained from the continuity equation in spherical coordinates, resulting in

\begin{equation}
\label{eqn:droplet_radial_velocity}
u_r(r,t) = 
    \begin{cases}
        0 & \text{if $r<R(t)$}\\
        u_G\bigg(\frac{R(t)}{r}\bigg)^2& \text{if $r\geq R(t)$}\\
    \end{cases}
\end{equation}

\noindent
where \(u_G=\dot{m}''(\rho_G^{-1}-\rho_L^{-1})\). Next, the pressure is calculated by substituting \(u_r\) into the momentum equation, also in spherical coordinates. Assuming that \(p_{r\rightarrow\infty}=0\) and a quasi-steady state evolution of the interface location, the pressure is given by

\begin{equation}
\label{eqn:droplet_radial_pressure}
p(r,t) = 
    \begin{cases}
        p_G + \sigma\kappa + (\dot{m}'')^2(\rho_G^{-1}-\rho_L^{-1}) + 4\frac{\mu_G}{R(t)}u_G & \text{if $r<R(t)$}\\
        -\frac{1}{2}\rho_G u_{G}^2\bigg(\frac{R(t)}{r}\bigg)^4& \text{if $r\geq R(t)$}\\
    \end{cases}
\end{equation}

\noindent
with \(p_G=-\frac{1}{2}\rho_G u_{G}^2\). In Eqs. (\ref{eqn:droplet_radial_velocity}) and (\ref{eqn:droplet_radial_pressure}), the droplet radius is given as a function of time by \(R(t)=R_0-\frac{\dot{m}''}{\rho_L}t\) where \(R_0\) is the initial radius. Note the pressure jump due to the normal viscous stress is included despite being negligible compared to the other terms. \par 

For the numerical test, a droplet with an initial diameter \(d_0=25\) mm is suspended at the center of the computational domain with a side length of 150 mm (i.e., \(6d_0\)). Open boundaries are considered in all directions, which have been approximated with homogeneous Neumann boundary conditions for all variables. The direct pressure solver imposes such condition in \(x\) and \(y\), while a reference pressure of 0 Pa is imposed on one point at the top \(z\) boundary. This treatment aims to recover the analytical solution for the pressure field given by \(p_{r\rightarrow\infty}=0\) under the assumption that the computational domain is large enough. \par

A range of \(\dot{m}''\) and \(\sigma\) values are considered (see Table \ref{tab:droplet_cases}) to test the accuracy of the one-fluid corrections and the interplay with surface tension. A non-dimensional time is defined as \(t^*=t/t_c\), where the characteristic time is given by \(t_c=\rho_L d_0/ \dot{m}''\). Note the scaling with the interface regression velocity \(\mathbf{u}_\Gamma\) for a static droplet. \par

\begin{table}
\begin{center}
\begin{tabular}{|c|c|c|} 
\hline
Case & $\dot{m}''$ [kg/(m\textsuperscript{2}s)] & $\sigma$ [mN/m] \\
\hline
1 & 0.1 & 1 \\ 
\hline
2 & 10 & 0.01 \\
\hline
3 & 10 & 1 \\
\hline
\end{tabular}
\caption{Mass flux per unit area, \(\dot{m}''\), and surface tension coefficient, \(\sigma\), for the different configurations of the static evaporating droplet.}
\label{tab:droplet_cases}
\end{center}
\end{table}

\subsubsection{Flow Field}
\label{subsubsec:staticdrop_solution}

\begin{figure}
\centering
\begin{subfigure}{0.33\textwidth}
  \centering
  \includegraphics[width=1.0\linewidth]{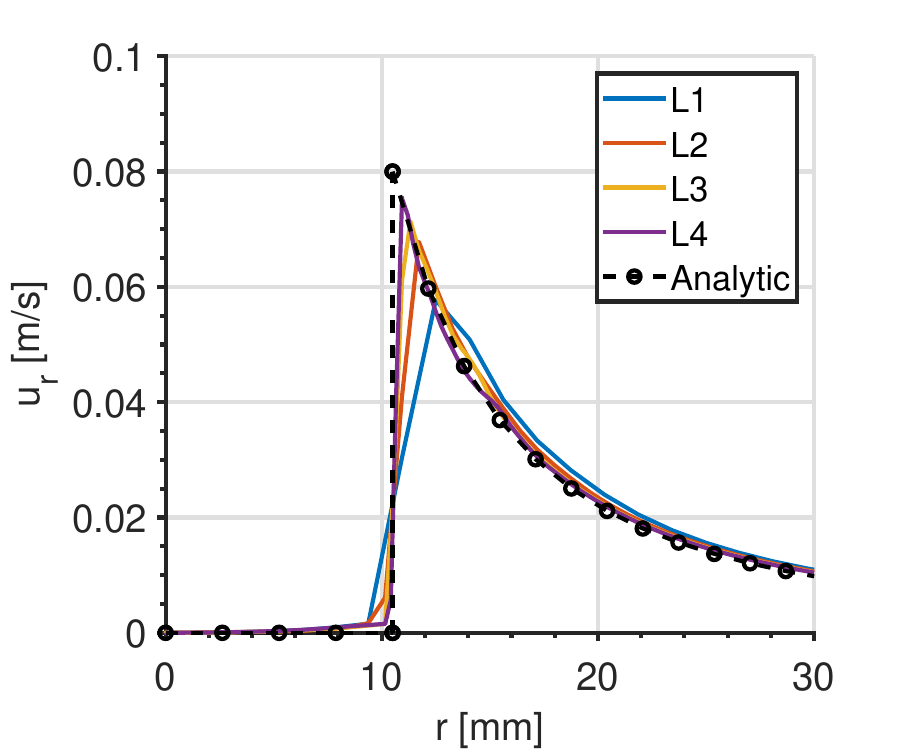}
  \caption{} % No text so (a) appears
  \label{subfig:droplet_ujump_100ms_C2}
\end{subfigure}%
\begin{subfigure}{0.33\textwidth}
  \centering
  \includegraphics[width=1.0\linewidth]{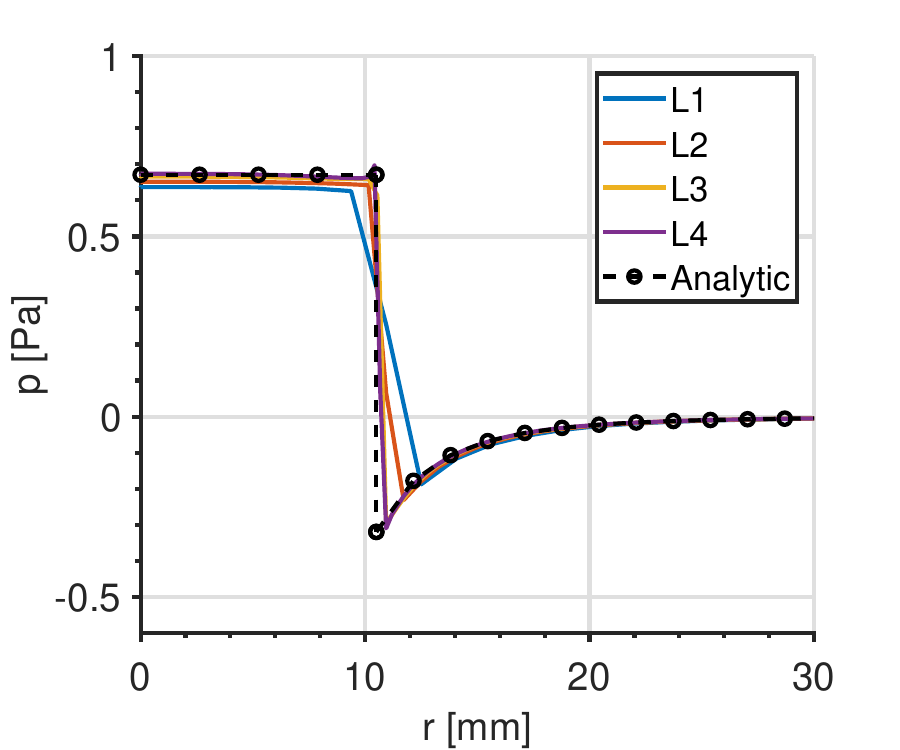}
  \caption{} % No text so (a) appears
  \label{subfig:droplet_pjump_100ms_C2}
\end{subfigure}%
\begin{subfigure}{0.33\textwidth}
  \centering
  \includegraphics[width=1.0\linewidth]{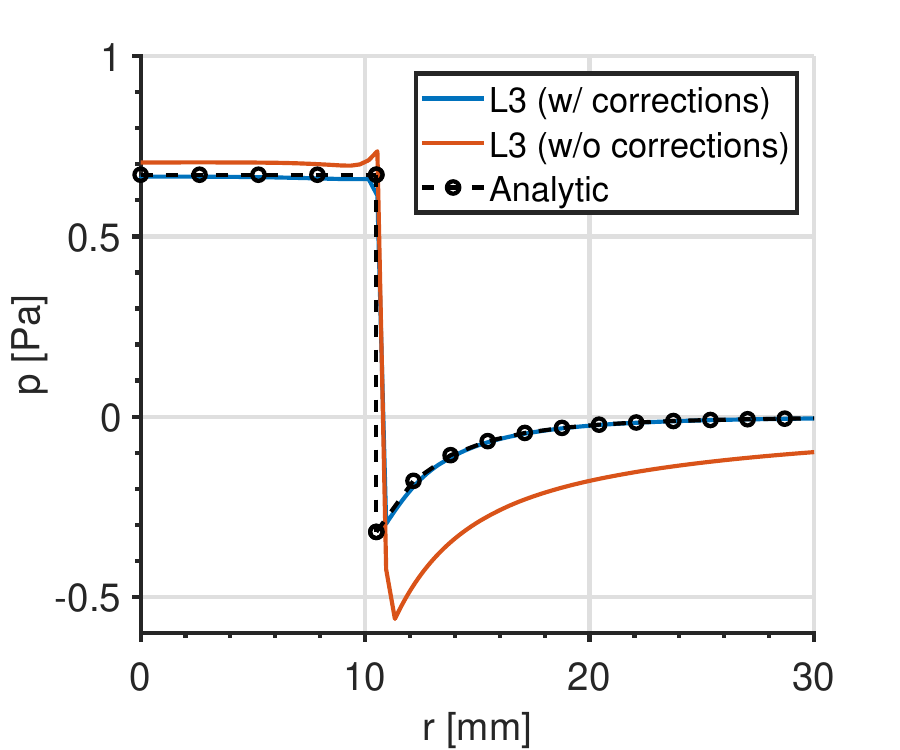}
  \caption{} % No text so (a) appears
  \label{subfig:droplet_pjump_100ms_C2_L3}
\end{subfigure}%
\caption{Velocity and pressure jumps across \(\Omega_\Gamma\) at \(t^*=0.08\) s in the static evaporating droplet for case 3. (a) velocity; (b) pressure; (c) pressure with and without flow solver corrections using mesh L3.}
\label{fig:droplet_jumps_100ms_C2}
\end{figure}

Similar to Section \ref{subsec:1D_liquid}, Figures \ref{subfig:droplet_ujump_100ms_C2} and \ref{subfig:droplet_pjump_100ms_C2} show the radial profiles of \(\mathbf{u}\) and \(p\) extracted along the \(z\) direction from the droplet's center for case 3 at \(t^*=0.08\) s. The same mesh resolutions L1, L2, L3 and L4 from Table \ref{tab:mesh_resolution} are considered, which show good convergence to the analytical solution given by Eqs. (\ref{eqn:droplet_radial_velocity}) and (\ref{eqn:droplet_radial_pressure}). Because of the three-dimensional nature of the problem, \(\mathbf{u}\) and \(p\) show some oscillations around \(\Omega_\Gamma\) affected by the grid resolution and spurious currents. \par

The impact of not considering the flow solver corrections is seen in Figure \ref{subfig:droplet_pjump_100ms_C2_L3} using L3. As expected from the discussion in Section \ref{subsec:1D_liquid}, a bigger pressure jump is introduced without the corrections, while the correct \(\Delta p\) is captured with the corrections. Compared to the one-dimensional results shown in Figure \ref{fig:1D_jumps200ms_L3}, the pressure field is unphysical in both cases. In particular, the pressure field in the gas phase without corrections may be more consistent with the unsteady analytical solution but overestimates \(\Delta p\), while the pressure field with corrections misses the unsteady effect of the shift in Stefan flow but reproduces the quasi-steady analytical solution and imposes the exact \(\Delta p\). \par 

\begin{figure}
\centering
\begin{subfigure}{0.33\textwidth}
  \centering
  \includegraphics[width=1.0\linewidth]{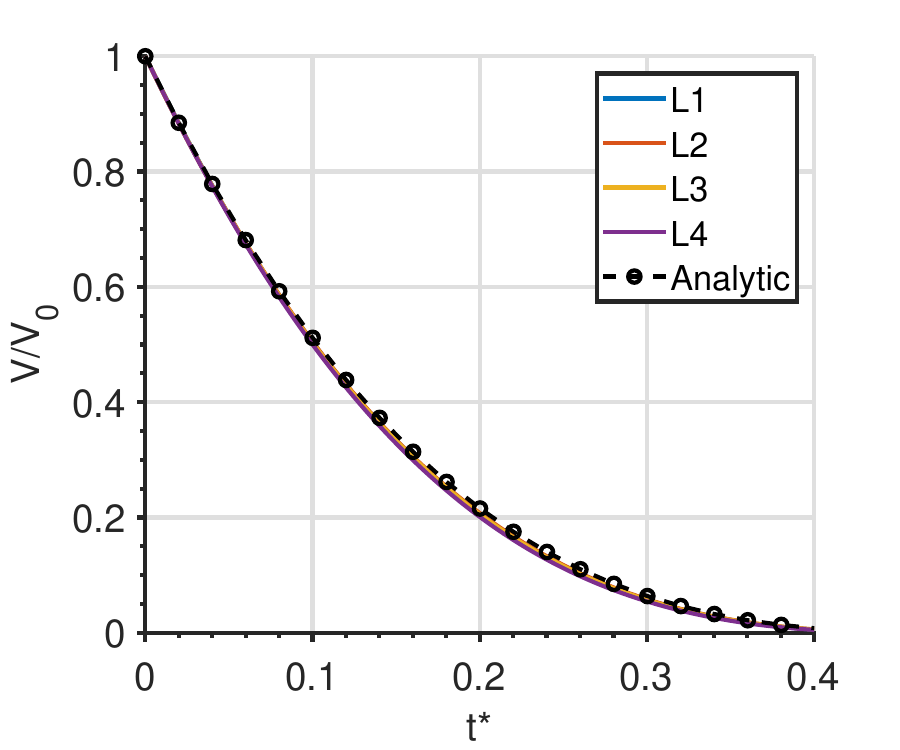}
  \caption{} % No text so (a) appears
  \label{subfig:droplet_volume_C2}
\end{subfigure}%
\begin{subfigure}{0.33\textwidth}
  \centering
  \includegraphics[width=1.0\linewidth]{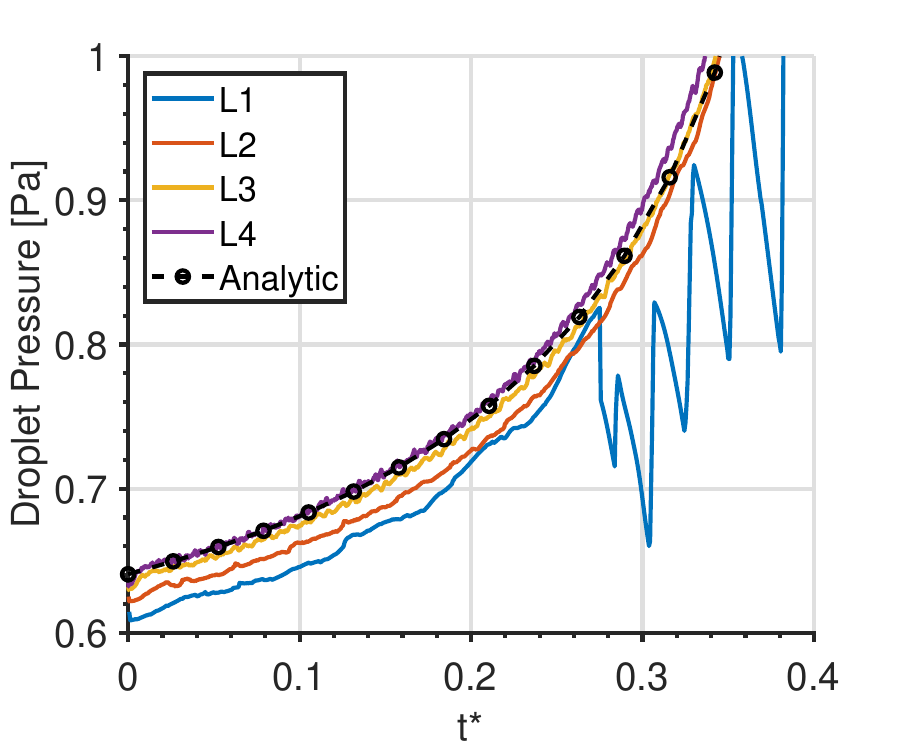}
  \caption{} % No text so (a) appears
  \label{subfig:droplet_pressure_C2}
\end{subfigure}%
\begin{subfigure}{0.33\textwidth}
  \centering
  \includegraphics[width=1.0\linewidth]{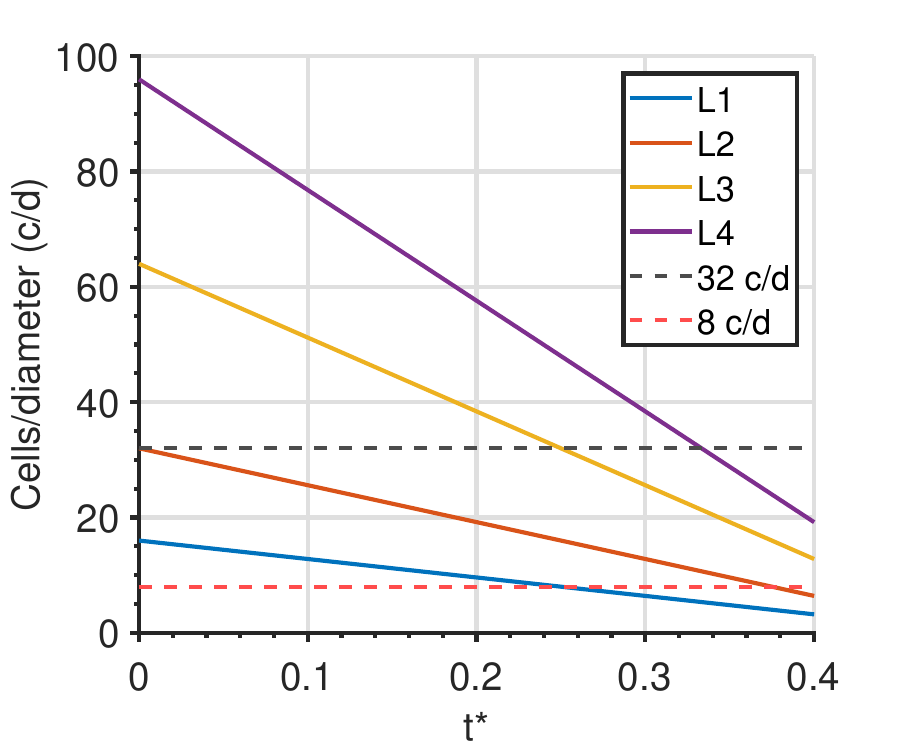}
  \caption{} % No text so (a) appears
  \label{subfig:droplet_resolution}
\end{subfigure}%
\caption{Temporal evolution of droplet volume, pressure and resolution in the static evaporating droplet for case 3. (a) normalized volume; (b) average internal droplet pressure; (c) grid resolution.}
\label{fig:droplet_C2_time}
\end{figure}

Figures \ref{subfig:droplet_volume_C2} and \ref{subfig:droplet_pressure_C2} show, respectively, the temporal evolution of the droplet volume normalized by the initial volume \(V_0=\tfrac{4}{3}\pi R_{0}^{3}\) and the droplet's internal pressure for case 3, taken as the average of the pressure value in all pure liquid cells (i.e., \(C=1\)). For reference, Figure \ref{subfig:droplet_resolution} presents the droplet resolution over time where the thresholds of 32 cells per diameter (c/d) and 8 c/d are emphasized. That is, 32 c/d is oftentimes assumed as a well-resolved droplet. However, lower resolutions are often used in configurations with multiple droplets or bubbles (e.g., 25 c/d in \cite{2020_IJMF_Cifani}). Note that L2 covers this resolution range with the initial droplet being resolved with 32 c/d. \par

All meshes capture the volume of the droplet accurately and maintain the spherical shape within resolution limits. A similar behavior is shown in \cite{2021_JCP_Malan}. However, a tendency to underpredict the droplet's volume is observed in Figure \ref{subfig:droplet_volume_C2}, especially for \(t^*>0.1\). This affects the pressure inside the droplet shown in Figure \ref{subfig:droplet_pressure_C2} as it shrinks and the pressure jump due to surface tension becomes more important, i.e., a larger curvature is predicted due to the smaller volume. A formal error analysis is presented in Section \ref{subsubsec:staticdrop_error}. \par 

\begin{figure}
\centering
\includegraphics[width=1.0\linewidth]{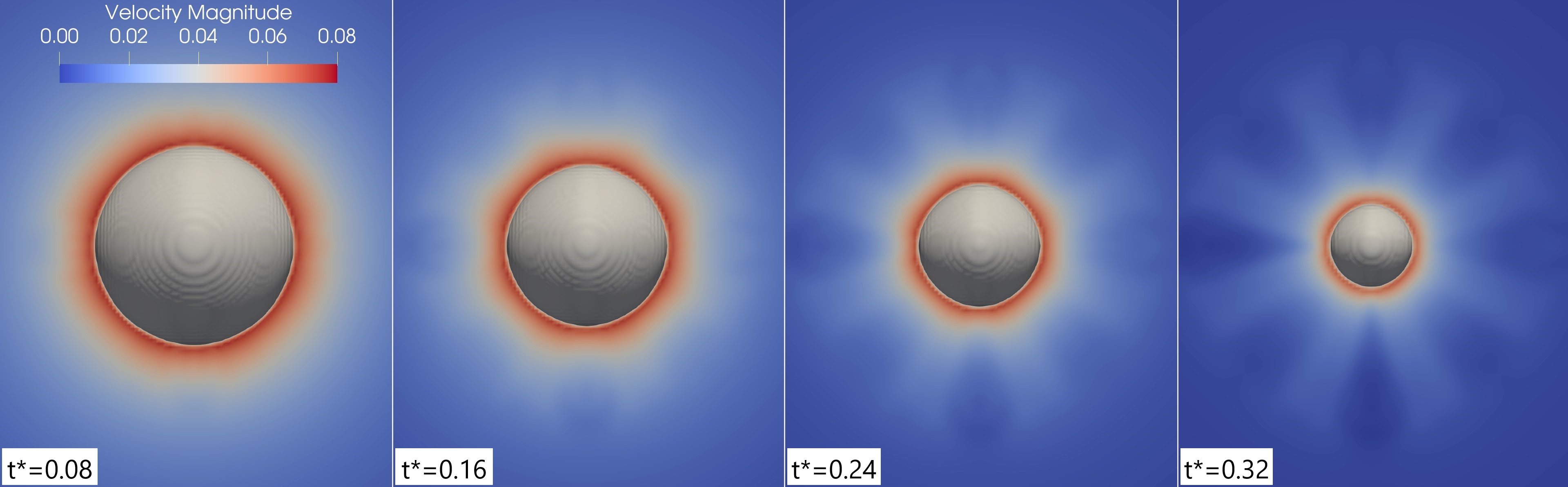}
\caption{Temporal evolution of the droplet represented by the iso-surface \(C=0.5\) for case 3 with mesh L3. The velocity magnitude resulting from the Stefan flow is shown on the \(xy\) plane across the center of the droplet. The length scale in each snapshot remains constant.}
\label{fig:droplet_C2}
\end{figure}

The evolution of the droplet shape is shown in Figure \ref{fig:droplet_C2} for case 3 with mesh L3 together with the magnitude of \(\mathbf{u}\), plotted on the \(xy\) plane at \(z=0\). The droplet remains nearly spherical despite the spurious currents that appear around the interface, which deteriorate over time as the droplet resolution decreases. The sphericity of the droplet is shown in Figure \ref{fig:sphericity}. These affect the Stefan flow in the gas phase and some degree of non-symmetry is seen. 8 c/d has been observed to be the limit where shape deterioration becomes evident, spurious currents rise considerably and numerical oscillations in, e.g., the internal droplet pressure, significantly increase (see L1 in Figure \ref{subfig:droplet_pressure_C2}). \par 

\subsubsection{Numerical Stability and Error Analysis}
\label{subsubsec:staticdrop_error}

The extension of the one-fluid formulation to two-phase flows undergoing phase change introduces numerical errors beyond the physical consistency of the results. The extensive use of non-smooth geometrical information (i.e., \(\kappa\) and \(\mathbf{n}_\Gamma\)) from the interface reconstruction in the calculation of \(\mathbf{f}_{\sigma}\), \(\mathbf{f}_{NC}\) or the interface shift in Eq. (\ref{eqn:VOFindicator}) to account for phase change in the VOF advection generate spurious currents around the interface that further deteriorate \(\mathbf{n}_\Gamma\) and \(\kappa\). Moreover, these numerical oscillations easily affect the extrapolation of \(\hat{p}\) and the stability of the flow solver. Partially affected by the low viscosities considered here, i.e., less dissipation, these issues intrinsic to geometric VOF frameworks are evident in the context of a static evaporating droplet where no convective flow exists and the velocities are solely a result of the Stefan flow and spurious currents. \par 

This motivates the use of PPIC to improve the calculation of \(\kappa\), which is a more sensitive parameter than \(\mathbf{n}_\Gamma\) under geometrical deterioration. In fact, from the cases outlined in Table \ref{tab:droplet_cases}, a direct benefit of using PPIC vs. PLIC is observed in achieving a smoother flow around the interface and stabilizing the solution throughout the entire droplet evaporation process (within resolution limits) for most grid sizes. With PLIC, case 1 eventually becomes unstable as spurious currents grow unbounded. That is, as \(\dot{m}''\) decreases, the spurious currents become of the order of the Stefan flow and dominate the dynamics in \(\Omega_\Gamma\). The use of energy-preserving schemes (i.e., balancing surface and kinetic energy) has been shown to guarantee the stability of the system \cite{2020_JCP_Valle} and may be pursued in future work. \par

\begin{figure}
\centering
\begin{subfigure}{0.33\textwidth}
  \centering
  \includegraphics[width=1.0\linewidth]{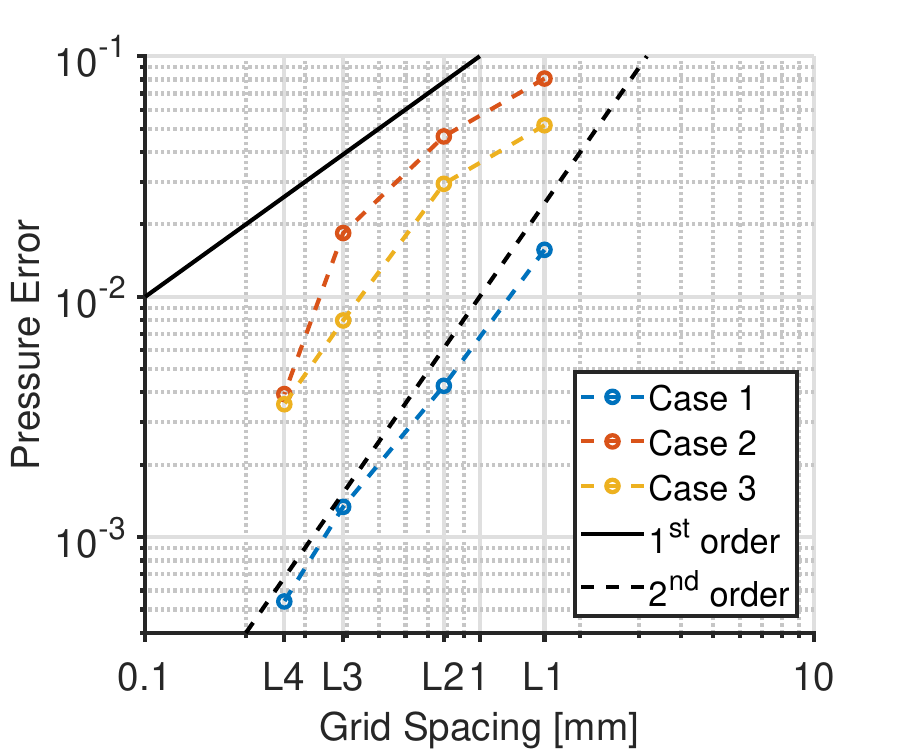}
  \caption{} % No text so (a) appears
  \label{subfig:droplet_perr_100ms}
\end{subfigure}%
\begin{subfigure}{0.33\textwidth}
  \centering
  \includegraphics[width=1.0\linewidth]{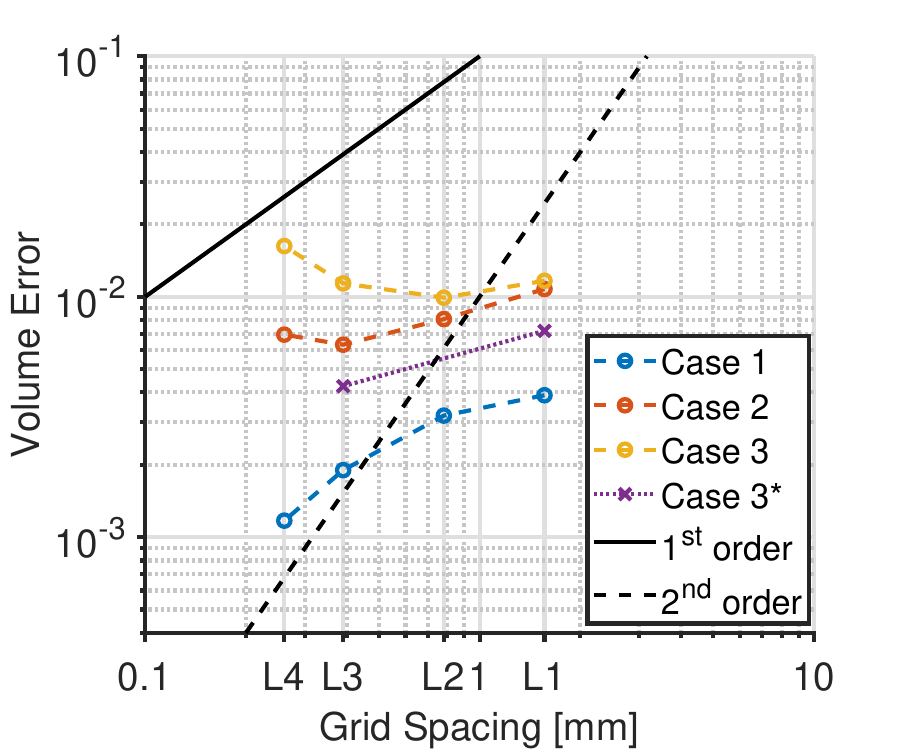}
  \caption{} % No text so (a) appears
  \label{subfig:droplet_volerr_100ms}
\end{subfigure}%
\begin{subfigure}{0.33\textwidth}
  \centering
  \includegraphics[width=1.0\linewidth]{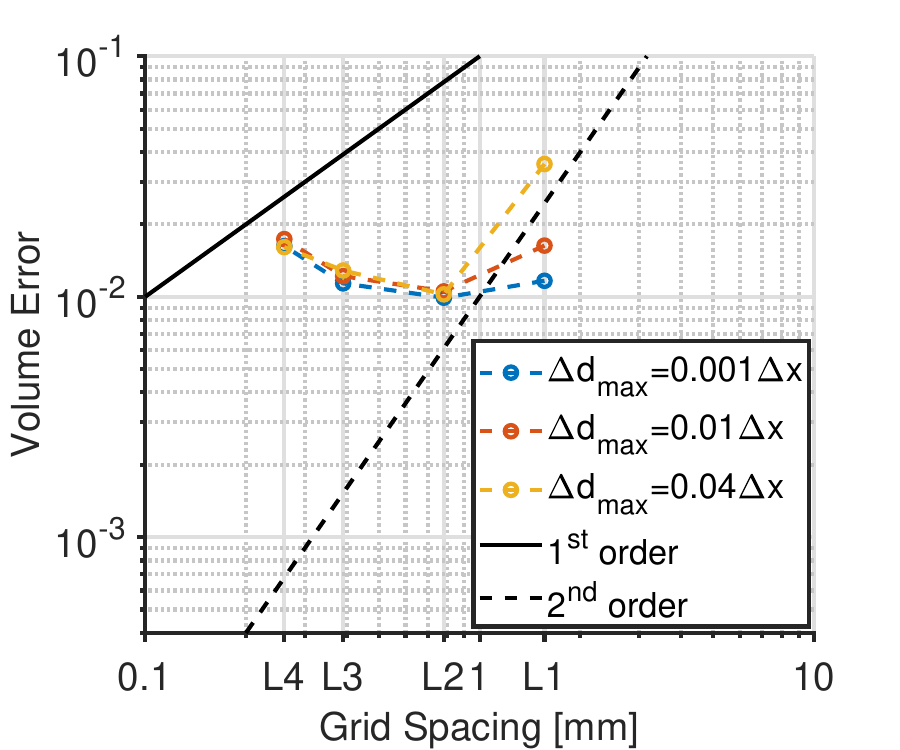}
  \caption{} % No text so (a) appears
  \label{subfig:droplet_volerr_dt_100ms_C2}
\end{subfigure}%
\caption{Grid convergence of the internal droplet pressure and droplet volume at \(t^*=0.08\) s in the static evaporating droplet. (a) pressure; (b) volume; (c) volume convergence for case 3 under the effect of the phase-change CFL constraint given by \(\Delta\text{d}_{\text{max}}\). Case 3* corresponds to case 3 but without solving the momentum equation.}
\label{fig:droplet_errors}
\end{figure}

Error metrics in terms of convergence with grid refinement are evaluated for the droplet's internal pressure and volume at \(t^*=0.08\), as shown in Figures \ref{subfig:droplet_perr_100ms} and \ref{subfig:droplet_volerr_100ms}, respectively. Here, the mesh resolutions from Table \ref{tab:mesh_resolution} are also used. At this \(t^*\), a large number of numerical integration steps have been taken and the volume has been reduced to \(V/V_0=0.592\). Two scenarios are considered by fixing \(\sigma=1\) mN/m and varying \(\dot{m}''\) (i.e., cases 1 and 3)  and, then, by fixing \(\dot{m}''=10\) kg/(m\textsuperscript{2}s) and varying \(\sigma\) (i.e., cases 2 and 3). The relative error at \(t^*=0.08\) is evaluated as \(E=|\phi_n-\phi_e|/\phi_e\), where \(\phi_n\) and \(\phi_e\) are the value of any variable \(\phi\) obtained, respectively, from the numerical solution and the exact (analytical) solution. \par

The convergence of the internal pressure error is nearly second order. This behavior is expected for a static droplet with \(\dot{m}''=0\) given the modeling of \(\mathbf{f}_\sigma\) in the context of the CSF approach and shows that the addition of \(\mathbf{f}_{\dot{m}''}\) and \(\mathbf{f}_{NC}\) behaves similarly. However, larger \(\dot{m}''\) values (cases 2 and 3) result in convergence rates varying between first and second order depending on the range of mesh sizes considered. Specifically, the convergence rate drops closer to first order when \(\sigma\rightarrow 0\) (case 2). Assuming that \(\mathbf{f}_{\dot{m}''}\) behaves similar to \(\mathbf{f}_\sigma\), it means the discretization of \(\mathbf{f}_{NC}\) is first-order at best. This is a result of the stability (i.e., source term smoothness) trade-off between a CSF-like discretization and the approach described in Section \ref{subsec:navierstokes_solve} using staggered volume fractions to obtain the interfacial terms in \(\mathbf{f}_{NC}\). \par 

Despite capturing the correct volume decrease and maintaining a spherical shape, grid convergence of the droplet's volume is hardly achieved with the plane shift strategy described in Section \ref{subsec:intcap}. This is a clear consequence of the lack of smoothness in \(\mathbf{n}_\Gamma\). For instance, Figure \ref{subfig:droplet_volerr_100ms} shows that the method benefits from the smoother \(\mathbf{n}_\Gamma\) distribution when the spurious currents in \(\Omega_\Gamma\) are less, such as when \(\dot{m}'' \rightarrow 0\) (case 1) or \(\sigma \rightarrow 0\) (case 2). However, since grid refinement demands a larger number of time steps to reach the same \(t^*=0.08\), the accumulation of geometrical errors during plane shifts prevents a clear converging trend. This is supported by the addition of case 3\(^*\) in Figure \ref{subfig:droplet_volerr_100ms}, which corresponds to case 3 (or case 2) but without solving the momentum equation in the algorithm. That is, the effects of surface tension are not included and the pressure is only used to impose the Stefan flow. This eliminates the spurious currents caused by surface tension and other CSF-like forces, and results in lower volume errors and better convergence. However, the convergence rate is still lower than first order, highlighting the limitations of plane shifting at high \(\dot{m}''\). Nonetheless, all mesh resolutions from Table \ref{tab:mesh_resolution} display volume errors around 1\% or less, in line with other works using different methodologies to handle phase change in VOF methods (e.g., advection of \(C\) with \(\mathbf{u}_\Gamma\) \cite{2020_JCP_Scapin,2023_hal_Cipriano}). \par 

\begin{figure}
\centering
\includegraphics[width=0.45\linewidth]{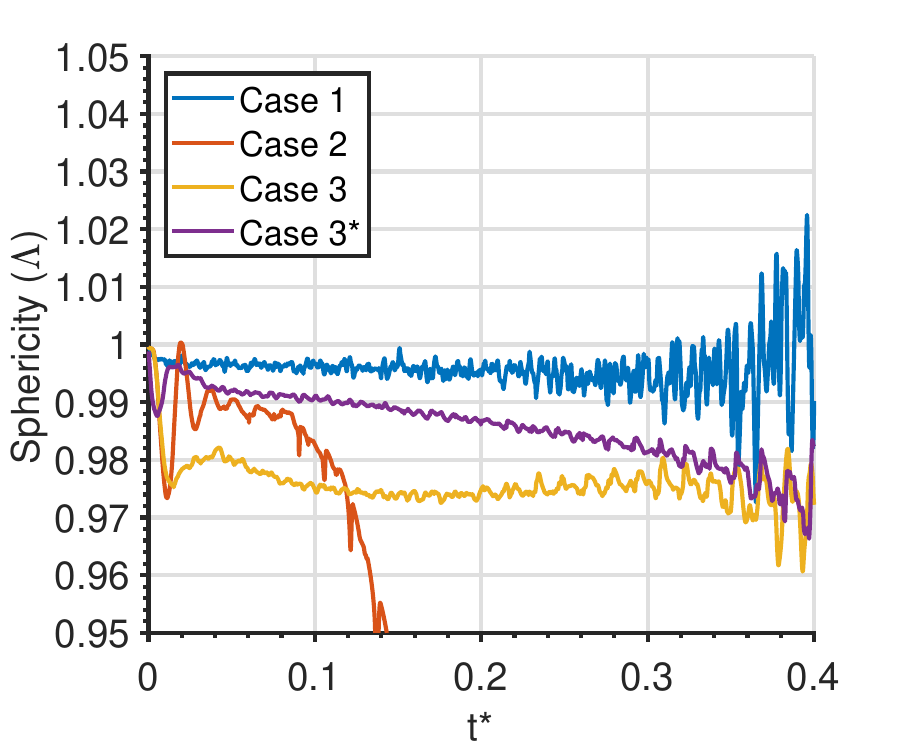}
\caption{Temporal evolution of droplet sphericity (\(\Lambda\)) in the static evaporating droplet with mesh L3. Case 3* corresponds to case 3 but without solving the momentum equation.}
\label{fig:sphericity}
\end{figure}

Geometry errors are also measured in terms of the sphericity of the vaporizing droplet, \(\Lambda=[\pi^{1/3}(6V)^{2/3}]/A\) where \(V\) is the droplet volume and \(A\) its surface area calculated as the sum of the area of all the local interface reconstruction planes. The temporal evolution of \(\Lambda\) is shown in Figure \ref{fig:sphericity} for all cases with mesh L3, including case 3\(^*\). The data extends beyond the time reported in Figure \ref{fig:droplet_errors} and clearly shows the enhanced accumulation of geometrical errors as \(\dot{m}''\) increases (case 1 vs. case 3). Without the restoring effect of surface tension (case 2), the geometrical errors may degrade the spherical shape completely. Further proof that this is a result of the spurious currents generated around the interface and not the plane shifting strategy is given by case 3\(^*\), which preserves the spherical shape of the vaporizing droplet better over time. Given these results, a clear connection between the deterioration of \(\Lambda\) and the volume convergence errors is observed. \par

Note that this validation test imposes a restrictive CFL condition based on a maximum allowed interface shift due to phase change in a given time step of \((\Delta\text{d}_{\text{max}}/\Delta x)=0.001\). However, no significant improvement in the volume errors is observed by varying \((\Delta\text{d}_{\text{max}}/\Delta x)\) or \(\theta_3\) from \(0.04\) to \(0.001\) (see \ref{apn:C} for the definition of \(\theta_3\)). Figure \ref{subfig:droplet_volerr_dt_100ms_C2} shows the volume errors for case 3 and, similar to Figure \ref{subfig:droplet_volerr_100ms}, grid convergence is not achieved. Except for mesh L1, volume errors are almost identical regardless of \(\Delta\text{d}_{\text{max}}\). Thus, a maximum plane shift around \((\Delta\text{d}_{\text{max}}/\Delta x)=0.01\) is more reasonable to avoid reducing the time step significantly if its calculation is limited by the evaporation of the interface. Beyond the reported trends, reducing the time step delays the onset of numerical instabilities because of two main reasons: (a) smaller plane shifts in the phase change correction of \(C\) deteriorate less the geometry; and (b) the errors in the temporal extrapolation of \(\hat{p}\), \(\hat{\mathbf{f}}_\sigma\) and \(\hat{\mathbf{f}}_{\dot{m}''}\) are reduced. Thus, spurious currents are smaller and geometrical calculations do not degrade as fast. \par

By pushing the solver to its limits in the context of a static configuration, some limitations of the proposed treatment of phase change in the geometric advection of \(C\) and the discretization of the various corrective source terms added in the momentum equation have been highlighted. Nonetheless, the magnitude of \(\dot{m}''\) is typically smaller, surface tension effects prevail (unless very close to the critical point of the fluid), and dynamic configurations aid in masking the spurious currents. \par

\section{Coupling with Energy Equation}
\label{sec:val_withenergy}

This section addresses the fully-coupled system involving the energy equation and the calculation of the interfacial mass flux. Section \ref{subsec:bubblegrowth} looks at the growth of a static vapor bubble in superheated liquid, while Section \ref{subsec:bubblerising} considers a dynamic setup where a bubble rises in superheated liquid due to the effects of buoyancy. These analyses look at the correct solution of the energy equation and the effects of the momentum imbalance and its corrections. Also, they involve a variety of fluid properties and show that the flow solver behaves well at, e.g., low and high density ratios and different surface tension coefficients. The relevant properties of the fluids considered here are listed in Table \ref{tab:rising_bubble_cases}. \par 

\begin{table}
\begin{center}
\begin{tabular}{|c|c|c|c|c|} 
\hline
 & Fluid A & Fluid B & Ethanol \\
\hline
\(p\) [bar] & - & - & 1.013 \\
\hline
\(\rho_L\) [kg/m\textsuperscript{3}] & 2.5 & 200 & 757 \\ 
\hline
\(\rho_G\) [kg/m\textsuperscript{3}] & 0.25 & 5 & 1.435 \\
\hline
\(\mu_L\) [\(\mu\)Pa\(\cdot\)s] & \(7 \times 10^3\) & \(1\times 10^5\) & 429 \\
\hline
\(\mu_G\) [\(\mu\)Pa\(\cdot\)s] & \(7 \times 10^2\) & \(5\times 10^3\) & 10.4 \\
\hline
\(k_L\) [W/(m\(\cdot\)K)] &  \(7 \times 10^{-2}\) & 1 & 0.154 \\
\hline
\(k_G\) [W/(m\(\cdot\)K)] & \(7 \times 10^{-3}\) & 1 & 0.02 \\
\hline
\(c_{p,L}\) [kJ/(kgK)] & 0.0025 & 0.2 & 3 \\
\hline
\(c_{p,G}\) [kJ/(kgK)] & 0.001 & 0.2 & 1.83 \\
\hline
\(\sigma\) [mN/m] & 1 & 100 & 18 \\
\hline
\(T_{sat}\) [K] & 1 & 1 & 351.45 \\
\hline
\(h_{LV}\) [kJ/kg] & 0.1 & 10 & 963 \\
\hline
\end{tabular}
\caption{Properties of the test fluids (A and B) and ethanol used in the bubble growth in superheated liquid studies and two-dimensional film boiling \cite{2013_JCP_Sato,2021_JCP_Malan,2023_CaF_Boyd}.}
\label{tab:rising_bubble_cases}
\end{center}
\end{table}

\subsection{Static Bubble Growth in Superheated Liquid}
\label{subsec:bubblegrowth}

The growth of a static saturated vapor bubble in superheated liquid without gravity is a common validation test performed in the literature for multiphase solvers with heat transfer \cite{2013_JCP_Sato,2021_JCP_Malan,2022_JFM_Gao,2023_CaF_Boyd}. Under spherical symmetry, an analytical solution with constant fluid properties and including the effects of radial convection is provided by Scriven \cite{1959_CES_Scriven}. The bubble radius is given by

\begin{equation}
\label{eqn:Scriven1}
R(t) = 2\beta\sqrt{\frac{k_L}{c_{p,L}\rho_L}t}
\end{equation}

\noindent
where the value of the growth constant \(\beta\) is obtained from solving

\begin{equation}
\label{eqn:Scriven2}
\frac{\rho_L c_{p,L}(T_\infty-T_{sat})}{\rho_G\big(h_{LV}+(c_{p,L}-c_{p,G})(T_\infty-T_{sat})\big)}=2\beta^2\int\limits_{0}^{1} \exp{\bigg[-\beta^2\bigg((1-\zeta)^{-2}-2\bigg(1-\frac{\rho_G}{\rho_L}\bigg)\zeta-1\bigg)\bigg]}d\zeta
\end{equation}

\noindent
with \(T_{sat}\) and \(T_\infty\) being, respectively, the saturation temperature and superheated temperature (i.e., \(T_\infty>T_{sat}\)). Furthermore, the temperature along the radial direction is obtained from

\begin{equation}
\label{eqn:Scriven3}
T(r,t)=
\begin{cases}
    T_{sat} & \text{if $r\leq R(t)$}\\
    T_\infty-2\beta^2\Bigg(\frac{\rho_G\big(h_{LV}+(c_{p,L}-c_{p,G})(T_\infty-T_{sat})\big)}{\rho_L c_{p,L}}\Bigg)\int\limits_{1-R(t)/r}^{1} \exp{\bigg[-\beta^2\bigg((1-\zeta)^{-2}-2\bigg(1-\frac{\rho_G}{\rho_L}\bigg)\zeta-1\bigg)\bigg]}d\zeta & \text{if $r>R(t)$}\\
\end{cases}
\end{equation}

Similar to other works \cite{2021_JCP_Malan,2023_CaF_Boyd}, a fluid with the properties of fluid A shown in Table \ref{tab:rising_bubble_cases} is considered, which is defined for numerical testing purposes and may not resemble any real fluid. The bubble is immersed in a superheated liquid with \(T_\infty=3\) K. These conditions may arise when a sudden decrease in system pressure causes the boiling of a previously saturated liquid at a higher pressure. The assumption that the change in \(T_{sat}\) is small holds, and the effect of temperature on the fluid properties is neglected, i.e., remain constant. In terms of the Jakob number, \(Ja=[\rho_L c_{p,L}(T_\infty-T_{sat})]/\rho_G h_{LV}\), this configuration with the properties of fluid A corresponds to \(Ja=0.5\). \par 

The computational domain is a cubic box of size 12\(d_0\) with the same boundary conditions used in Section \ref{subsec:droplet}. The reason for the bigger domain is to limit the effect of the boundary conditions on the solution as the bubble grows and gets closer to the domain boundaries. The problem is initialized with the analytical solution at \(t=0.5\) s with the bubble of diameter \(d_0\approx0.23408\) m centered in the computational domain. Then, the simulation runs until \(t=2\) s, when the bubble diameter roughly doubles. Three mesh sizes are considered: V1, V2 and V3 (see Table \ref{tab:mesh_resolution_bubble}). c/d ratios like those in Section \ref{subsubsec:staticdrop_solution} are used, with V1 initially resolving the bubble with about 15 c/d, V2 about 30 c/d, and V3 about 45 c/d. \par 

% \begin{table}
% \begin{center}
% \begin{tabular}{|c|c|} 
% \hline
% Mesh & Grid size [m] \\
% \hline
% V1 & 1/64 \\ 
% \hline
% V2 & 1/128 \\
% \hline
% V3 & 1/256 \\
% \hline
% \end{tabular}
% \caption{Grid size for each level of mesh refinement V1, V2, and V3 used in the static growing bubble.}
% \label{tab:mesh_resolution_bubble}
% \end{center}
% \end{table}

\begin{table}
\begin{center}
\begin{tabular}{|c|c|} 
\hline
Mesh & Grid size [m] \\
\hline
V1 & 1/64 \\ 
\hline
V2 & 1/128 \\
\hline
V3 & 1/192 \\
\hline
\end{tabular}
\caption{Grid size for each level of mesh refinement V1, V2, and V3 used in the static growing bubble.}
\label{tab:mesh_resolution_bubble}
\end{center}
\end{table}

\begin{figure}
\centering
\begin{subfigure}{0.33\textwidth}
  \centering
  \includegraphics[width=1.0\linewidth]{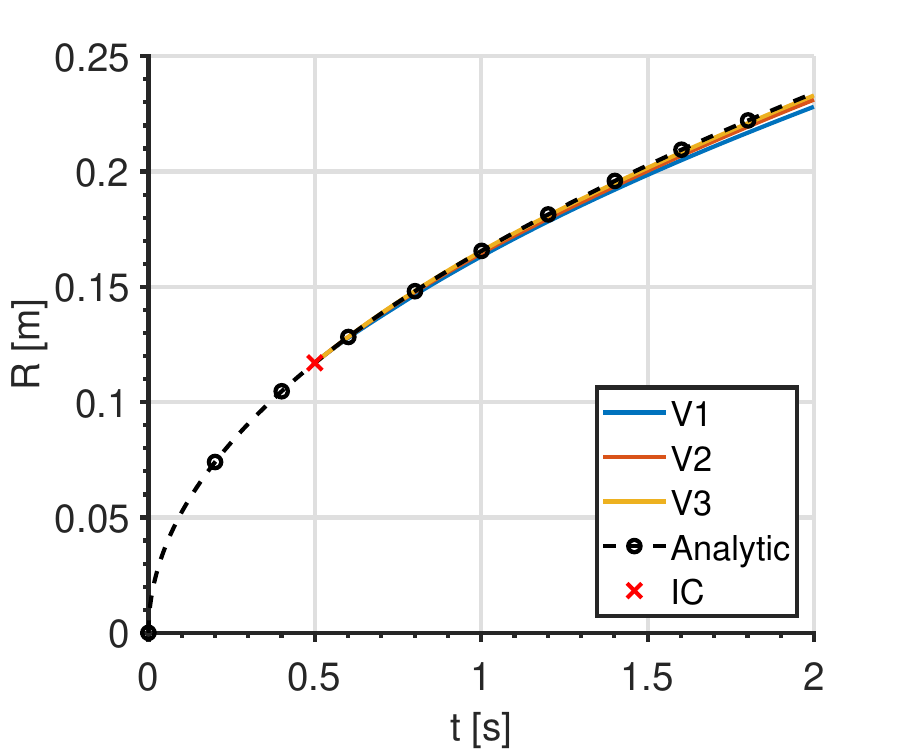}
  \caption{} % No text so (a) appears
  \label{subfig:bubblegrowth_radius}
\end{subfigure}%
\begin{subfigure}{0.33\textwidth}
  \centering
  \includegraphics[width=1.0\linewidth]{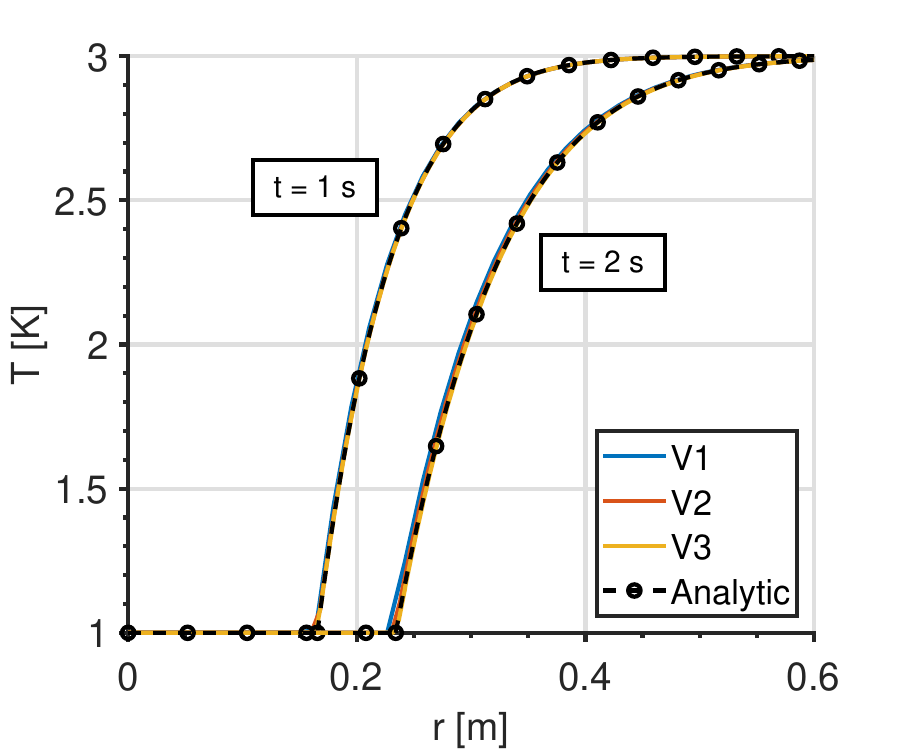}
  \caption{} % No text so (a) appears
  \label{subfig:bubblegrowth_Tprofile}
\end{subfigure}%
\begin{subfigure}{0.33\textwidth}
  \centering
  \includegraphics[width=1.0\linewidth]{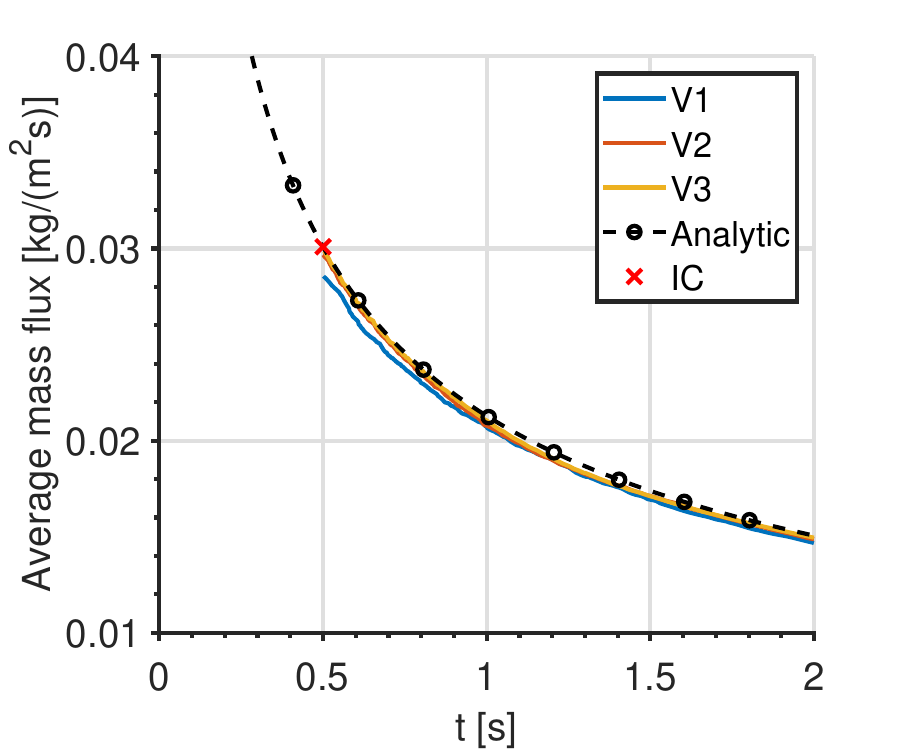}
  \caption{} % No text so (a) appears
  \label{subfig:bubblegrowth_mflux}
\end{subfigure}%
\caption{Temporal evolution of the bubble radius, temperature radial profiles and average mass flux in the static growing bubble. The initial conditions (IC) are shown. (a) radius; (b) temperature profiles; (c) average mass flux.}
\label{fig:bubble_growth}
\end{figure}

The evolution of the bubble radius, temperature radial profiles at \(t=1\) s and \(t=2\) s (extracted along the \(z\) direction), and average mass flux are shown in Figure \ref{fig:bubble_growth}. The analytical solution is captured and converging trends are observed. Table \ref{tab:errors_bubble} reports convergence rates of the bubble radius and average mass flux between first and second order, which are in line with previous works \cite{2021_JCP_Malan,2023_CaF_Boyd}. Still, the bubble radius is directly related to the volume and the limitations of the plane shift strategy shown in Section \ref{subsec:droplet} (i.e., low volume convergence rates with fixed mass flux) may resurface once the mass flux errors become very small. That is, the volume error clearly decreases with mesh refinement as long as the calculated \(\dot{m}''\) improves significantly. \par

% \begin{table} % with original periodic BCs and finer mesh at V3
% \begin{center}
% \begin{tabular}{|c|c|c|c|c|c|c|} 
% \hline
% Mesh & Radius [m] & Error [\%] & Rate & Mass flux [kg/(m\textsuperscript{2}s)] & Error [\%] & Rate  \\
% \hline
% V1 & 0.216419 & 2.494 & - & 0.015467 & 2.521 & - \\ 
% \hline
% V2 & 0.219580 & 1.115 & 1.16 & 0.015681 & 1.172 & 1.11 \\
% \hline
% V3 & 0.224289 & 1.006 & 0.15 & 0.015822 & 0.283 & 2.05 \\
% \hline
% \end{tabular}
% \caption{Relative errors, \(|\phi_n-\phi_e|/\phi_e\), and convergence rate of the bubble radius and average mass flux at \(t=1.8\) s. The analytical radius and mass flux are, respectively, \(R_e=0.222056\) m and \(\dot{m}''_e=0.015867\) kg/(m\textsuperscript{2}s).}
% \label{tab:errors_bubble}
% \end{center}
% \end{table}

\begin{table}
\begin{center}
\begin{tabular}{|c|c|c|c|c|c|c|} 
\hline
Mesh & Radius [m] & Error [\%] & Rate & Mass flux [kg/(m\textsuperscript{2}s)] & Error [\%] & Rate  \\
\hline
V1 & 0.228099 & 2.556 & - & 0.0146752 & 2.535 & - \\ 
\hline
V2 & 0.231191 & 1.235 & 1.05 & 0.0148696 & 1.244 & 1.03 \\
\hline
V3 & 0.232995 & 0.464 & 2.41 & 0.0149492 & 0.715 & 1.37 \\
\hline
\end{tabular}
\caption{Relative errors, \(|\phi_n-\phi_e|/\phi_e\), and convergence rate of the bubble radius and average mass flux at \(t=2\) s. The analytical radius and mass flux are, respectively, \(R_e=0.234081\) m and \(\dot{m}''_e=0.0150569\) kg/(m\textsuperscript{2}s).}
\label{tab:errors_bubble}
\end{center}
\end{table}

When accounting for viscosity, the error introduced by the jump in viscous stresses is significant given the relatively high viscosity of fluid A. If viscosity is neglected as in the original work by Scriven \cite{1959_CES_Scriven}, the correct pressure jump is obtained. In other words, the pressure jump and pressure field are inconsistent with the proposed modeling neglecting corrections to the viscous jump, and a pressure spike is observed across the interface. Moreover, the pressure solution suffers from the same limitations discussed in Section \ref{subsec:droplet} regarding the quasi-steady interface behavior assumption. Nonetheless, a consistent velocity field, i.e., Stefan flow, is still obtained due to the static configuration. The bubble remains centered in the computational domain, as indicated by the results presented in Figure \ref{fig:bubble_growth}. In Section \ref{subsec:2dfilmboiling}, however, it is shown that the inconsistent viscous jump is critical in a dynamic setup, emphasizing the need to develop consistent momentum balance frameworks in future works. \par

\subsection{Rising Bubble in Superheated Liquid}
\label{subsec:bubblerising}

The impact of the momentum imbalance in a dynamical setup is assessed for a three-dimensional rising bubble in a superheated liquid. Following \cite{2013_JCP_Sato,2021_IJHMT_Bures}, saturated ethanol at a pressure of 1.013 bar is used (see fluid properties in Table \ref{tab:rising_bubble_cases}) with a liquid superheat of 3.1 K, for which experimental data is available \cite{1969_IJHMT_Florschuetz}. Due to gravity (i.e., \(\mathbf{g}=(0,0,-9.81)\) m/s\textsuperscript{2}), buoyancy drives the bubble's upward motion. The computational cost is alleviated by using the symmetry of the problem. The computational domain is a box with size \([0,4]\times[0,4]\times[0,16]\) mm. The bubble is initially spherical with \(d_0=0.42\) mm and is centered at \(\mathbf{x}_0=(0,0,1)\) mm. Temperature is initialized with Scriven's analytical solution \cite{1959_CES_Scriven} for the given bubble size at \(t=0.0056\) s. Symmetry is considered at the \(x=0\) and \(y=0\) boundaries while no-slip boundary conditions are imposed at \(x=4\) mm, \(y=4\) mm and \(z=0\). Lastly, outflow boundary conditions are imposed at \(z=16\) mm. For the pressure solver, homogeneous Neumann boundary conditions are imposed at all boundaries with a reference pressure of 0 Pa at \(\mathbf{x}_p=(0,0,16)\) mm. \par 

\begin{figure}
\centering
\includegraphics[width=0.8\linewidth]{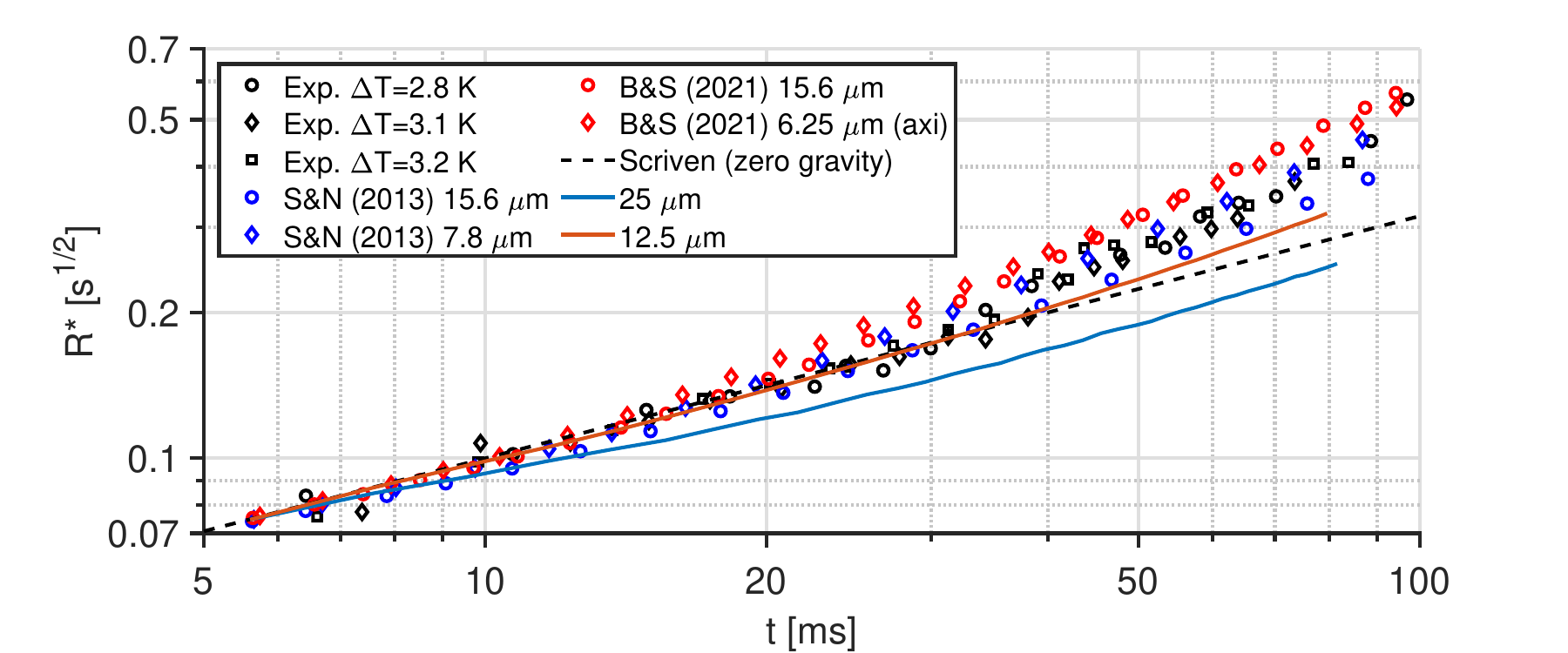}
\caption{Temporal evolution of the normalized radius, \(R^*\), in the ethanol bubble rising in superheated liquid for different mesh resolutions. The results are compared against experimental data \cite{1969_IJHMT_Florschuetz}, computational data \cite{2013_JCP_Sato,2021_IJHMT_Bures}, and Scriven's analytical solution \cite{1959_CES_Scriven} for zero gravity, Eq. (\ref{eqn:Scriven1}).}
\label{fig:bubble_rise_radius}
\end{figure}

The bubble growth is measured in terms of the average radius from the bubble diameters in each direction computed as \(R_b=(d_x+d_y+2d_z)/8\). Figure \ref{fig:bubble_rise_radius} reports the normalized radius \(R^*=R_b/(2\beta\sqrt{\alpha_L})\) where \(\beta=5.39869\) is the growth constant from Scriven's solution, i.e., Eq. (\ref{eqn:Scriven2}), and \(\alpha_L=k_L/(\rho_L c_{p,L})\) the liquid's thermal diffusivity. Results for different mesh resolutions are shown and are compared to available data from numerical \cite{2013_JCP_Sato,2021_IJHMT_Bures} and experimental \cite{1969_IJHMT_Florschuetz} works, as well as Scriven's solution for bubble growth under zero gravity. The sensitivity to mesh resolution is evident. That is, if the thermal boundary layer is under-resolved, the evaluation of \(\dot{m}''\) affects the bubble growth and the overall dynamics of the problem. For the mesh with a grid spacing of \(\Delta x = 25\) \(\mu\)m, equivalent to an initial bubble resolution of 16 c/d, the evaporation of the bubble is significantly underestimated. In contrast, a resolution of \(\Delta x = 12.5\) \(\mu\)m (i.e., initially 32 c/d) is more in line with previous numerical studies and follows the reference data better. Due to limitations in the available computational resources, finer mesh resolutions have not been considered. Thus, certain parameters shown in this section, such as \(R^*\), are not guaranteed to have converged, probably requiring a mesh resolution closer to \(\Delta x = 7\) \ \(\mu\)m based on previous works \cite{2013_JCP_Sato,2021_IJHMT_Bures}. \par 

\begin{figure}
\centering
\includegraphics[width=0.45\linewidth]{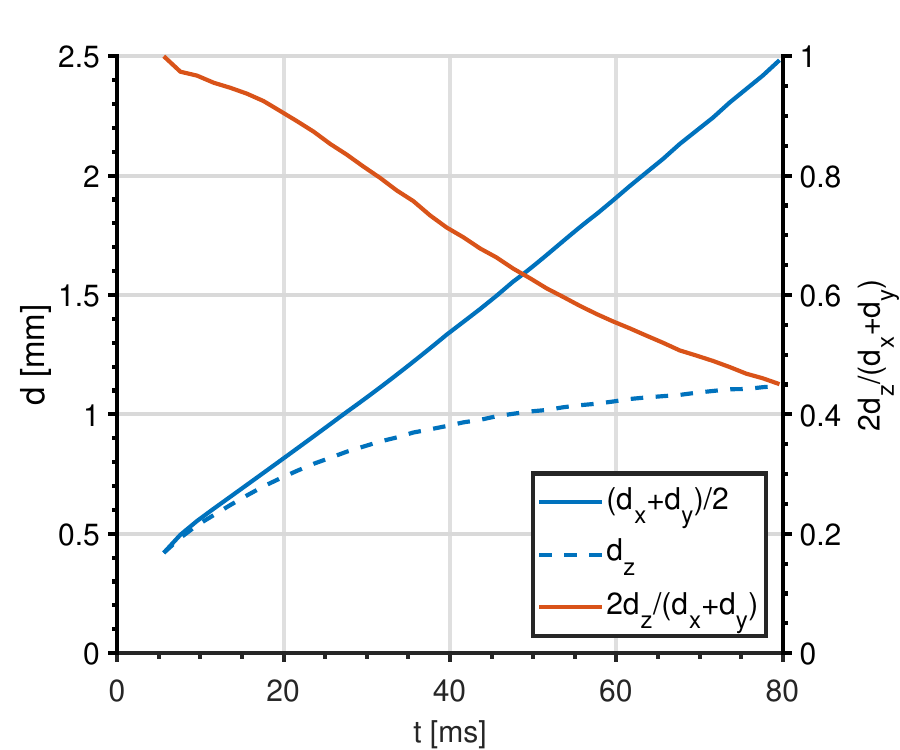}
\caption{Temporal evolution of the diameter in each direction of the rising ethanol bubble given by \(d_z\) and \((d_x+d_y)/2\) (left axis) and the aspect ratio \(2d_z/(d_x+d_y)\) (right axis) obtained with a grid spacing of 12.5 \(\mu\)m.}
\label{fig:bubble_rise_diameters}
\end{figure}

\begin{figure}
\centering
\includegraphics[width=0.35\linewidth]{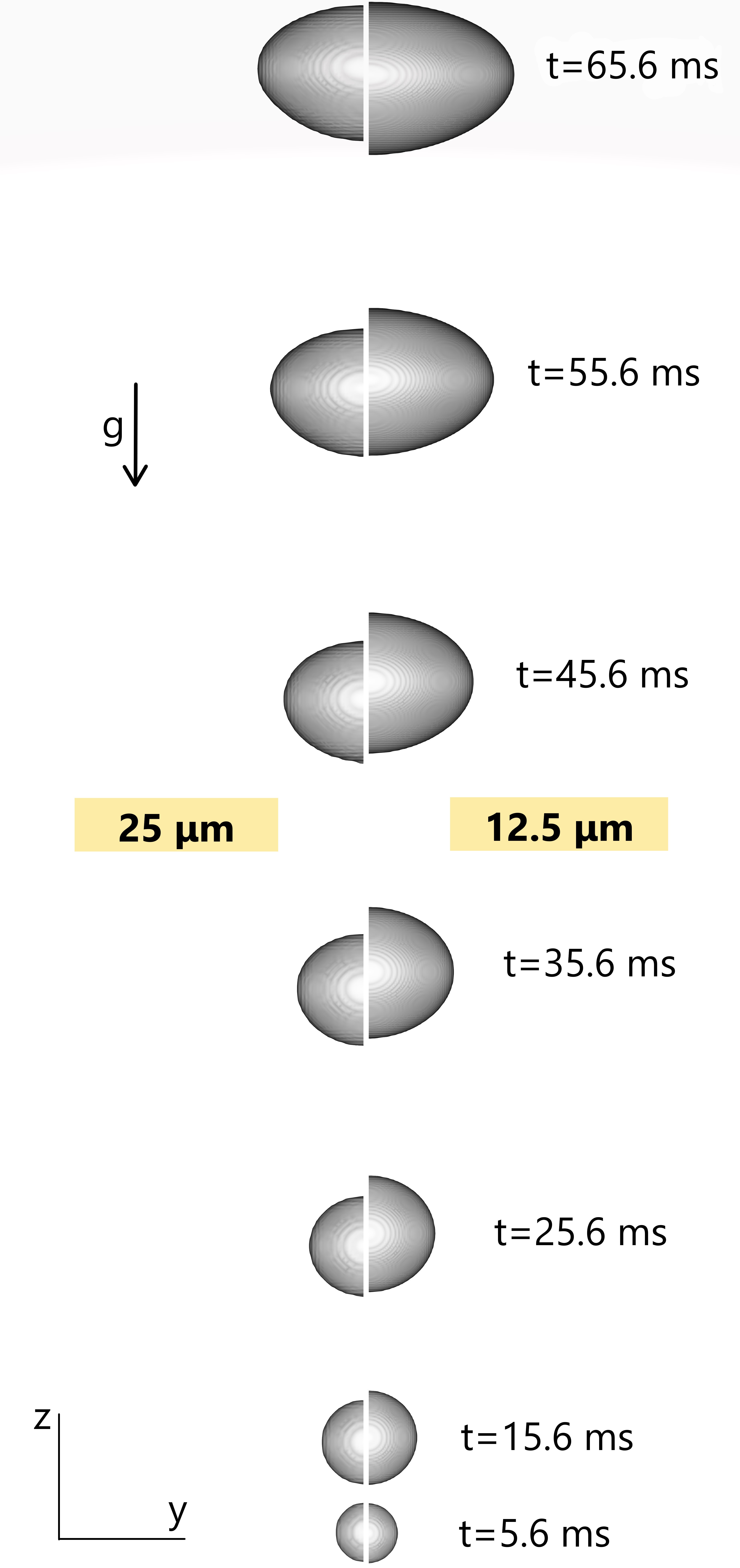}
\caption{Temporal evolution of the ethanol bubble shape as it rises due to buoyancy. The results obtained with a grid spacing of 25 \(\mu\)m (left) and 12.5 \(\mu\)m (right) are compared. The interface location is visualized by the iso-surface with \(C=0.5\).}
\label{fig:bubble_shape_annoted2}
\end{figure}

In Figure \ref{fig:bubble_rise_radius}, substantial differences can be seen in the evolution of \(R^*\) between three different works with very similar resolutions, namely the recent results using a grid spacing of 12.5 \(\mu\)m and two previous works using a resolution of 15.6 \(\mu\)m \cite{2013_JCP_Sato,2021_IJHMT_Bures}. While these differences can partially be justified by the thermal boundary layer resolution and the accuracy in evaluating \(\dot{m}''\) in each work, uncertainties remain that may be linked to the different code frameworks. This highlights the intrinsic difficulties that arise in two-phase solvers once phase change is involved. Figure \ref{fig:bubble_rise_diameters} presents the evolution of \((d_x+d_y)/2\), \(d_z\) and the aspect ratio \(2d_z/(d_x+d_y)\) obtained with the resolution of 12.5 \(\mu\)m. After a brief growth stage where the bubble diameter increases almost uniformly in all directions, it begins to grow faster in the radial direction outward of the \(z\) axis, evolving into a flat ellipsoid or disk (see Figure \ref{fig:bubble_shape_annoted2}). \par 

\begin{figure}
\centering
\includegraphics[width=0.45\linewidth]{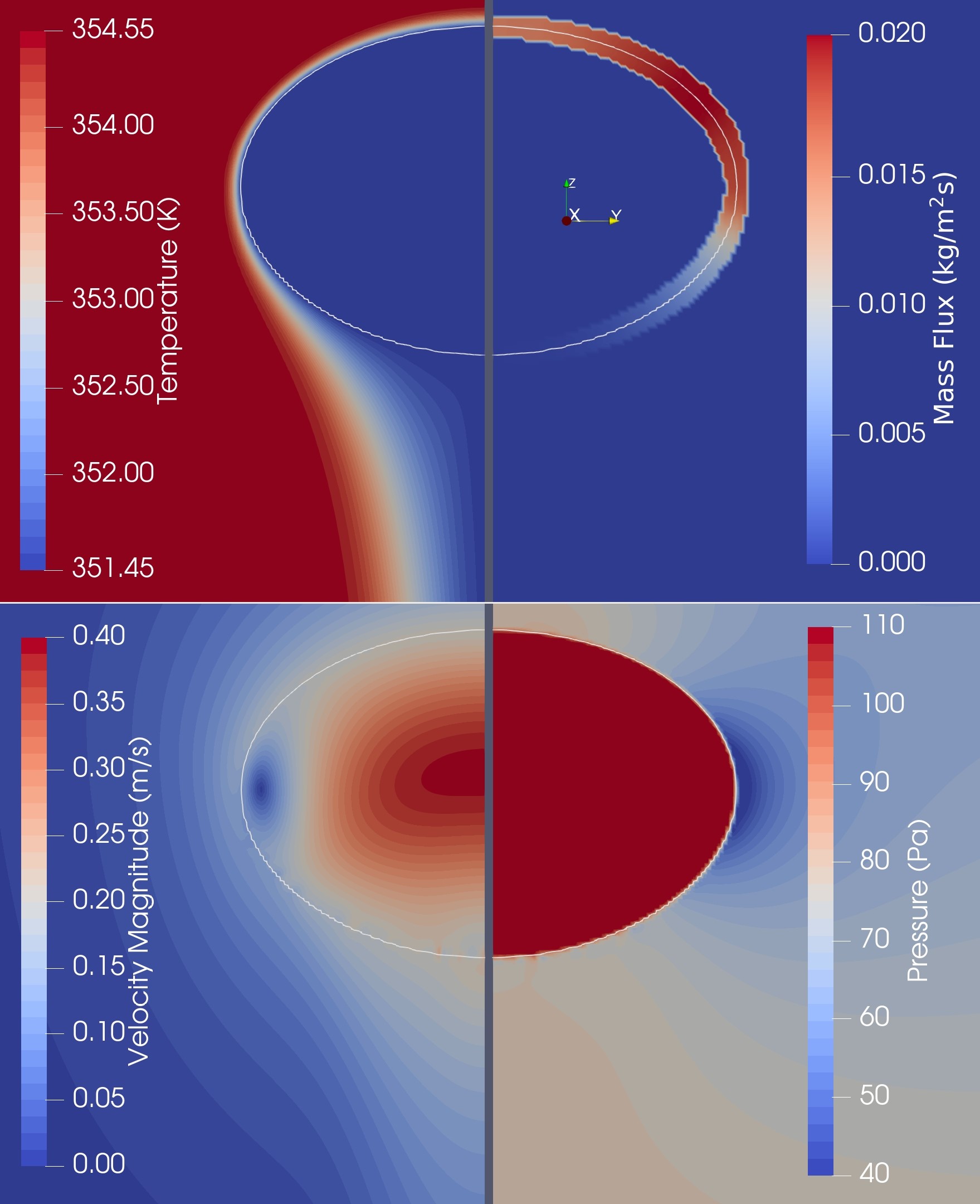}
\caption{Contours on the \(yz\) plane at \(x=0\) of temperature (top left), mass flux across the interface (top right), velocity magnitude (bottom left) and pressure (bottom right) of the rising ethanol bubble at \(t=45.6\) ms obtained with a grid spacing of 12.5 \(\mu\)m. The interface location is visualized by the iso-contour with \(C=0.5\).}
\label{fig:contours_0p0456s}
\end{figure}

The impact of mesh resolution on the evolution of the bubble's shape is also visualized in Figure \ref{fig:bubble_shape_annoted2}, which compares side by side the bubble location and shape obtained with the resolutions of 25 \(\mu\)m and 12.5 \(\mu\)m. Similar to the discussion around Figure \ref{fig:bubble_rise_radius}, the 12.5 \(\mu\)m mesh resolves the thermal boundary layer better, enhancing the calculation of \(\dot{m}''\) and increasing the bubble volume faster than with 25 \(\mu\)m. For reference, Figure \ref{fig:contours_0p0456s} depicts the contours on the \(yz\) plane at \(x=0\) of temperature, mass flux, velocity magnitude and pressure around the bubble at \(t=45.6\) ms obtained with the grid spacing of 12.5 \(\mu\)m. The thin thermal boundary layer along the bubble's upper side contrasts sharply with the bubble size. The vaporization mass flux is stronger there, reaching a peak value around 0.02 kg/(m\textsuperscript{2}s) near the edge of the bubble. Instead, almost no vaporization occurs at the bottom of the bubble due to the low temperature region in the tail. Indicated by the lower pressure and the velocity magnitude contours, a recirculation region is identified along the bubble's edge as in previous works \cite{2013_JCP_Sato}. \par

\begin{figure}
\centering
\begin{subfigure}{0.45\textwidth}
  \centering
  \includegraphics[width=1.0\linewidth]{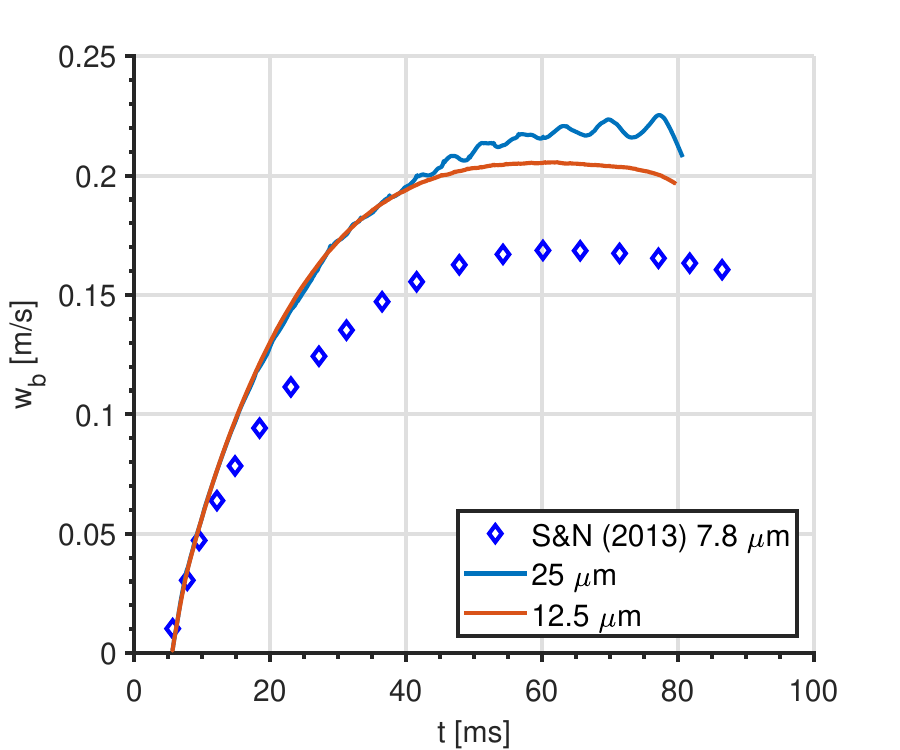}
  \caption{} % No text so (a) appears
  \label{subfig:rise_velocities_mesh}
\end{subfigure}%
\begin{subfigure}{0.45\textwidth}
  \centering
  \includegraphics[width=1.0\linewidth]{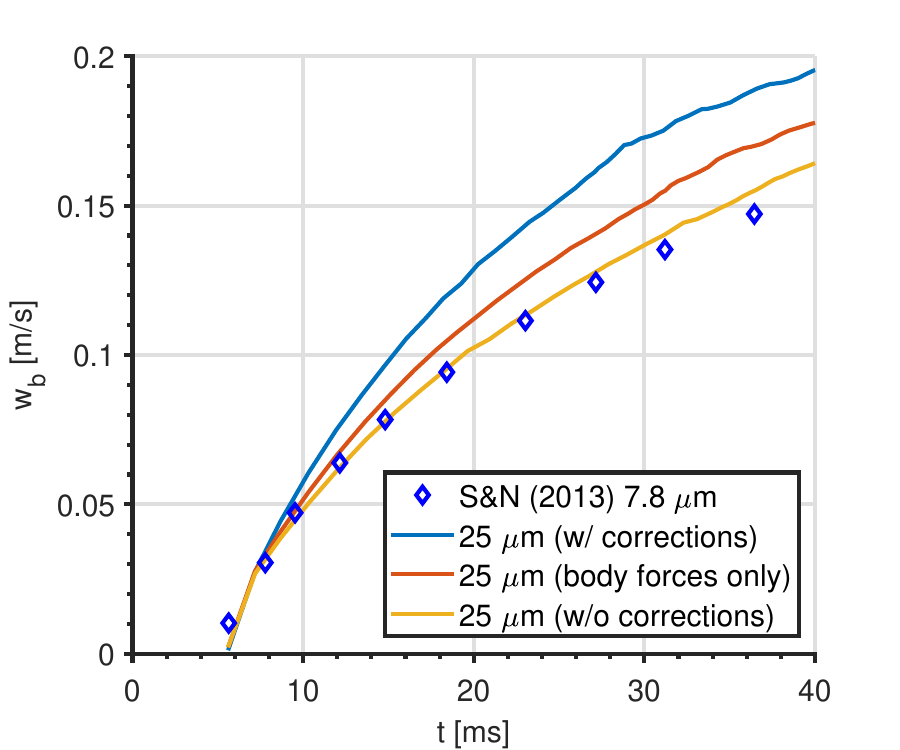}
  \caption{} % No text so (a) appears
  \label{subfig:rise_velocities_corrections}
\end{subfigure}%
\caption{Effect of mesh resolution and momentum balance corrections on the rise velocity \(w_b\) of the ethanol bubble rising in superheated liquid. The results are compared against computational data \cite{2013_JCP_Sato}. (a) grid convergence of \(w_b\); (b) variation of \(w_b\) depending on the momentum balance corrections.}
\label{fig:bubble_rise_velocity}
\end{figure}

Since the bubble displacement shown in Figure \ref{fig:bubble_shape_annoted2} obtained with both grid resolutions is nearly identical, the bubble rise velocity \(w_b\) is compared against available data \cite{2013_JCP_Sato}, which is calculated as a volume average of the \(w\) velocity component inside the bubble. Figure \ref{subfig:rise_velocities_mesh} shows that minor differences are observed in \(w_b\) between using a grid resolution of 25 \(\mu\)m or 12.5 \(\mu\)m until about 40 ms. That is, despite the thermal boundary layer is captured more accurately and the average radius of the bubble is larger, the balance between the drag and buoyancy forces is similar, suggesting that \(w_b\) is a parameter less sensitive to the mesh resolution. After 40 ms, the velocities start to deviate. With the finer mesh, the bubble reaches a terminal velocity as it grows into a flat ellipsoid, while the coarser mesh displays oscillations mainly as a result of spurious currents growing unbounded. Regardless, a clear offset with respect to the reference data is observed. \par

The shift in rise velocity is caused by the momentum balance corrections, as shown in Figure \ref{subfig:rise_velocities_corrections}. Following the previous discussion on the grid sensitivity of \(w_b\), the coarse mesh is used to analyze the early \(w_b\) trends. Three cases are shown together with the reference data \cite{2013_JCP_Sato}: (a) with all corrections, (b) only adding \(\mathbf{f}_{NC}\) and \(\mathbf{f}_{\dot{m}''}\), and (c) with no corrections. The latter is more aligned with the numerical approach from \cite{2013_JCP_Sato} to solve the one-fluid Navier-Stokes equations and reproduces more closely the \(w_b\) trends shown in their work. As the momentum balance corrections are progressively considered, first by only adding \(\mathbf{f}_{NC}\) and \(\mathbf{f}_{\dot{m}''}\) and then by also adding the first predictor step to shift the Stefan flow, an increase in \(w_b\) occurs. In any case, all the reported velocities fall within expected values \cite{1969_IJHMT_Florschuetz,1971_IJHMT_Ruckenstein}. This result is somewhat surprising given that the maximum value of \(\dot{m}''\) is around 0.02 kg/(m\textsuperscript{2}s) throughout the simulation, leading to believe that the momentum balance corrections are negligible, i.e., \((\dot{m}'')^2(\rho_{G}^{-1}-\rho_{L}^{-1})\) is of \(\mathcal{O}(10^{-4})\) Pa. However, they still impact the pressure field around the bubble and the flow dynamics. \par

\begin{figure}
\centering
\includegraphics[width=1.0\linewidth]{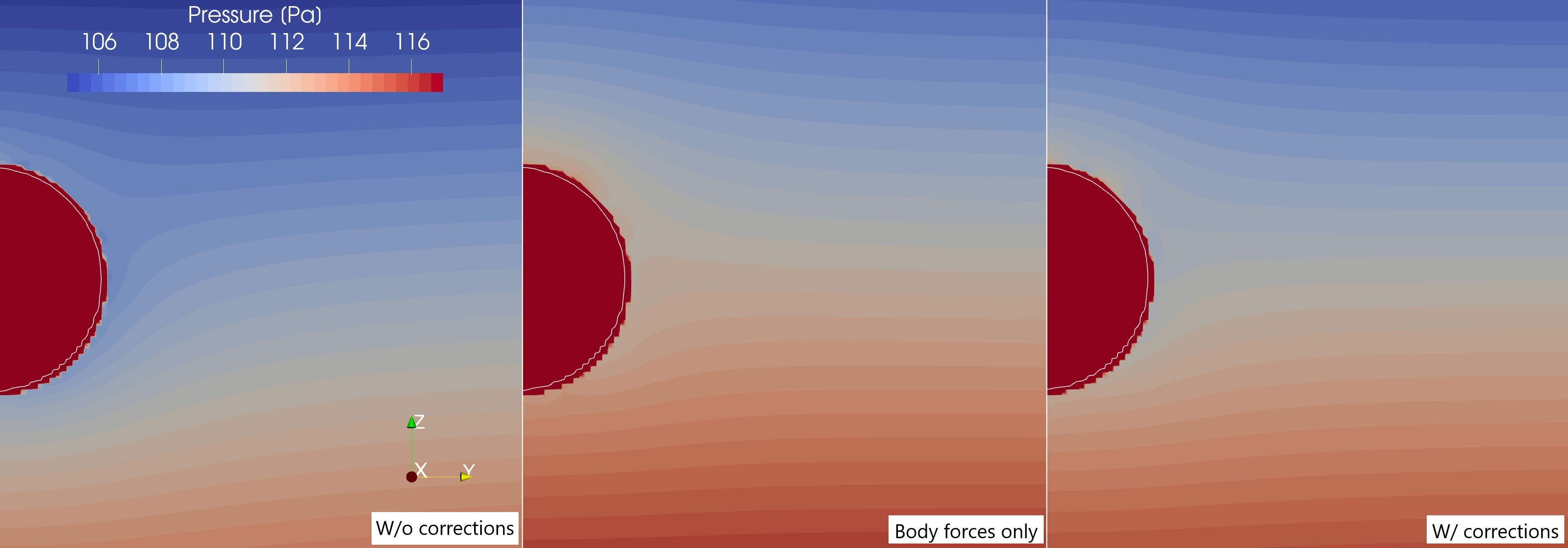}
\caption{Pressure field on the \(yz\) plane with \(x=0\) at \(t=7.6\) ms of the rising ethanol bubble obtained with a grid spacing of 12.5 \(\mu\)m. The results without momentum balance corrections (left), only with the body forces \(\mathbf{f}_{NC}\) and \(\mathbf{f}_{\dot{m}''}\) (center), and with all corrections (right) are shown. The interface location is visualized by the iso-contour with \(C=0.5\).}
\label{fig:pressure_field_7p6ms_annoted}
\end{figure}

\begin{figure}
\centering
\includegraphics[width=0.45\linewidth]{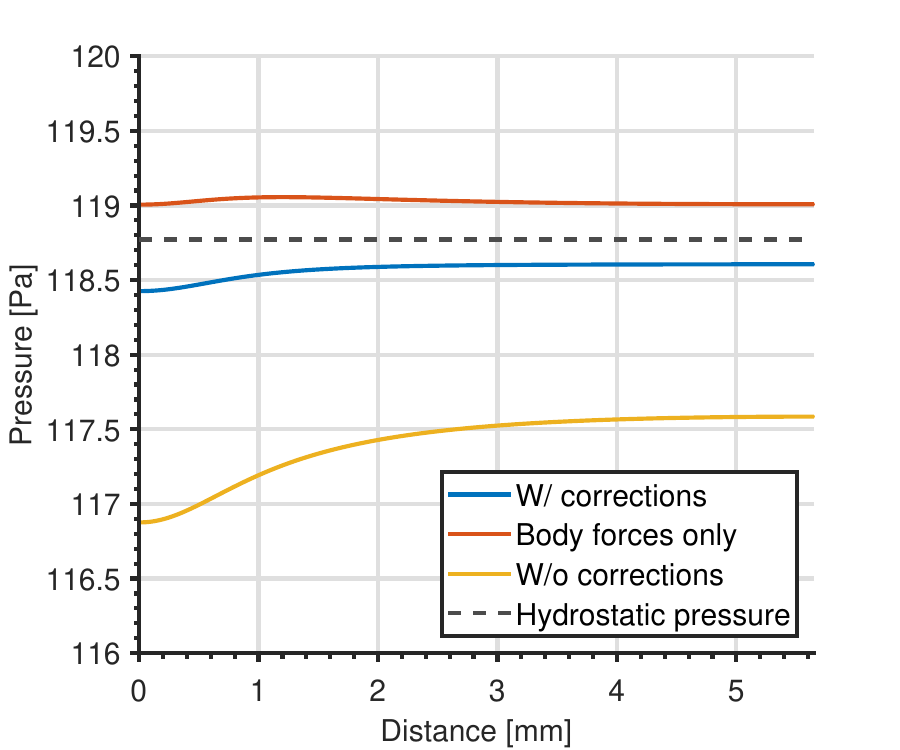}
\caption{Pressure field on the \(xy\) plane with \(z=0.00625\) mm at \(t=7.6\) ms of the rising ethanol bubble obtained with a grid spacing of 12.5 \(\mu\)m. The data is extracted along the diagonal of the plane, i.e., from \(\mathbf{x}_i=(0,0,0.00625)\) mm to \(\mathbf{x}_f=(4,4,0.00625)\) mm. Additionally, the hydrostatic pressure of liquid ethanol at a depth of \(h_d=15.99375\) mm is shown and is given by \(p_h=\rho_L||\mathbf{g}||h_d=118.772\) Pa.}
\label{fig:bubble_rise_hydrostatic_pressure_bottom}
\end{figure}

Additional insights can be gained by looking at various details of the pressure field. Figure \ref{fig:pressure_field_7p6ms_annoted} shows the pressure contours at \(t=7.6\) ms on a \(yz\) plane at \(x=0\), focusing on the pressure around the bubble and considering the three momentum balance modeling frameworks shown in Figure \ref{subfig:rise_velocities_corrections}. In this snapshot, the bubble size and position is almost identical for all frameworks. Nonetheless, different pressure distributions and magnitudes are observed, with the simulation without momentum balance corrections seemingly predicting a lower pressure in the liquid overall. This is corroborated in Figure \ref{fig:bubble_rise_hydrostatic_pressure_bottom}, which shows the pressure extracted along the diagonal of an \(xy\) plane located at \(z=0.00625\) mm, i.e., from \(\mathbf{x}_i=(0,0,0.00625)\) mm to \(\mathbf{x}_f=(4,4,0.00625)\) mm, at \(t=7.6\) ms. Under the assumption that the bubble and its motion have negligible influence on the pressure at the considered depth, the hydrostatic pressure given the liquid column is 118.772 Pa. Indeed, the omission of the momentum balance corrections induces a lower pressure, while the addition of \(\mathbf{f}_{NC}\), \(\mathbf{f}_{\dot{m}''}\) and the Stefan flow shift improves the agreement. \par 

\begin{figure}
\centering
\begin{subfigure}{0.33\textwidth}
  \centering
  \includegraphics[width=1.0\linewidth]{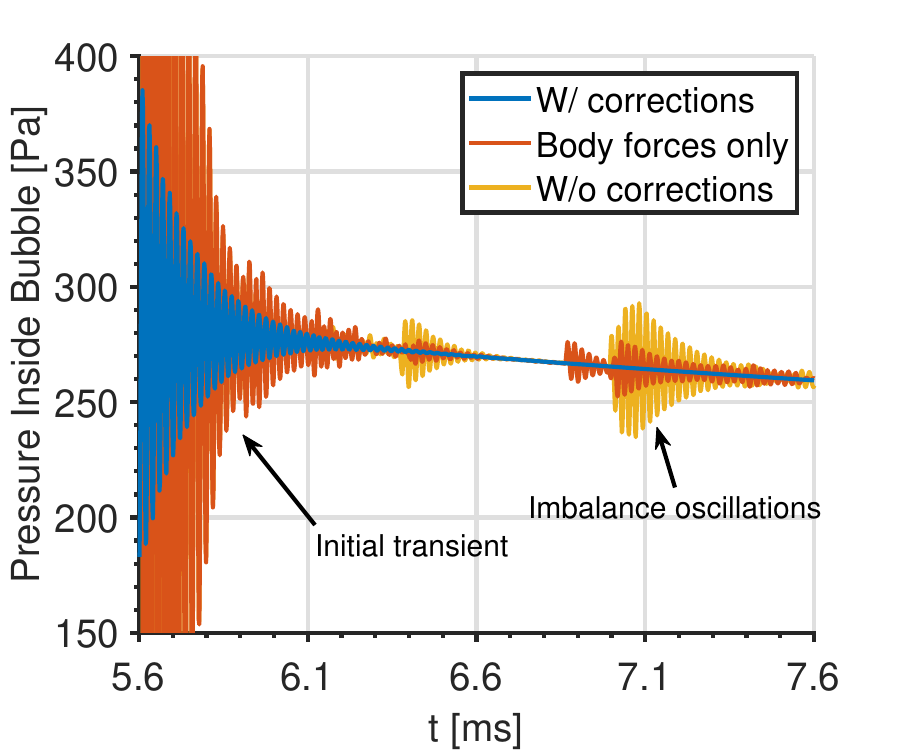}
  \caption{} % No text so (a) appears
  \label{subfig:pressure_inisde_bubble_192}
\end{subfigure}%
\begin{subfigure}{0.33\textwidth}
  \centering
  \includegraphics[width=1.0\linewidth]{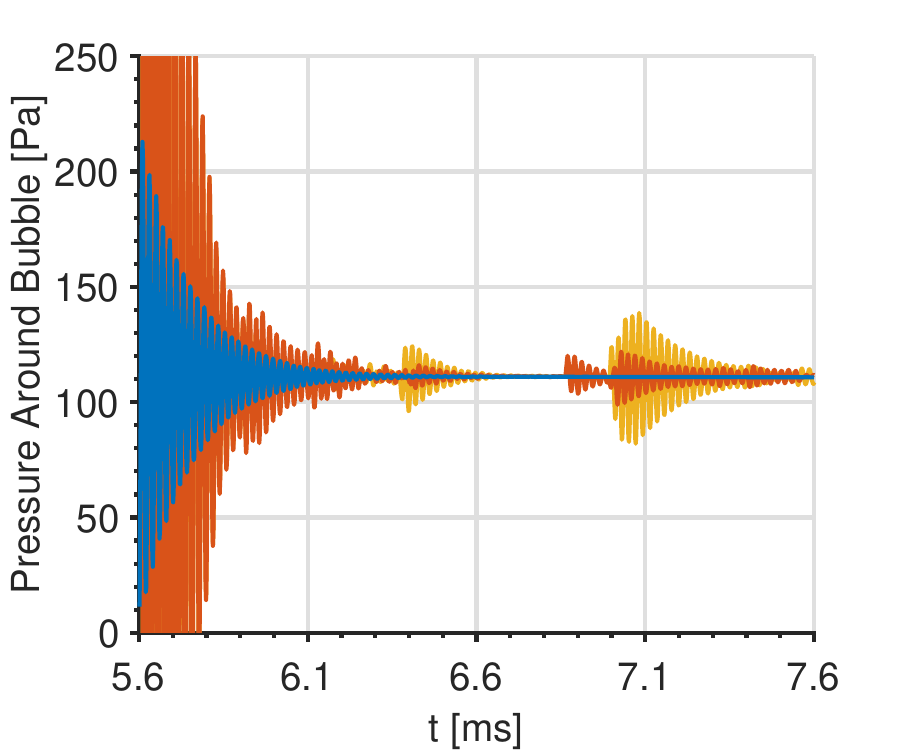}
  \caption{} % No text so (a) appears
  \label{subfig:pressure_around_bubble_192}
\end{subfigure}%
\begin{subfigure}{0.33\textwidth}
  \centering
  \includegraphics[width=1.0\linewidth]{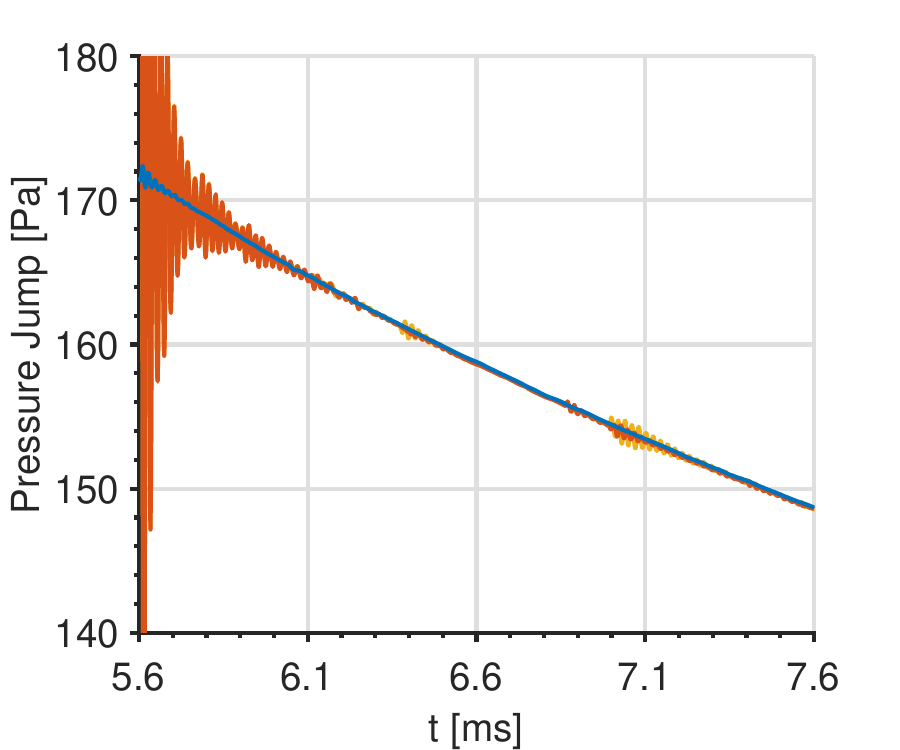}
  \caption{} % No text so (a) appears
  \label{subfig:pressure_jump_bubble_192}
\end{subfigure}%
\caption{Effect of momentum balance corrections on the pressure field inside and in the vicinity of the ethanol bubble rising in superheated liquid obtained with a grid spacing of 12.5 \(\mu\)m. (a) average internal bubble pressure; (b) average pressure in the immediate vicinity of the bubble; (c) average pressure jump.}
\label{fig:pressure_bubble_time}
\end{figure}

However, further inspection reveals that the pressure field is affected by considerable oscillations if no momentum balance corrections are implemented, similar to Figure \ref{subfig:1D_pjump_vs_time_L3}. Figure \ref{fig:pressure_bubble_time} shows the average pressure inside the bubble (i.e., from all pure gas cells), the average pressure around the bubble (i.e., from all pure liquid cells directly neighboring the interface) and the average pressure jump calculated by subtracting the two previous quantities. After an initial transient due to the relaxation of the initial conditions, the pressure inside and around the bubble obtained with all corrections implemented behaves smoothly over time. In contrast, frequent pressure oscillations are observed if the momentum balance is affected by the numerical approach. Here, the addition of only the body forces \(\mathbf{f}_{NC}\) and \(\mathbf{f}_{\dot{m}''}\) helps reducing the pressure oscillations after the initial transient although nearly complete mitigation is only possible once the Stefan flow shift is considered. It can be observed that at \(t=7.6\) ms, the solution without corrections is predicting a smaller pressure than the trend, i.e., the observations from Figures \ref{fig:pressure_field_7p6ms_annoted} and \ref{fig:bubble_rise_hydrostatic_pressure_bottom}. Moreover, the imbalance oscillations translate into pressure jump oscillations over time. During the initial transient, the magnitudes of the oscillations obtained without corrections and by only adding the body forces are very similar. This shows that the addition of the Stefan flow shift is responsible for improving the behavior of the pressure field during the adjustment phase too. Notably, taking proper care of the momentum balance is also reflected in the relatively smooth evolution of the average pressure jump. \par

Lastly, it is worth noting that the trends of the pressure field quantities reported in Figure \ref{fig:pressure_bubble_time} are very similar during the early stages despite the oscillations introduced by the formulation without momentum balance corrections. This is more aligned with the expectation that the corrections are, in terms of magnitude, negligible for this problem. However, different rise velocities are reported. Since the interface dynamics strongly depend on the pressure jump between liquid and gas, it is expected that the bubble shape evolves differently if the solution of the momentum equation is according to the corrected pressure field. This results in a different balance between buoyancy and drag; thus, a different rise velocity is observed. \par

% \subsection{Two-Dimensional Film Boiling}
\section{Assessment of the Viscous Stress Jump with Evaporation}
\label{subsec:2dfilmboiling}

A two-dimensional film boiling configuration is considered as an additional benchmark \cite{2011_NHT_Guo,2014_NHT_Sun,2023_CaF_Boyd}. Another test fluid with the properties of fluid B shown in Table \ref{tab:rising_bubble_cases} is used. This highly viscous fluid causes a momentum imbalance across the interface due to the regularization approach of the viscous term assumed in Section \ref{subsec:onefluid}, which is a more challenging task to address in this kind of multiphase solvers. As discussed below, the untreated discretization of the viscous term modifies the flow dynamics due to the sharp one-fluid formulation. \par

Following Rayleigh-Taylor instability theory, the vapor film is initialized with a perturbation of wavelength \(\lambda_d\) to trigger the most unstable mode. The evaporation of the liquid increases the vapor volume and the dynamics are dominated by the combined effect of the momentum balance (e.g., surface tension) and buoyancy, forming a bubble that rises due to gravity (i.e., \(\mathbf{g}=(0,-9.81)\) m/s\textsuperscript{2}). The computational domain is a box of size \([0,\lambda_d/2]\times[0,\lambda_d]\) with symmetric boundaries at \(x=0\) and \(x=\lambda_d/2\), a wall with no-slip boundary conditions at \(y=0\) and outflow boundary conditions at \(y=\lambda_d\). For the pressure solver, homogeneous Neumann boundary conditions are imposed at the bottom and top boundaries with a reference pressure of 0 Pa at \(\mathbf{x}_p=(0,\lambda_d)\). For this problem, PLIC is used since no major differences are observed in the curvature computation and interface evolution with respect to PPIC; thus, favoring computational savings. The liquid remains at \(T_{sat}\) and the wall temperature is fixed at \(T_{wall}=T_{sat}+5\) K. The vapor film, which sits on top of the wall, is initialized with a linear temperature profile, \(T_0\), between the initial interface location, \(y_0\), and the wall. \(\lambda_d\), \(T_0\) and \(y_0\) are given, respectively, by Eqs. (\ref{eqn:RTwavelength}), (\ref{eqn:RTinitialtemperature}) and (\ref{eqn:RTinitialwave}) (see \cite{2023_CaF_Boyd} for more details). For the given fluid properties, \(\lambda_d=78.6844\) mm. \par

\begin{equation}
\label{eqn:RTwavelength}
\lambda_d = 2\pi\sqrt{\frac{3\sigma}{(\rho_L-\rho_G)||\mathbf{g}||}}
\end{equation}

\begin{equation}
\label{eqn:RTinitialtemperature}
T_0(x,y)=
\begin{cases}
    T_{sat} & \text{if $y\geq y_0(x)$}\\
    T_{wall}-(T_{wall}-T_{sat})\frac{y}{y_0(x)} & \text{if $y<y_0(x)$}\\
\end{cases}
\end{equation}

\begin{equation}
\label{eqn:RTinitialwave}
y_0(x) = \frac{\lambda_d}{128}\bigg[4+\cos{\bigg(\frac{2\pi x}{\lambda_d}\bigg)}\bigg]
\end{equation}

The results are compared against the reportedly converged solutions from Guo et al. \cite{2011_NHT_Guo}, Sun et al. \cite{2014_NHT_Sun} and Boyd and Ling \cite{2023_CaF_Boyd}. However, each work uses different forms of the one-fluid governing equations and different numerical approaches, resulting in different results of the same configuration, mainly in terms of bubble formation time scales. Note that the viscous jump for fluid B cannot be neglected due to its high viscosity, and the numerical jump induced by the one-fluid formulation remains in the center of the discussion as neither this work nor the aforementioned works perform any attempt to address it. \par 

\begin{table}
\begin{center}
\begin{tabular}{|c|c|c|c|c|c|} 
\hline
Mesh & Grid size [cells] & Grid size [mm] & Mesh & Grid size [cells] & Grid size [mm] \\
\hline
M1 & 50x100 & 0.786844 & M4 & 300x600 & 0.131141 \\ 
\hline
M2 & 100x200 & 0.393422 & M5 & 400x800 & 0.098356\\ 
\hline
M3 & 200x400 & 0.196711 & M6 & 500x1000 & 0.078684 \\ 
\hline
\end{tabular}
\caption{Grid size for each level of mesh refinement M1, M2, M3, M4, M5 and M6 used in the two-dimensional film boiling.}
\label{tab:mesh_resolution_film}
\end{center}
\end{table}

\begin{equation}
\label{eqn:averageNu}
\text{Nu} = \frac{2}{\lambda_d} \int_{0}^{\lambda_d/2} \bigg(\frac{1}{T_{wall}-T_{sat}}\bigg) \sqrt{\frac{\sigma}{(\rho_L-\rho_G)||\mathbf{g}||}} \bigg(\frac{\partial T}{\partial y}\bigg)_{y=0} dx
\end{equation}

\begin{figure}
\centering
\begin{subfigure}{0.5\textwidth}
  \centering
  \includegraphics[width=1.0\linewidth]{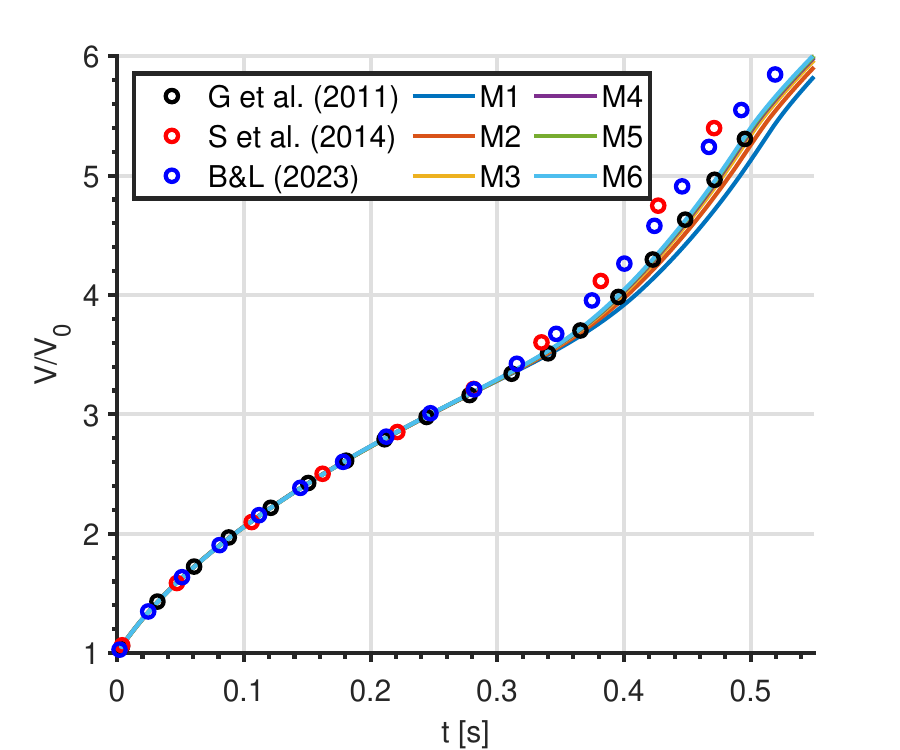}
  \caption{} % No text so (a) appears
  \label{subfig:void_increase}
\end{subfigure}%
\begin{subfigure}{0.5\textwidth}
  \centering
  \includegraphics[width=1.0\linewidth]{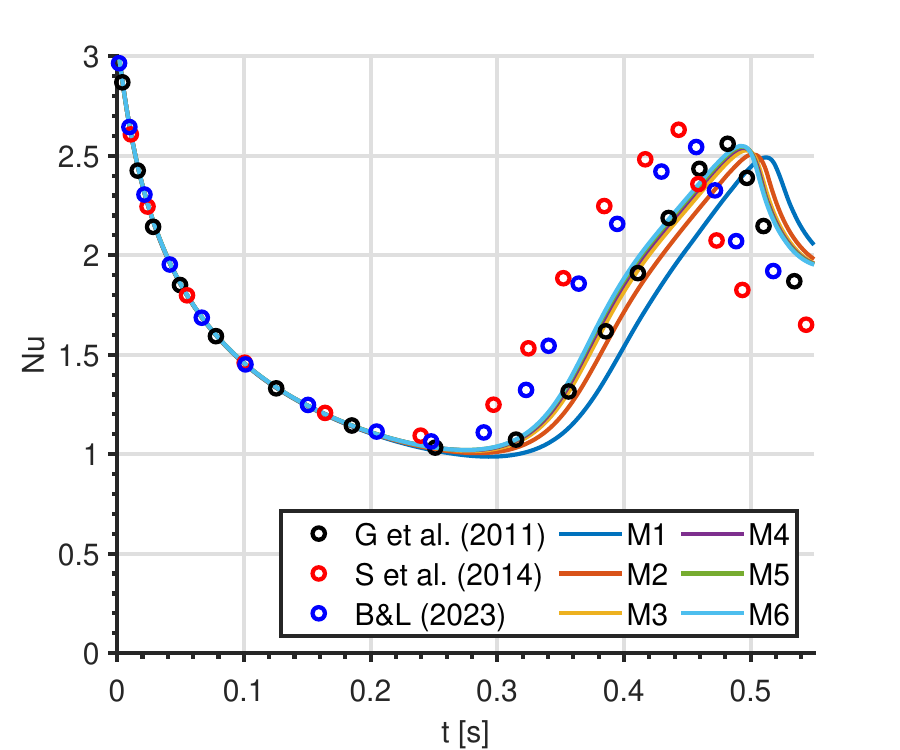}
  \caption{} % No text so (a) appears
  \label{subfig:Nu_average}
\end{subfigure}%
\caption{Convergence of the increase in vapor volume and the average Nusselt number at the wall of the two-dimensional film boiling case compared against benchmark solutions from Guo et al. \cite{2011_NHT_Guo}, Sun et al. \cite{2014_NHT_Sun} and Boyd and Ling \cite{2023_CaF_Boyd}. (a) vapor volume ratio \(V/V_0\); (b) average Nusselt number.}
\label{fig:void_Nu}
\end{figure}

\begin{figure}
\centering
\begin{subfigure}{0.33\textwidth}
  \centering
  \includegraphics[width=1.0\linewidth]{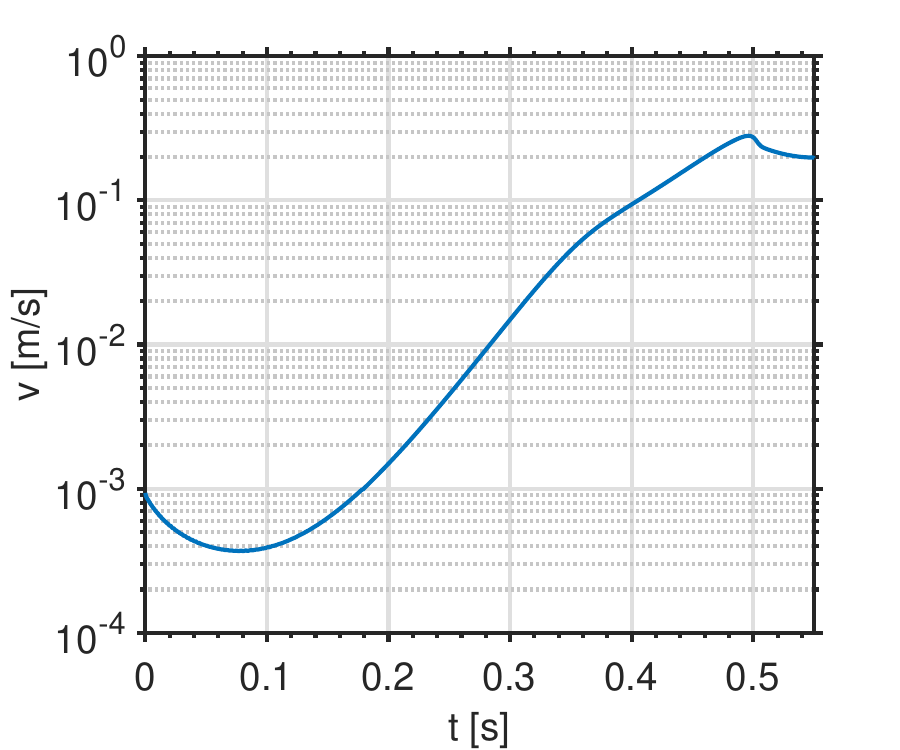}
  \caption{} % No text so (a) appears
  \label{subfig:v_centermass}
\end{subfigure}%
\begin{subfigure}{0.33\textwidth}
  \centering
  \includegraphics[width=1.0\linewidth]{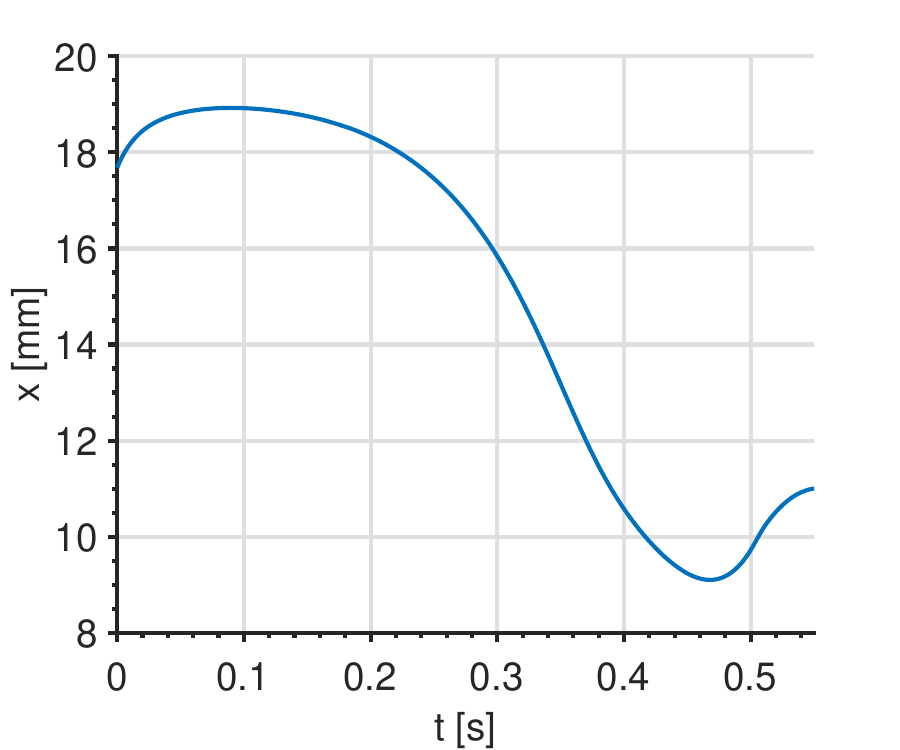}
  \caption{} % No text so (a) appears
  \label{subfig:x_centermass}
\end{subfigure}%
\begin{subfigure}{0.33\textwidth}
  \centering
  \includegraphics[width=1.0\linewidth]{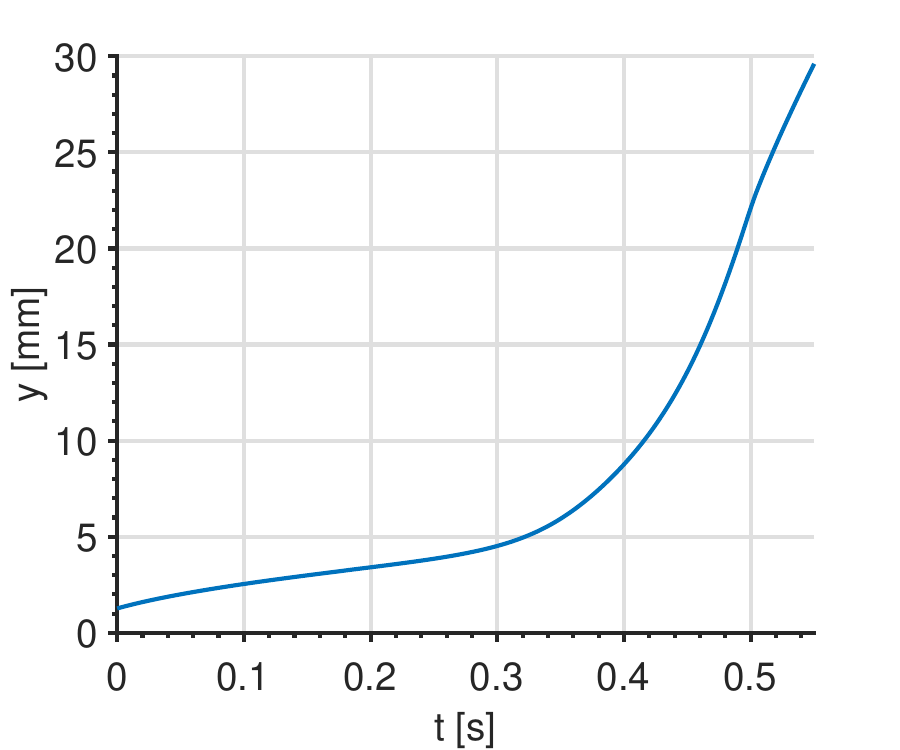}
  \caption{} % No text so (a) appears
  \label{subfig:y_centermass}
\end{subfigure}%
\caption{Evolution of the mass-averaged rise velocity of the two-dimensional vapor film and its center of mass obtained with mesh M6. Note that the center of mass (CM) of half the film is reported; thus, \(x\neq0\). (a) rise velocity; (b) \(x\) coordinate of the CM; (c) \(y\) coordinate of the CM.}
\label{fig:centermass_film}
\end{figure}

The increase in vapor volume \(V\) normalized by the initial vapor volume \(V_0\) and the average Nusselt number, Nu, on the bottom wall, defined as in Eq. (\ref{eqn:averageNu}) \cite{2023_CaF_Boyd}, are shown in Figure \ref{fig:void_Nu}. Moreover, grid convergence is assessed with six mesh sizes (i.e., M1 to M6, with M6 corresponding to 1000 cells across \(\lambda_d\)), summarized in Table \ref{tab:mesh_resolution_film}. The reference data shown here corresponds to a resolution across \(\lambda_d\) of 128 cells for Guo et al. \cite{2011_NHT_Guo}, 400 cells for Sun et al. \cite{2014_NHT_Sun}, and 2048 cells for Boyd and Ling \cite{2023_CaF_Boyd} (the latter using adaptive mesh refinement near the interface). Up to 0.2 s, the results agree well with the reference data and are fairly independent of mesh resolution. This corresponds to the initial growth of the vapor film before the instability accelerates and the bubble forms. For reference, Figure \ref{fig:centermass_film} shows the evolution of the mass-averaged vapor rise velocity and its center of mass obtained with M6 (i.e., the quantities are averaged over the vapor volume due to \(\rho_G\) being constant). After that, substantial differences in \(V/V_0\) and Nu are observed across different mesh resolutions and, more importantly, across different numerical frameworks. \par

\begin{figure}
\centering
\begin{subfigure}{0.5\textwidth}
  \centering
  \includegraphics[width=1.0\linewidth]{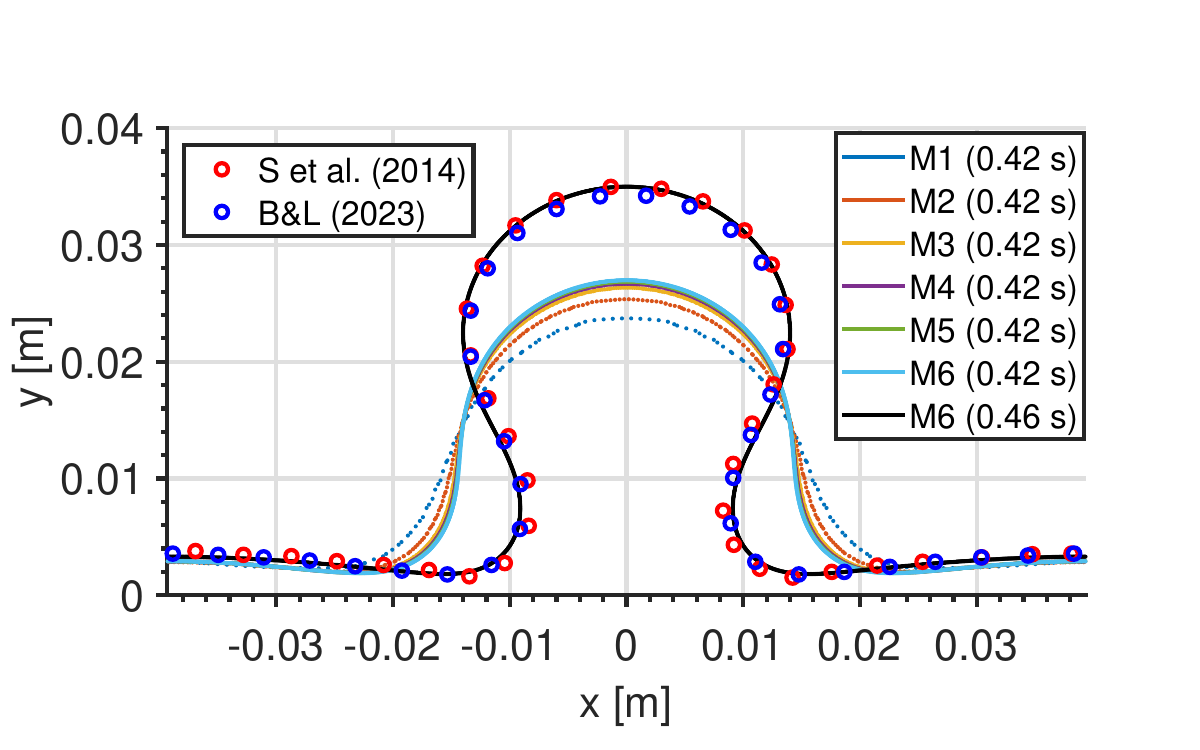}
  \caption{} % No text so (a) appears
  \label{subfig:bubbleshape_0p42s}
\end{subfigure}%
\begin{subfigure}{0.5\textwidth}
  \centering
  \includegraphics[width=1.0\linewidth]{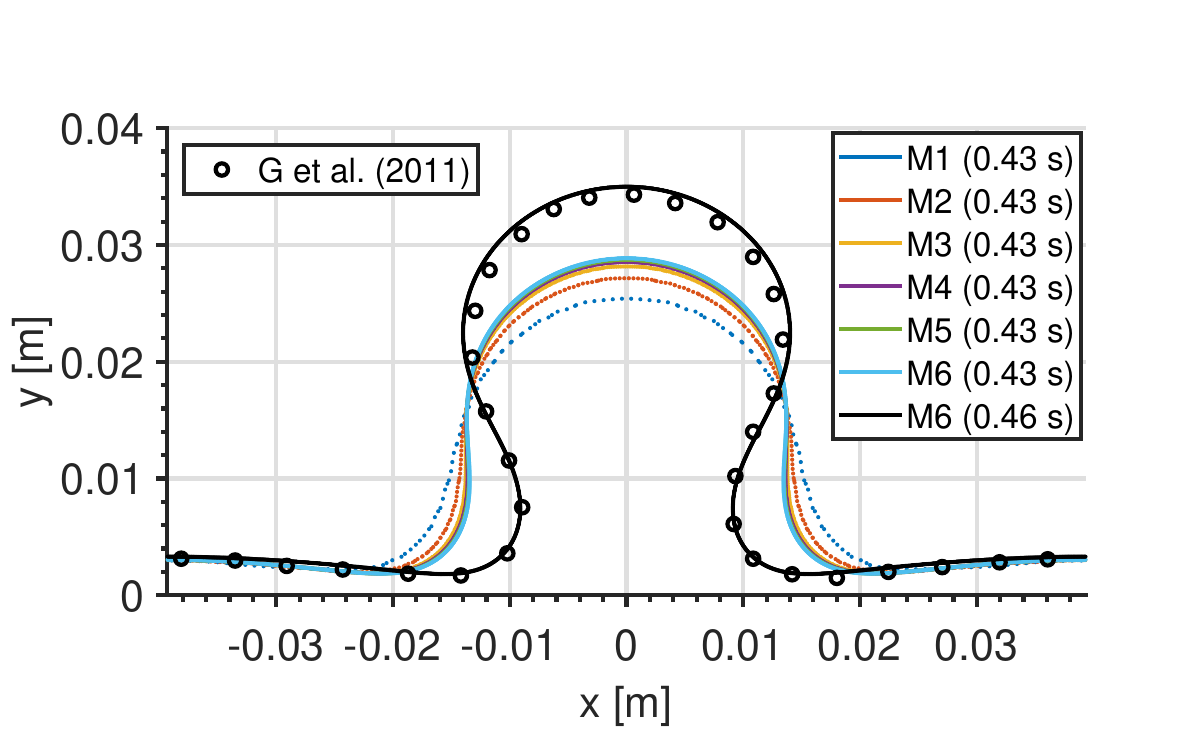}
  \caption{} % No text so (a) appears
  \label{subfig:bubbleshape_0p43s}
\end{subfigure}%
\caption{Convergence of the bubble shape in the two-dimensional film boiling case and comparison against benchmark solutions from Guo et al. \cite{2011_NHT_Guo}, Sun et al. \cite{2014_NHT_Sun} and Boyd and Ling \cite{2023_CaF_Boyd}. (a) \(t=0.42\) s; (b) \(t=0.43\) s. A later snapshot at \(t=0.46\) s with mesh M6 is also shown. The interface location is visualized by the iso-contour with \(C=0.5\).}
\label{fig:2dfilm_shape}
\end{figure}

At first glance, one may claim that the solution from Guo et al. \cite{2011_NHT_Guo} is reproduced. However, upon closer inspection of the bubble shape, this is not the case. Instead, a shift in the time scale linked to the bubble formation, i.e., instability growth, is observed. The shape of the bubble is compared in Figure \ref{fig:2dfilm_shape} against benchmark solutions at \(t=0.42\) s \cite{2014_NHT_Sun,2023_CaF_Boyd} and at \(t=0.43\) s \cite{2011_NHT_Guo}, and its growth is clearly slower. The plots also show the results with mesh M6 at \(t=0.46\) s, which agree much better with the reference data. This delay between 0.03 s and 0.04 s is consistent with the shift in the \(V/V_0\) and Nu profiles shown in Figure \ref{fig:void_Nu}. When it comes to the numerical approach, the present work is close in methodology to Boyd and Ling \cite{2023_CaF_Boyd} and considers the recent state-of-the-art in multiphase flow modeling. Some differences revolve around the conservative/non-conservative treatment of the momentum equation and the implementation of the volume dilatation due to phase change, but convergence of the numerical solution is observed with similar grid resolutions to M6. In fact, one could argue that earlier works are somewhat under-resolved \cite{2011_NHT_Guo,2014_NHT_Sun}.

\begin{figure}
\centering
\begin{subfigure}{1.0\textwidth}
  \centering
  \includegraphics[width=1.0\linewidth]{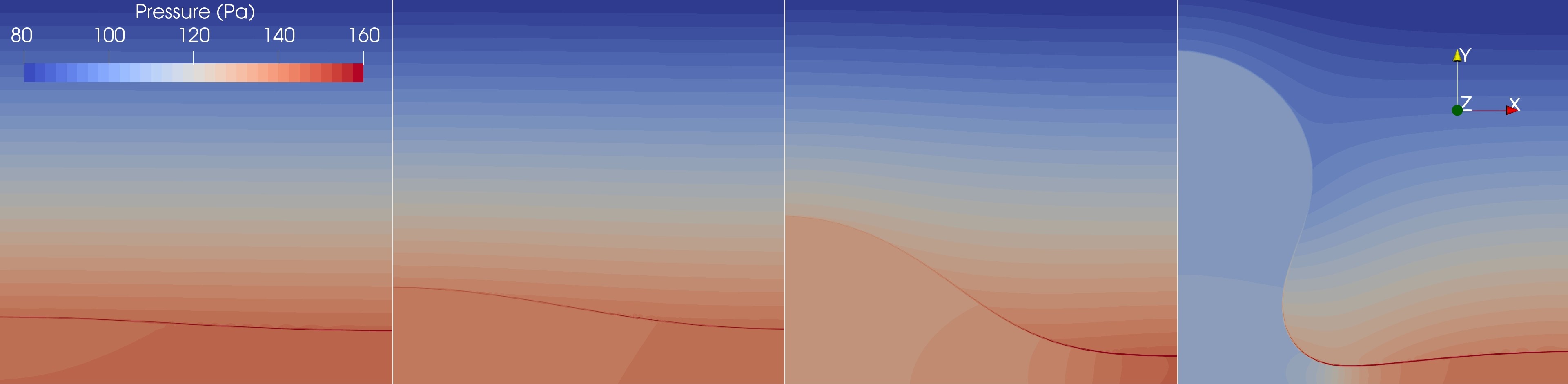}
  % \caption{} % No text so (a) appears
  \label{subfig:pressure_contours}
\end{subfigure}%
\\[-2.5ex]
\begin{subfigure}{1.0\textwidth}
  \centering
  \includegraphics[width=1.0\linewidth]{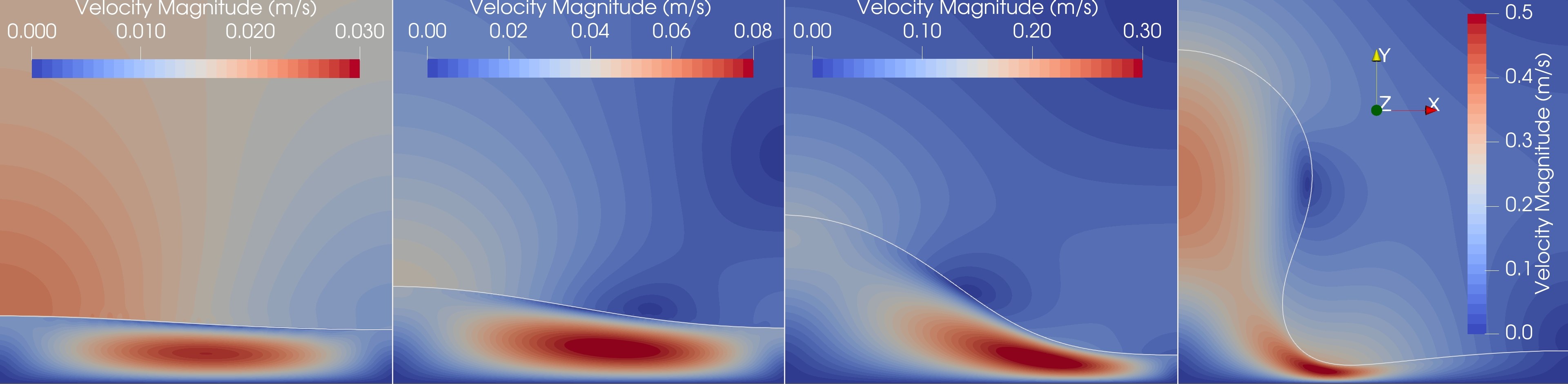}
  % \caption{} % No text so (a) appears
  \label{subfig:velocity_contours}
\end{subfigure}%
\\[-2.5ex]
\begin{subfigure}{1.0\textwidth}
  \centering
  \includegraphics[width=1.0\linewidth]{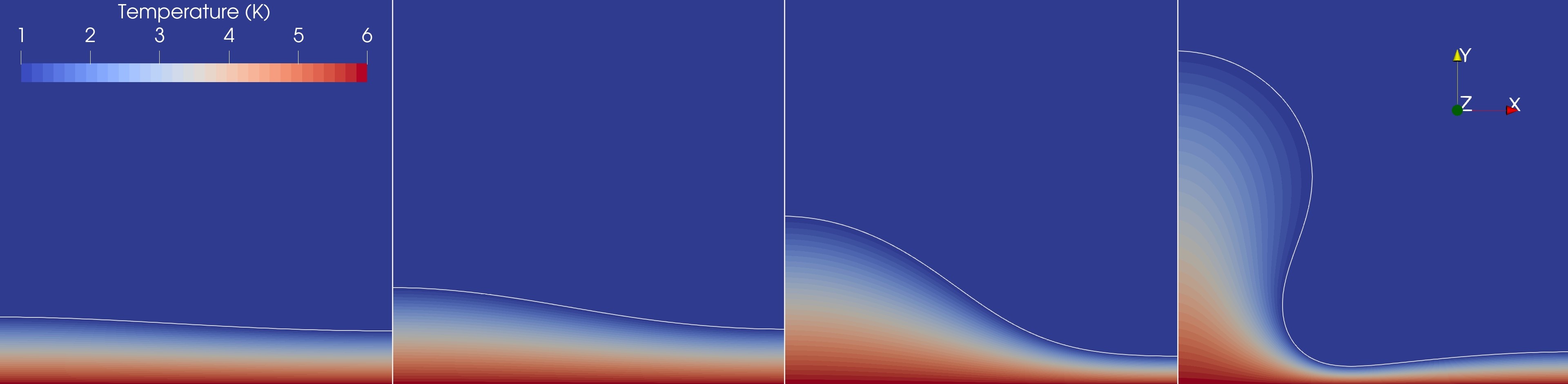}
  % \caption{} % No text so (a) appears
  \label{subfig:temperature_contours}
\end{subfigure}%
\\[-2.5ex]
\begin{subfigure}{1.0\textwidth}
  \centering
  \includegraphics[width=1.0\linewidth]{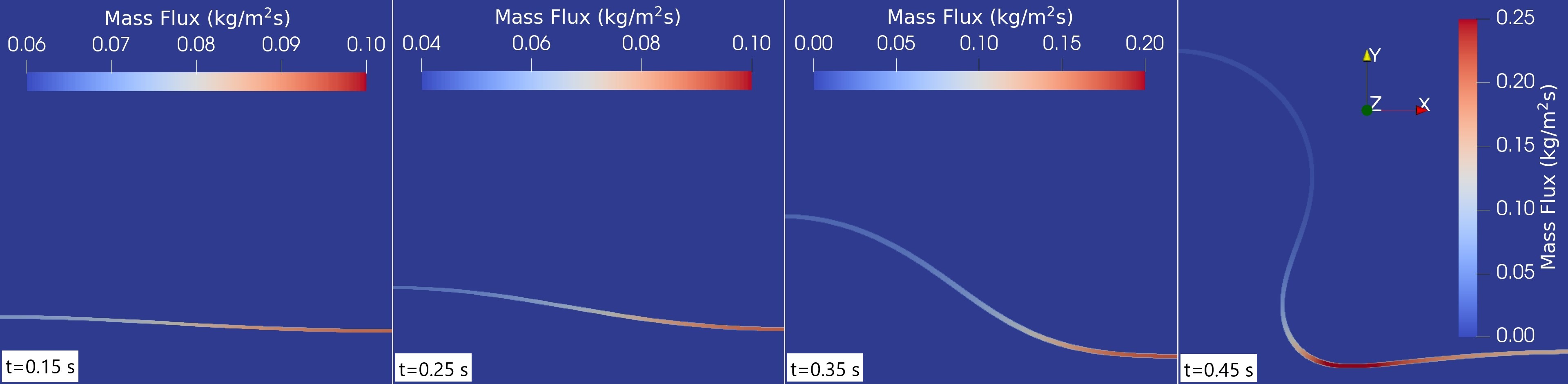}
  % \caption{} % No text so (a) appears
  \label{subfig:massflux_contours}
\end{subfigure}%
\caption{Contour plots of pressure (top), velocity magnitude (center top), temperature (center bottom) and mass flux across the interface (bottom) of the two-dimensional film boiling case obtained with mesh M6 at various times. The interface location is visualized by the iso-contour with \(C=0.5\).}
\label{fig:contours_film}
\end{figure}

Regardless of the exact onset, the numerical solution is very similar to previous literature. Thus, the differences must arise from the flow dynamics during the growth of the vapor film and the necking process leading to the bubble formation. Beyond the possibility that slight different initial conditions are affecting the time scales, the flow dynamics are affected by the specific numerical framework. These issues are not related to lack of convergence or to capturing the thermal layer as in Section \ref{subsec:bubblerising}. Figure \ref{fig:contours_film} presents various snapshots of the solution at \(t=0.15\) s, \(t=0.25\) s, \(t=0.35\) s and \(t=0.45\) s obtained with M6, focusing on the contours of pressure, velocity magnitude, temperature and mass flux across the interface. It becomes immediately clear that a pressure spike occurs across the interface, which scales with \(\dot{m}''\). This is a result of a momentum imbalance introduced by the discrete viscous term whereby ``numerical regularization" fails. That is, a numerical pressure jump is introduced due to the calculation of \(\nabla\mathbf{u}\) in the one-fluid framework, augmented by a high fluid viscosity. The thin region of the vapor film closer to the wall shows the largest momentum imbalance, while the error is much less significant along the bubble's edge due to vapor cooling and the decrease in \(\dot{m}''\). This regularization failure occurs independently of the type of mean used to average the viscosity, i.e., arithmetic vs. harmonic. \par

\begin{figure}
\centering
\includegraphics[width=0.45\linewidth]{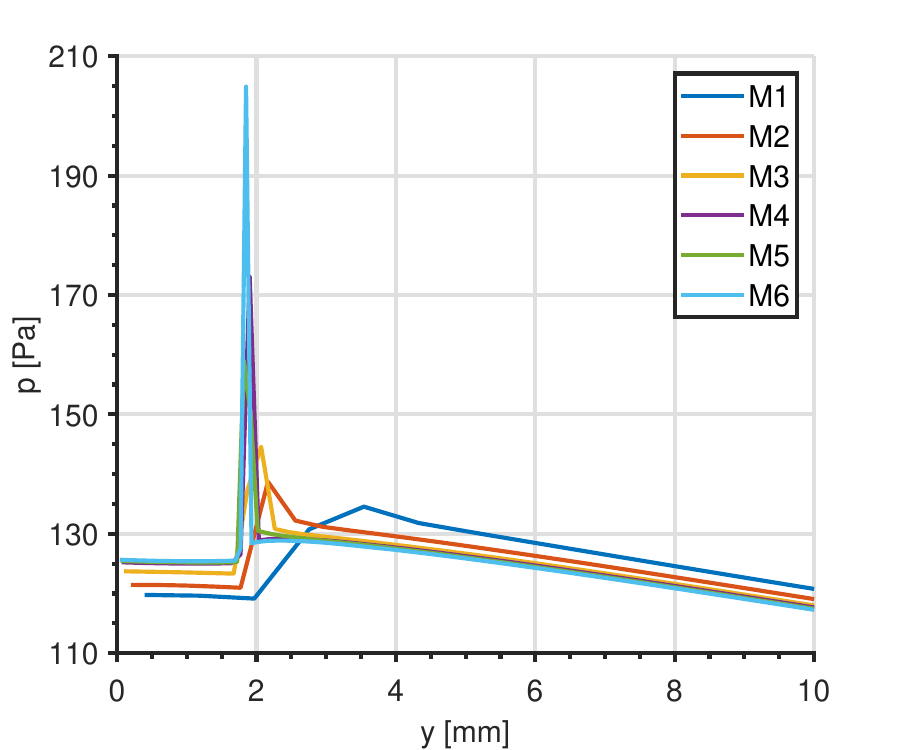}
\caption{Pressure field extracted along \(x=17\) mm at \(t=0.45\) s in the two-dimensional film boiling case. The impact of mesh resolution on the pressure spike across the interface is shown.}
\label{fig:pressure_x0p017_t0p45}
\end{figure}

As depicted in Figure \ref{fig:contours_film}, the lateral pressure gradient that drives the vapor flow and bubble growth is a result of the varying film thickness and neck formation process. Because \(\dot{m}''\) varies along the interface, the error introduced by the imbalance also does, modifying the pressure jump along the interface non-uniformly. Thus, the dynamics are easily affected, such as the spurious currents around the interface, and could explain the slower bubble growth with the current formulation. Note that despite the pressure spike across the interface scales with \(\sim(\Delta x)^{-1}\), the solution seems to converge but to a wrong pressure field. This is visualized in Figure \ref{fig:pressure_x0p017_t0p45}, showing the pressure extracted along \(x=17\) mm at \(t=0.45\) s. \par 

In contrast, previous works with one-fluid formulations might suffer less from this issue, which might explain the different solutions. Boyd and Ling \cite{2023_CaF_Boyd} work with a sharp VOF approach but shift the volume dilatation caused by phase change to the vapor cells adjacent to the interface. Although this is done to work with a divergence-free velocity field across the interface for the purpose of the VOF advection, it inadvertently mitigates the numerical pressure jump induced by the discrete viscous term since the cells with an ill-defined \(\nabla\mathbf{u}\) are only weighted by \(\mu_G\), which is two orders of magnitude lower than \(\mu_L\). Sun et al. \cite{2014_NHT_Sun} use the commercial solver Fluent and no explicit mention of the software version is provided, thus making it difficult to analyze. However, from the few details provided on the numerical method, it is not clear how the one-fluid velocity and the local volume dilatation are treated in the VOF advection and governing equations. In terms of the discretization of the viscous term, a sharp approach is used similar to the present work. Lastly, Guo et al. \cite{2011_NHT_Guo} also consider a divergence-free velocity for the interface advection but diffuse the interface discontinuity like in the Level-Set method, which results in a lower numerical jump. Thus, the growth of the bubble is also delayed to some extent with respect to Boyd and Ling \cite{2023_CaF_Boyd}. \par 

\begin{figure}
\centering
\includegraphics[width=0.45\linewidth]{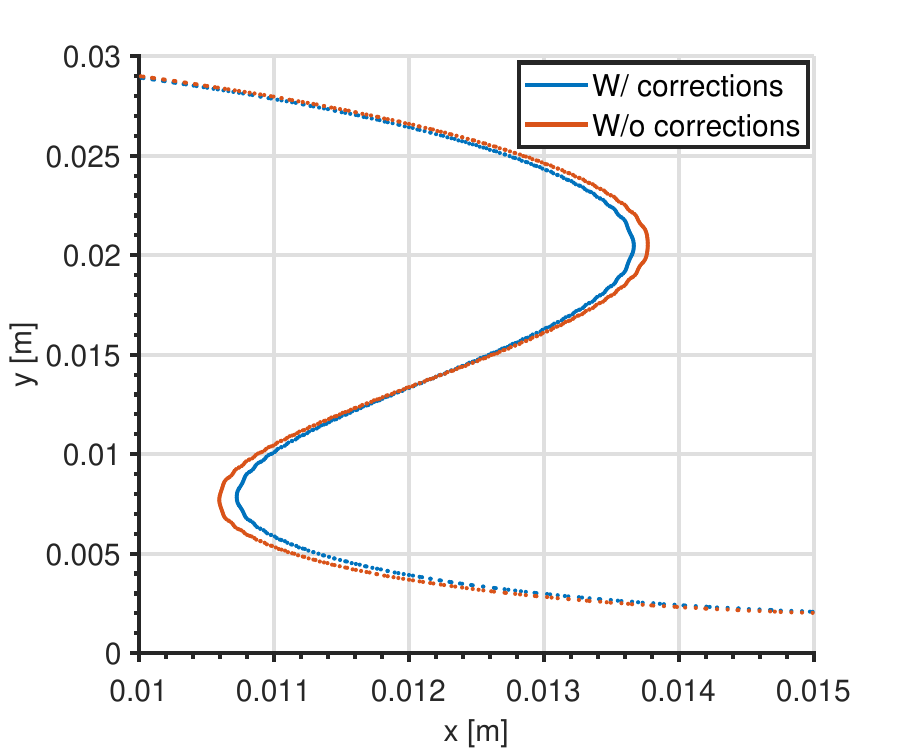}
\caption{Shape along the bubble edge at \(t=0.45\) s in the two-dimensional film boiling case obtained with mesh M6. The impact of considering the momentum balance corrections discussed in this work is shown. The interface location is visualized by the iso-contour with \(C=0.5\).}
\label{fig:shape_corrections_t0p45}
\end{figure}

In this configuration, the impact of the momentum balance corrections from the Stefan flow shift and the addition of \(\mathbf{f}_{NC}\) and \(\mathbf{f}_{\dot{m}''}\) are minimal given that the viscous imbalance dominates. However, some differences are still observed that suggest that corrections in the treatment of the viscous term to improve consistency may help bring the solution closer to recent reference data \cite{2023_CaF_Boyd} and reduce uncertainty. For example, Figure \ref{fig:shape_corrections_t0p45} shows a slight change in the shape of the bubble along its edge by considering or not the momentum balance corrections. The shape corresponds to the snapshot at \(t=0.45\) s obtained with M6. \par

\begin{figure}
\centering
\begin{subfigure}{0.5\textwidth}
  \centering
  \includegraphics[width=1.0\linewidth]{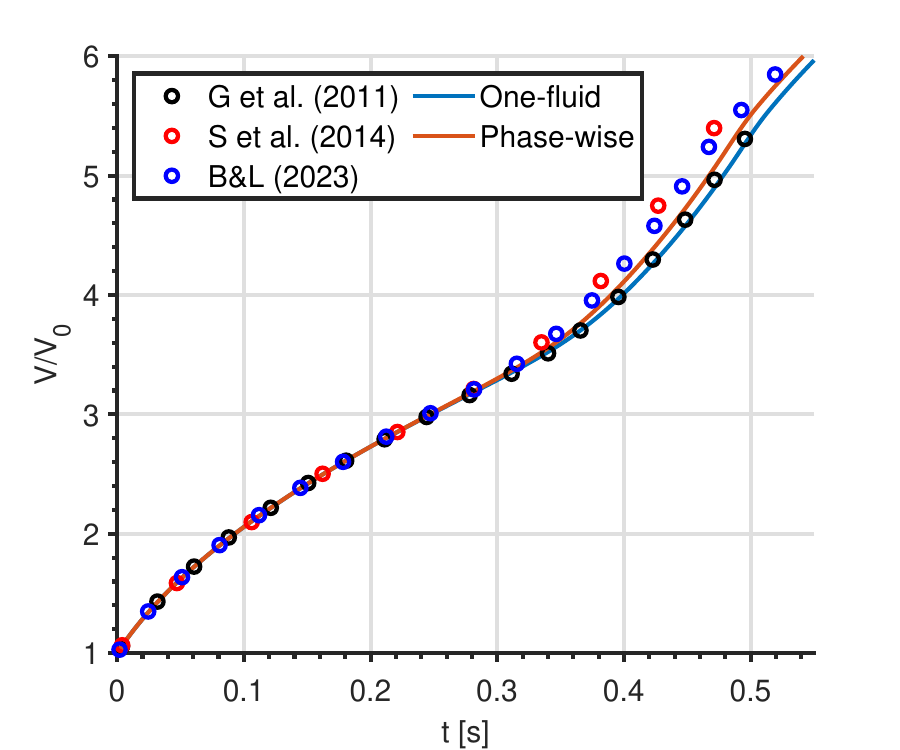}
  \caption{} % No text so (a) appears
  \label{subfig:void_increase_phasewiseM3}
\end{subfigure}%
\begin{subfigure}{0.5\textwidth}
  \centering
  \includegraphics[width=1.0\linewidth]{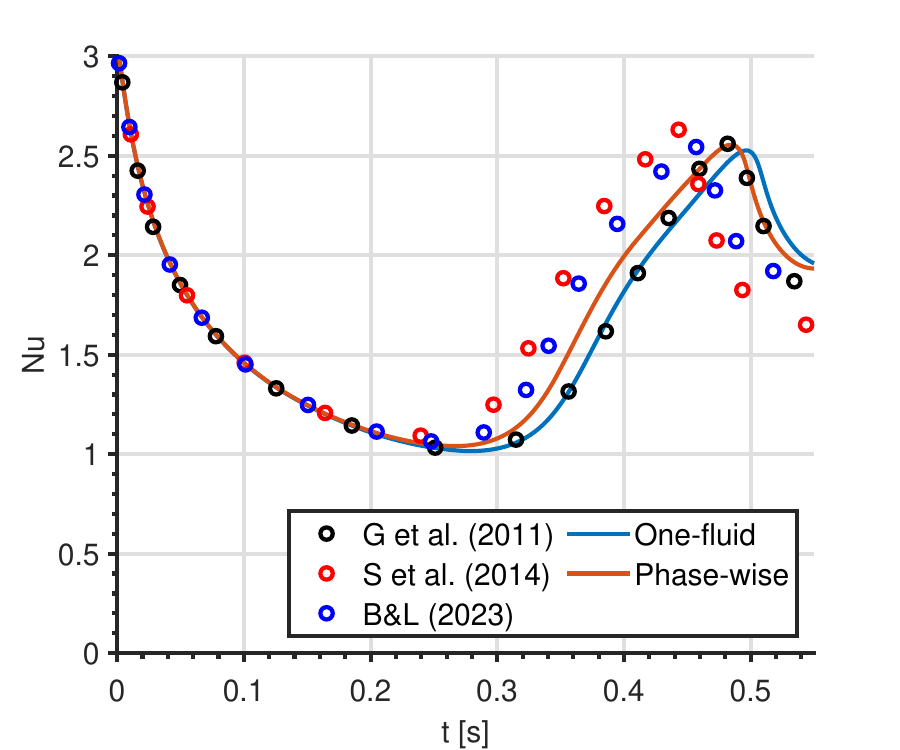}
  \caption{} % No text so (a) appears
  \label{subfig:Nu_average_phasewiseM3}
\end{subfigure}%
\caption{Effects of the one-fluid vs. phase-wise discretization of the viscous term on the increase in vapor volume and the average Nusselt number at the wall of the two-dimensional film boiling case compared against benchmark solutions from Guo et al. \cite{2011_NHT_Guo}, Sun et al. \cite{2014_NHT_Sun} and Boyd and Ling \cite{2023_CaF_Boyd}. Mesh M3 is used. (a) vapor volume ratio \(V/V_0\); (b) average Nusselt number.}
\label{fig:void_Nu_phasewiseM3}
\end{figure}

A preliminary attempt to correct the viscous imbalance follows a phase-wise discretization of \(\nabla\cdot\big(\mu\big[\nabla\mathbf{u}+(\nabla\mathbf{u})^T\big]\big)\). Here, phase-wise velocities are obtained by solving a Poisson equation in a narrow band around the interface to satisfy \(\nabla\cdot\mathbf{u}_l=0\) and \(\nabla\cdot\mathbf{u}_g=0\) \cite{2021_JCP_Malan}. Then, phase-wise values of velocity and viscosity are used at each face of the staggered velocity cell when discretizing the viscous term according to the predominant phase at the face. For that, the volume fraction in a cell centered around each face, \(C_\text{face}\), is calculated geometrically. Liquid values are used if \(C_\text{face}\geq 0.5\) and gas values are used if \(C_\text{face}<0.5\). This discretization mitigates the momentum imbalance caused by the viscous term and improves the agreement with the instability timescale shown in previous work that may not suffer from the issue, as discussed in previous lines. This is shown in Figure \ref{fig:void_Nu_phasewiseM3}, which plots again the evolution of the normalized vapor volume and the average Nu at the bottom wall obtained with the one-fluid and the phase-wise approaches using mesh M3. \par

The implications of the momentum imbalance due to the treatment of the viscous term in this problem are many. Not only the growth of the first bubble is delayed, but the release frequency of bubbles during the boiling process may be altered. Therefore, it becomes necessary to develop a comprehensive one-fluid formulation with consistent momentum balancing to ensure the solution is not compromised by the arbitrarities resulting from the choice of numerical framework. \par

\section{Conclusions}
\label{sec:concl}

This work has shown that a direct extension of multiphase codes based on the non-conservative one-fluid formulation of the momentum equation to simulations with phase change introduces a momentum imbalance. This originates from the treatment of convective terms \cite{2021_IJMF_Trujillo} and the implementation of classic discretization techniques, e.g., central differences, due to the local volume dilatation at the interface and the induced velocity jump. \par

A two-step correction has been proposed consisting of the addition of: (1) two body forces within the context of the CSF model \cite{1992_JCP_Brackbill}, namely \(\mathbf{f}_{NC}\) to correct or cancel the effects of an ill-defined convective term and \(\mathbf{f}_{\dot{m}''}\) to recover the exact momentum jump due to phase change \cite{2021_IJHMT_Dodd}, and (2) a shift of the Stefan flow after advecting the interface to obtain an intermediate velocity field \(\mathbf{u}^*\) before solving the momentum equation. The latter effectively addresses the momentum jump induced by the interface displacement in the non-conservative formulation \cite{2021_IJMF_Trujillo}, but imposes the assumption that the interface regression relative to the fluid is a quasi-steady process. Various tests have shown that these modifications successfully impose the correct momentum balance in viscous flows with low viscosity, recovering analytical solutions and reducing oscillations in the pressure field as the interface moves across grid cells. Moreover, the improvements are also evident in dynamic flows where the momentum balance corrections are seemingly negligible and have direct implications on the dynamic behavior of the flow. For example, bubble rise velocities may differ as shown in Section \ref{subsec:bubblerising}. However, the sharp discretization of the viscous term in highly viscous flows can induce pressure spikes across the interface and affect the momentum balance and dynamics of the flow, regardless of whether the conservative or non-conservative form of the momentum equation is used (see Section \ref{subsec:2dfilmboiling}). \par

It is in these scenarios that inconsistencies among different flow solvers are highlighted. The introduction of phase change is not trivial and requires a thorough examination of the impacts of a particular interface capturing scheme or discretization of the governing equations. Otherwise, different solvers will produce solutions similar in nature but with critical differences that may affect the particular analysis, e.g., the shift in the instability growth time scales seen in film boiling (see Section \ref{subsec:2dfilmboiling}). Therefore, future work will focus on developing a comprehensive momentum balance correction model for the one-fluid formulation in a sharp VOF framework to include the correct treatment of the pressure jump due to viscous stresses and reduce the solution uncertainty. \par

Additionally, this work has shown that the required resolution to achieve convergence can be very demanding in terms of cells per diameter of the bubble or droplet. This can be a result of having a very thin thermal boundary layer due to the properties of the fluid or it can be related to the resolution of surface tension. For practical simulations of multiple bubbles or droplets, realistic resolutions are much lower and require the development and integration of sub-models. Note that turbulence scales may be well-resolved to be considered a Direct Numerical Simulation framework, but under-resolution may still be an issue when it comes to interfacial phenomena. Thus, future work will consider heat and mass transfer sub-models for \(\dot{m}''\) and sub-models to accelerate, e.g., the bubble growth and necking formation in film boiling. Likewise, the momentum balance corrections might need some enhancement to converge faster to the expected pressure jump. Lastly, the proposed methodology must be extended to handle multiple objects, which includes the coupling with coalescence and breakup models to prevent, e.g., numerical coalescence. \par

\section*{Declaration of Interests}

The authors report no conflict of interest.

\section*{Acknowledgements}

The authors acknowledge the use of computational resources of DelftBlue supercomputer, provided by Delft High Performance Computing Centre (https://www.tudelft.nl/dhpc). The first author would also like to express his gratitude to Dr. Pedro Sim\~{o}es Costa and Dr. Giandomenico Lupo for the useful discussions during the elaboration of this work.

\newpage
\appendix

\section{Calculation of Phase-wise Velocities for Energy Transport}
\label{apn:A}

The phase-wise velocity \(\mathbf{u}_f\) is estimated by using an efficient fast marching method \cite{2014_JCP_McCaslin} to extrapolate the velocity of each phase across the interface. A constant extrapolation normal to the interface is performed. This strategy has been previously used in the literature with a PDE-based extrapolation instead \cite{2004_JCP_Aslam}, and has been shown to mitigate numerical instability issues that may arise from higher-order \(\mathbf{u}_f\) calculations in regions affected by spurious currents \cite{2019_JCP_Palmore}. Since \(\mathbf{n}_\Gamma\) is only defined in interface cells, a narrow band of cells around the interface is populated with a weighted average of the normal unit vectors calculated at interface cells. The weights are defined based on the projection of the distance vector between cells onto \(\mathbf{n}_\Gamma\) and the inverse of the squared distance to emphasize locality and directionality in the averaging process (see \cite{2023_CaF_Boyd} for more details on a similar implementation). \par 

The resulting phase-wise velocities are not divergence-free but provide a close approximation of the velocity at interface cells without the expense of solving two Poisson-type equations in the narrow band to satisfy \(\nabla\cdot\mathbf{u}_l=0\) and \(\nabla\cdot\mathbf{u}_g=0\), such as in \cite{2021_JCP_Malan}. Further, the extrapolation of \(\mathbf{u}_f\) is performed with cell-centered values which are later re-staggered to cell faces. Thus, the averages involved in this process effectively act as a spatial filter that smooths \(\mathbf{u}_f\) across the interface, improving stability. \par

\section{Interface Embedding in the Discretization of the Energy Transport}
\label{apn:B}

\setcounter{figure}{0}

\begin{figure}
\centering
\begin{subfigure}{0.5\textwidth}
  \centering
  \includegraphics[width=0.8\linewidth]{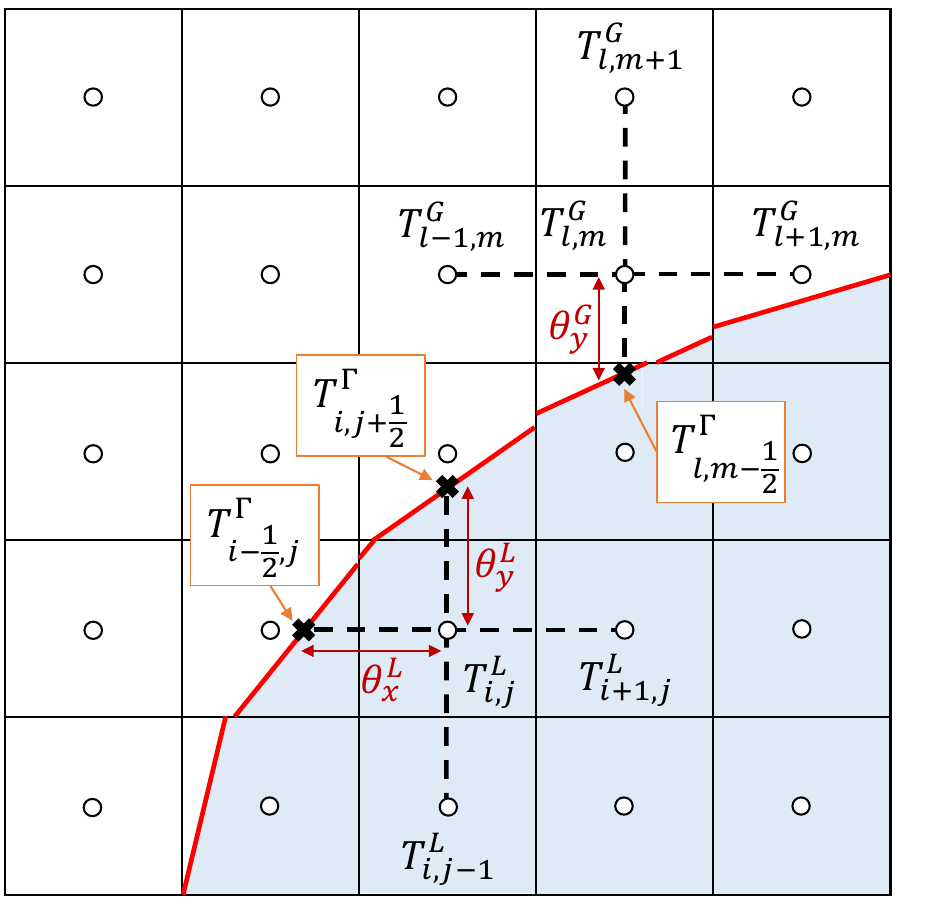}
  \caption{} % No text so (a) appears
  \label{subfig:interface_embedding}
\end{subfigure}%
\begin{subfigure}{0.5\textwidth}
  \centering
  \includegraphics[width=1.0\linewidth]{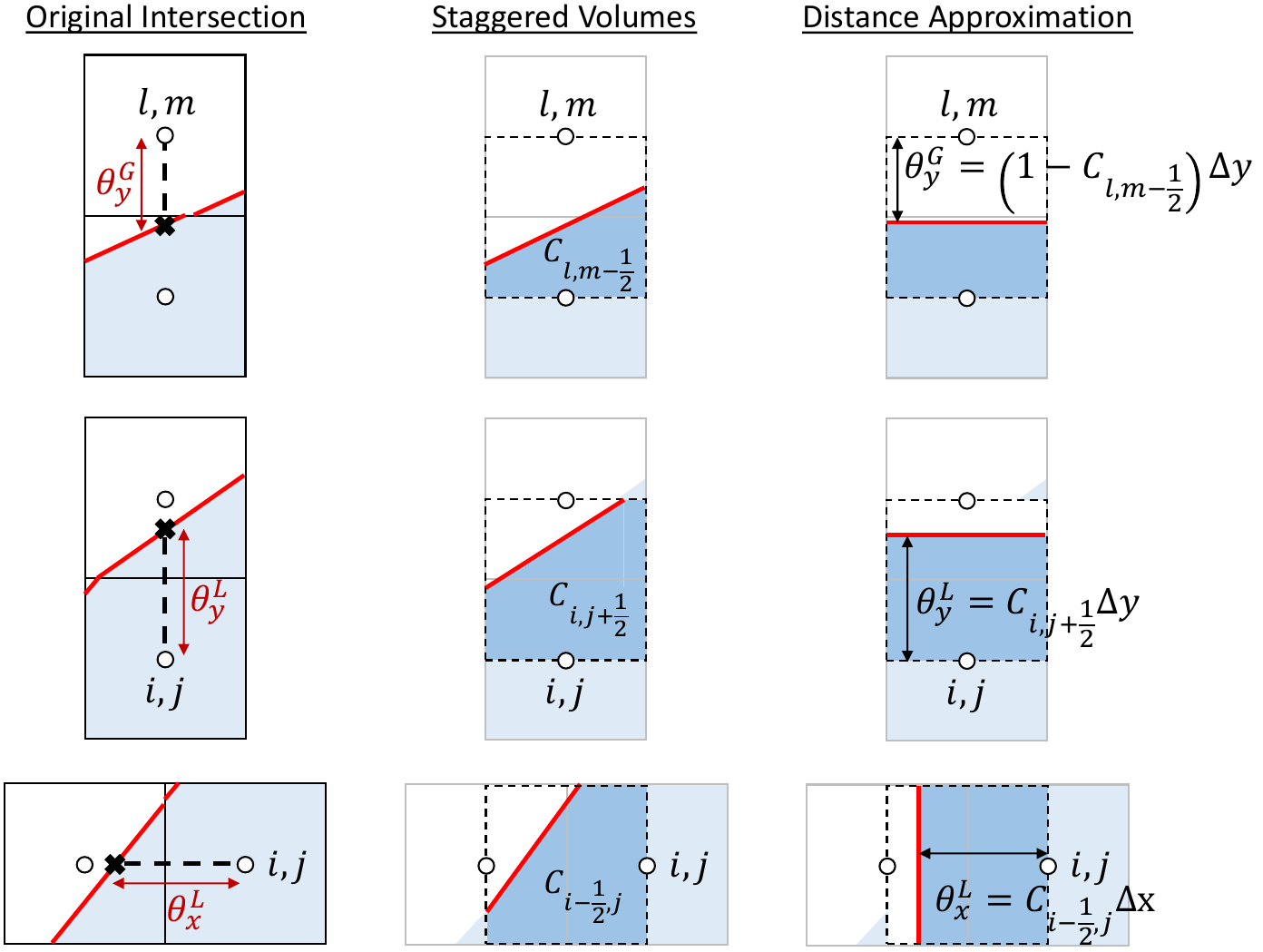}
  \caption{} % No text so (a) appears
  \label{subfig:distances_sketch}
\end{subfigure}%
\caption{Two-dimensional sketch showing (a) the stencils used in the discretization of the energy equation and (b) the steps to calculate distances for the interface embedding in the discretization of the phase-wise energy equation (from left to right: original intersections, calculation of staggered volume fractions, and distance approximation from the staggered volume fraction). The PLIC interface is represented by a solid red line and the liquid volume is colored in blue. The extension to three dimensions is straightforward.}
\label{fig:energy_discretization}
\end{figure}

A two-dimensional illustration of some possible interface intersections with the numerical stencils for the energy equation, Eq. (\ref{eqn:ene_pw}), is given in Figure \ref{subfig:interface_embedding}. The extension to three dimensions is straightforward. Rather than calculating the exact distance between a node and the intersection of the stencil with the interface, \(\theta_{y}^{G}\), \(\theta_{y}^{L}\) and \(\theta_{x}^{L}\) are estimated from staggered volume fractions obtained geometrically by evaluating the enclosed volumes of two adjacent cells overlapping with the staggered cell \cite{2017_PhD_Dodd,2019_JCP_Palmore,2022_PoF_Poblador}. For the gas cell \((l,m)\) in Figure \ref{subfig:interface_embedding}, the distance to the interface is given by \(\theta_{y}^{G}\approx(1-C_{l,m-\frac{1}{2}})\Delta y\) and, for the liquid cell \((i,j)\), the distances are given by \(\theta_{y}^{L}\approx C_{i,j+\frac{1}{2}}\Delta y\) and \(\theta_{x}^{L}\approx C_{i-\frac{1}{2},j}\Delta x\). The steps to approximate \(\theta_{y}^{G}\), \(\theta_{y}^{L}\) and \(\theta_{x}^{L}\) are visualized in Figure \ref{subfig:distances_sketch}. \par

If the components of the phase-wise velocity in cell \((i,j)\) are \(u_{i,j}>0\) and \(v_{i,j}<0\) in Figure \ref{subfig:interface_embedding}, the discretization of Eq. (\ref{eqn:energy_RT}) is given by

\begin{equation}
\label{eqn:energy_RT_discr}
\begin{split}
    &\mathbf{u}_{f}^{n}\cdot\nabla T^n = u_{i,j} \frac{T_{i,j}^{L}-T_{i-\frac{1}{2},j}^{\Gamma}}{\theta_{x}^{L}} + v_{i,j} \frac{T_{i,j+\frac{1}{2}}^{\Gamma}-T_{i,j}^{L}}{\theta_{y}^{L}}\\
    &\nabla\cdot\big(k_{f}^{n}\nabla T^n\big) = \frac{1}{h_x}\Bigg(k_{i+\frac{1}{2},j}^{L}\frac{T_{i+1,j}^{L}-T_{i,j}^{L}}{\Delta x} - k_{i-\frac{1}{2},j}^{L}\frac{T_{i,j}^{L}-T_{i-\frac{1}{2},j}^{\Gamma}}{\theta_{x}^{L}}\Bigg) + \frac{1}{h_y}\Bigg(k_{i,j+\frac{1}{2}}^{L}\frac{T_{i,j+\frac{1}{2}}^{\Gamma}-T_{i,j}^{L}}{\theta_{y}^{L}} - k_{i,j-\frac{1}{2}}^{L}\frac{T_{i,j}^{L}-T_{i,j-1}^{L}}{\Delta y}\Bigg)\\
\end{split}
\end{equation}

\noindent
where upwinding is used in the convective term and \(h_x=0.5\Delta x + \min(0.5\Delta x,\theta_{x}^{L})\) and \(h_y=0.5\Delta y + \min(0.5\Delta y,\theta_{y}^{L})\). Note \(h_x=\Delta x\) and \(h_y=\Delta y\) in the given example. Fluid properties and temperatures belong to \(t^n\). Although the interface temperature and properties are constant for the problems considered in this work, a generalized averaging of these variables is proposed for completeness, e.g., \(k_{i+\frac{1}{2},j}^{L}=\tfrac{1}{2}\big(k_{i+1,j}^{L}+k_{i,j}^{L}\big)\) and \(k_{i-\frac{1}{2},j}^{L}=\tfrac{1}{2}\big(k_{i,j}^{L}+k_{i-\frac{1}{2},j}^{\Gamma}\big)\) where \(k_{i-\frac{1}{2},j}^{\Gamma}\) is obtained from averaging interface values at the cell face \cite{2014_JCP_Dodd}, i.e., the thermal conductivities on the liquid side of the interface. \(T_{i-\frac{1}{2},j}^{\Gamma}\) and \(T_{i,j+\frac{1}{2}}^{\Gamma}\) are calculated analogously. This approach aims to use a characteristic value of the interface properties belonging to the PLIC reconstructions involved in the calculation of, e.g., \(\theta_{y}^{L}\) and \(\theta_{x}^{L}\). \par 

Singularities may arise when the interface gets too close to a cell node, e.g., \(\theta_{x}^{L}\rightarrow 0\) or \(\theta_{y}^{L}\rightarrow 0\). Instead of re-defining the numerical stencil, the solution is stabilized by considering a region of influence of the interface where, e.g., \(T_{i,j}^{n+1}=T_{i,j}^{\Gamma}\) if \(\theta_{x}^{L}<\varepsilon\Delta x\) or \(\theta_{y}^{L}<\varepsilon\Delta y\), with \(\varepsilon=0.01\) in this work. A smaller \(\varepsilon\) value may be considered to the detriment of the CFL conditions presented in \ref{apn:C}, requiring smaller time steps to maintain a bounded temperature field upon numerical integration. Further, the interface temperature is also assigned to any cell that ``changes phase" from \(t^n\) to \(t^{n+1}\), e.g., \(C_{i,j}^{n}<0.5\) and \(C_{i,j}^{n+1}\geq 0.5\). \par 

\section{Time Step Calculation}
\label{apn:C}

The time step \(\Delta t\) is determined by a CFL condition for multiphase flows \cite{2000_JSC_Kang}, which has been implemented in other works using explicit solvers \cite{2018_JCP_Anumolu,2018_IJMF_Duret,2020_JCP_Scapin,2022_PoF_Poblador}. \par

A time step for the momentum equation is defined as

\begin{equation}
\label{eqn:timestep1}
\Delta t_\mathbf{u} = \frac{2}{\delta t_\mathbf{u}+\delta t_\mu+\sqrt{(\delta t_\mathbf{u}+\delta t_\mu)^2+4\delta t_\mathbf{g}^2+4\delta t_\sigma^2}}
\end{equation}

\noindent
where

\begin{equation}
\label{eqn:timestep2}
\begin{split}
    &\delta t_\mathbf{u} = \frac{|u|_\text{max}}{\Delta x}+ \frac{|v|_\text{max}}{\Delta y}+ \frac{|w|_\text{max}}{\Delta z}\\
    &\delta t_\mu = 2\bigg(\frac{1}{\Delta x^2}+\frac{1}{\Delta y^2}+\frac{1}{\Delta z^2}\bigg)\bigg(\frac{\mu}{\rho}\bigg)_\text{max}\\
    &\delta t_\mathbf{g} = \sqrt{\frac{||\mathbf{g}||}{\Delta z}}\\
    &\delta t_\sigma = \sqrt{\frac{(\sigma|\kappa|)_\text{max}}{\text{min}(\rho_G,\rho_L)\text{min}(\Delta x^2,\Delta y^2,\Delta z^2)}}
\end{split}
\end{equation}

Next, a time step for the energy equation is given by

\begin{equation}
\label{eqn:timestep3}
\Delta t_T = \frac{\text{min}(\Delta x_{e}^{2},\Delta y_{e}^{2},\Delta z_{e}^{2})}{2\alpha_\text{max}}
\end{equation}

\noindent
with \(\alpha=k/(\rho c_p)\) being the thermal diffusivity, and \(\Delta x_e\), \(\Delta y_e\) and \(\Delta z_e\) being effective grid spacings based on the numerical stencil used to calculate the diffusion term in Eq. (\ref{eqn:energy_RT}) with the interface embedding. In the example provided in Eq. (\ref{eqn:energy_RT_discr}), \(\Delta x_e=\tfrac{1}{2}(\theta_{x}^{L}+\Delta x)\) and \(\Delta y_e=\tfrac{1}{2}(\theta_{y}^{L}+\Delta y)\). A similar time step could be defined to account for diffusion of species in multi-component systems. \par

Additional time-step restrictions are given by the interface displacement due to phase change, \(\Delta t_{\dot{m}''}\), and the limitation of geometrical errors during advection of \(C\) with split-advection solvers, \(\Delta t_C\). These are given by limiting the maximum shift of the interface plane under phase change, \(\Delta \text{d}_\text{max}\), as

\begin{equation}
\label{eqn:timestep4}
\Delta t_{\dot{m}''} = \Delta \text{d}_\text{max}\frac{\rho_L}{\dot{m}''}
\end{equation}

\noindent
and the advection of \(C\) as

\begin{equation}
\label{eqn:timestep5}
\Delta t_C = \theta_2\text{min}\bigg(\frac{\Delta x}{|u_L|_\text{max}},\frac{\Delta y}{|v_L|_\text{max}},\frac{\Delta z}{|w_L|_\text{max}}\bigg)
\end{equation}

\noindent
using the non-zero components of \(\mathbf{u}_L\). To be consistent with the advection of \(C\) described in Section \ref{subsec:intcap}, Eq. (\ref{eqn:timestep4}) uses \(\rho_L\) for droplet-laden flows and \(\rho_G\) for bubbly flows, and Eq. (\ref{eqn:timestep5}) uses \(\mathbf{u}_L\) or \(\mathbf{u}_G\) accordingly. \par 

Lastly, the time step is obtained from

\begin{equation}
\label{eqn:timestep6}
\Delta t = \text{min}\bigg(\theta_1\text{min}(\Delta t_\mathbf{u},\Delta t_T),\Delta t_{\dot{m}''},\Delta t_C\bigg)
\end{equation}

The coefficients for the time step calculations are given as follows. \(\theta_1\) is the classic CFL parameter that varies among solvers. In this work, a value of \(\theta_1\) between 0.1 and 0.2 has been shown to provide a stable and consistent numerical integration when \(\Delta t_\mathbf{u}\) or \(\Delta t_T\) dominate. However, other works in the literature report \(\theta_1\) values in the range of 0.35 to 0.5 \cite{2018_JCP_Anumolu,2020_JCP_Scapin}. Then, \(\theta_2=0.01\) following the geometrical errors under advection reported in \cite{2014_CaF_Baraldi}. \(\Delta \text{d}_\text{max}\) can be defined similarly, but it set to \(\Delta \text{d}_\text{max}=0.001\Delta x\) for the purpose of the validation of the code. Note that one can express \(\Delta \text{d}_\text{max}\equiv\theta_3 \Delta x\) for a uniform mesh; thus, the value of \(\theta_3\) can be adjusted (see Section \ref{subsubsec:staticdrop_error} for more details). \par

\newpage
\typeout{}
\bibliography{journal_bib}

\end{document}